\documentclass[12pt]{article}
\pdfoutput=1

\usepackage{amssymb}
\usepackage{amsmath}
\usepackage{bbold}
\usepackage{color}
\usepackage{colordvi}
\usepackage{fancybox}
\usepackage[footnotesize]{caption2}
\usepackage{graphicx}
\usepackage[center,footnotesize,hang]{subfigure}
\usepackage{bbm}
\usepackage{multirow}
\usepackage{array}
\usepackage{arydshln}
\usepackage{ulem}
\usepackage{enumitem}  %  control item indent
\newcommand{\PreserveBackslash}[1]{\let\temp=\\#1\let\\=\temp}
\newcolumntype{C}[1]{>{\PreserveBackslash\centering}p{#1}}
\newcolumntype{R}[1]{>{\PreserveBackslash\raggedleft}p{#1}}
\newcolumntype{L}[1]{>{\PreserveBackslash\raggedright}p{#1}}
\newcommand{\cleqn}{\setcounter{equation}{0}}
\allowdisplaybreaks \allowdisplaybreaks[2]

\begin{document}

\title{\textbf{Lepton Mixing Parameters from $\Delta(48)$ Family Symmetry and Generalised CP}}
\date{}

\author{\\[1mm]Gui-Jun Ding$^{a}$\footnote{E-mail: {\tt dinggj@ustc.edu.cn}}~,~Ye-Ling Zhou$^b$\footnote{E-mail: {\tt zhouyeling@ihep.ac.cn}}\\ \\
{\it\small $^a$ Department of Modern Physics,}\\
{\it\small University of Science and Technology of China, Hefei, Anhui 230026, China}\\[4mm]
{\it\small $^b$ Institute of High Energy Physics, Chinese Academy of Sciences, Beijing 100049, China}
}
\maketitle

\begin{abstract}

We provide a systematic and thorough exploration of the $\Delta(48)$ family symmetry and the consistent generalised CP symmetry. A model-independent analysis of the achievable lepton flavor mixing is performed by considering all the possible remnant symmetries in the neutrino and the charged lepton sectors. We find a new interesting mixing pattern in which both lepton mixing angles and CP phases are nontrivial functions of a single parameter $\theta$. The value of $\theta$ can be fixed by the measured reactor mixing angle $\theta_{13}$, and the excellent agreement with the present data can be achieved. A supersymmetric model based on $\Delta(48)$ family symmetry and generalised CP symmetry is constructed, and this new mixing pattern is exactly reproduced.

\end{abstract}
\thispagestyle{empty}
\vfill
%\tableofcontents
\newpage

\section{Introduction}

\cleqn

It is well-known that CP (CP is a combination of charge conjugation symmetry and parity symmetry) is not an exact symmetry of the nature. CP violation in the quark sector has been firmly established in oscillations and decays of $K$ and $B$ mesons~\cite{pdg}. In the lepton sector, the precise measurement of the reactor mixing angle $\theta_{13}$~\cite{An:2012eh} opens the door to measure the leptonic CP violation. Measurement of the Dirac CP-violating phase $\delta_{\rm{CP}}$ has become one of the primary physical goals of the next-generation neutrino oscillation experiments. The origin of CP violation is a longstanding fundamental question in particle physics. In the Standard Model, violation of CP occurs in the
flavor sector. It is conceivable that promoting CP to a symmetry at high energies which is then broken allows to impose constraints on the neutrino and charged lepton mass matrices~\cite{Ecker:1981wv,Grimus:1995zi}.

In the past years, non-abelian discrete groups have been widely used to explain the structure of lepton mixing angles,  please see Refs.~\cite{Altarelli:2010gt} for recent reviews. It seems natural to combine the discrete family symmetry with the CP symmetry to predict both lepton mixing angles and CP phases simultaneously. However, the interplay between family and CP symmetries should be carefully treated~\cite{Feruglio:2012cw,Holthausen:2012dk,Chen:2014tpa}. In the presence of a family symmetry, in many cases it is impossible to define CP in the naive way, i.e., $\phi\rightarrow\phi^*$, but rather a nontrivial transformation in flavor space is needed~\cite{Feruglio:2012cw,Holthausen:2012dk}. A typical example is the so-called  $\mu-\tau$ reflection symmetry ~\cite{Harrison:2002kp,Grimus:2003yn,Farzan:2006vj}, which interchanges a muon (tau) neutrino with a tau (muon) antineutrino in the charged lepton mass basis. In order to consistently define CP transformations in the context of non-abelian discrete family groups, certain consistency conditions must be satisfied~\cite{Ecker:1981wv,Grimus:1995zi,Feruglio:2012cw,Holthausen:2012dk}. In fact, it has been established that all generalized CP symmetries are outer automorphisms of the symmetry group~\cite{Grimus:1995zi,Holthausen:2012dk}.

Combining family symmetry with generalised CP symmetry, one can obtain lepton mixing angles compatible with the current experimental data and in the meantime predict CP-violating phases.
This interesting idea has been explored to some extent in the past. The generalised CP symmetry has been implemented within $S_4$~\cite{Ding:2013hpa,Feruglio:2013hia,Luhn:2013lkn,Li:2013jya} , $A_4$~\cite{Ding:2013bpa}, and $T^{\prime}$~\cite{Girardi:2013sza} family symmetries, and some concrete models have also been constructed. In these models, the full symmetry is generally spontaneously broken down to a cyclic subgroup in the charged lepton sector and to $Z_2\times H^\nu_{\rm{CP}}$ in the neutrino sector.
The surviving symmetries constrain the neutrino mass matrix and charged lepton mass matrix, leading to predictions for CP-violating phases as well as constraints on mixing angles. Typically, the Dirac CP-violating phase is predicted to take simple values such as $0$, $\pi$, or $\pm\pi/2$. A comprehensive analysis of the generalised CP within $\Delta(96)$ family symmetry is recently performed in the semi-direct approach~\cite{Ding:2014ssa}, and some new interesting mixing patterns are found. The generalised CP has also been investigated for an infinite series of finite groups $\Delta(6n^2)$~\cite{King:2014rwa}, where the full Klein symmetry is assumed to be preserved in the neutrino sector such that the Dirac CP phase can only be $0$ or $\pi$. There are also other approaches in which family symmetries and CP violation appear together~\cite{Mohapatra:2012tb,Branco:1983tn,Chen:2009gf,Antusch:2011sx}.

In our recent paper~\cite{Ding:2013nsa}, we propose to use $\Delta(48)$ as the family symmetry and extend it to include the generalized CP symmetry. As we shall see later, the group $\Delta(48)$, which has been overlooked in the literature, has a large automorphism group of order 384. Hence $\Delta(48)$ provides us more choices for generalised CP transformations than some popular family symmetries $A_4$, $S_4$, etc. As a consequence, we find a new interesting mixing pattern, which is denoted as patter D in~\cite{Ding:2013nsa}, is admissible by neutrino oscillation experiments. This mixing texture can fit the experimental data quit well and predict the Dirac-type CP violation neither vanishing nor maximal, i.e., $\delta_{\rm{CP}} \neq 0$, $\pi$, $\pm \pi/2$.

This paper is devoted to a comprehensive analysis of lepton flavor mixing within the context of the $\Delta(48)$ family symmetry combined with generalized CP symmetry. In section~\ref{sec:consistent_GCP}, we discuss the structure of the automorphism group of $\Delta(48)$ and present the generalized CP transformations consistent with the $\Delta(48)$ family symmetry. In section~\ref{sec:model_independent_analysis}, we perform a systematic scan of lepton mixing within the framework of $\Delta(48)\rtimes H_{\rm{CP}}$ by analyzing all possible residual symmetries in the neutrino and the charged lepton sectors. We find 10 different cases, and subsequently we investigate the corresponding phenomenological predictions for the lepton mixing parameters which depend on one single free parameter $\theta$. In particular, a new interesting mixing pattern (pattern D in Ref.~\cite{Ding:2013nsa}) is found. Both mixing angles and CP phases are nontrivial functions of $\theta$, and three leptonic mixing angles in the experimentally preferred range can be achieved for certain values of the parameter $\theta$. In section~\ref{sec:model_construction}, we construct a supersymmetric model with both $\Delta(48)$ family symmetry and generalised CP symmetry. This model gives rise to the new mixing pattern we mentioned above, and its phenomenological predictions are discussed. In addition, the required vacuum alignment is justified. We summarize the main results of our paper in section~\ref{sec:conclusions}.  Some details of the group theory of $\Delta(48)$ are contained in Appendix~\ref{sec:appendix_A_group_theory}, and the Clebsch-Gordan coefficients in the chosen basis are reported. Appendix~\ref{sec:appendix_B} lists the vacuum alignment invariant under the remnant family and CP symmetries in cases II, IV, VI and VIII.

%%%%%%%%%%%%%%%%%%%%%%%%%%%%%%%%
%%%%%%%%%%%%%%%%%%%%%%%%%%%%%%%%
\section{\label{sec:consistent_GCP}Generalized CP transformations consistent with $\Delta(48)$ }

\cleqn

In this section, we are going to discuss all the possible generalized CP transformations which are consistent with $\Delta(48)$. Before that, we
give a brief review of the consistent definition of the generalized CP transformation in the context of discrete family symmetries. It is nontrivial to impose the generalized CP symmetry on a theory in the presence of a family symmetry $G_f$. For a field multiplet
\begin{equation}
\Phi=\left(\phi_R, \phi_P, \phi^{*}_P,\phi_C, \phi^{*}_C\right)^{T}\,,
\end{equation}
where the subscript $R$, $P$ and $C$ denote that the fields $\phi$ are in the real, pseudo-real and complex representations of $G_f$, respectively. Under the action of the family symmetry, the field $\Phi$ transforms as
\begin{equation}
\Phi\stackrel{g}{\longrightarrow}\rho(g)\Phi,\qquad g\in G_f\,,
\end{equation}
where $\rho$ is a representation of the group element $g$. Depending on the component fields $\phi_R$, $\phi_P$, $\phi^{*}_P$, $\phi_C$ and $\phi^{*}_C$, the representation $\rho$ is generally reducible with
\begin{equation}
\rho(g)=\left(\begin{array}{ccccc}
\rho_R(g) &&&&\\
&\rho_P(g)&&&\\
&&\rho^{*}_P(g)&&\\
&&&\rho_C(g)&\\
&&&&\rho^{*}_C(g)\\
\end{array}\right)\;.
\end{equation}
The generalized CP symmetry acts on $\phi$ as
\begin{equation}
\Phi(x)\stackrel{CP}{\longrightarrow}X\Phi^{*}(x_P)\,,
\end{equation}
where $x_P=(t,-\mathbf{x})$. Here we have omitted the action of CP on spinor indices for the case that $\phi$ is a spinor. $X$ is a unitary matrix which represents a generalized CP transformation, and it is referred to as the generalized CP transformation matrix. In contrast with conventional CP transformation, $X$ is not necessarily block-diagonal, and it generally interchanges different representations. In order to combine the generalized CP symmetry and the family symmetry, one has to satisfy the so-called consistency equation~\cite{Grimus:1995zi,Holthausen:2012dk,Feruglio:2012cw}
\begin{equation}
\label{eq:consistence_equation}X\rho^{*}(g)X^{-1}=\rho(g'),\qquad g,g'\in G_f\,,
\end{equation}
which implies that the generalized CP transformation $X$ maps the group element $g$ into another element $g'$, and it is remarkable that the family group structure is preserved under this mapping~\cite{Holthausen:2012dk,Ding:2013hpa}.
Since $X$ is unitary and invertible, the kernel of this mapping is identity, and therefore the generalized CP transformation matrix $X$ corresponds to an automorphism of $G_f$.

Now we turn to the concerned $\Delta(48)$ family symmetry group. From the multiplication rule listed in Eq.~\eqref{eq:group_rules}, we know that only the identity element commutes with all other elements of $\Delta(48)$. Hence $\Delta(48)$ has a trivial center $\mathrm{Z}(\Delta(48))\cong Z_1$, and therefore the inner automorphism group $\mathrm{Inn}(\Delta(48))$ is isomorphic to $\Delta(48)$ itself. The automorphism group of $\Delta(48)$ is quite involved, it is of order 384, and its structure can be summarized as follows:
\begin{equation}
\begin{array}{ll}
\mathrm{Z}(\Delta(48))\cong Z_1,& \qquad \mathrm{Aut}(\Delta(48))\cong\left(\left(\left(\left(Z_4\times Z_4\right)\rtimes Z_3\right)\rtimes Z_4\right)\rtimes Z_2\right),\\
\mathrm{Inn}(\Delta(48))\cong \Delta(48),& \qquad \mathrm{Out}(\Delta(48))\cong D_8\,,
\end{array}
\end{equation}
where $\mathrm{Aut}(\Delta(48))$ denotes the automorphism group of $\Delta(48)$, and $\mathrm{Out}(\Delta(48))$ is the outer automorphism group of $\Delta(48)$ with $\mathrm{Out}(\Delta(48))\equiv\mathrm{Aut}(\Delta(48))/\mathrm{Inn}(\Delta(48))$. We find that $\mathrm{Out}(\Delta(48))$ is isomorphic to the dihedral group of order eight, which is the group of all symmetries of the square, and it can be generated by two generators $u_1$ and $u_2$ with
\begin{equation}
\left\{\begin{array}{l}
a\stackrel{u_1}{\longrightarrow}a^2\\
c\stackrel{u_1}{\longrightarrow}cd^2
\end{array}\right.,\qquad
\left\{\begin{array}{l}
a\stackrel{u_2}{\longrightarrow}a\\
c\stackrel{u_2}{\longrightarrow}cd^2
\end{array}\right.\,.
\end{equation}
It is straightforward to check that they satisfy the following relations
\begin{equation}
u^4_1=u^2_2=\left(u_1u_2\right)^2=id\,,
\end{equation}
where $id$ denotes the trivial map $id(g)=g,~\forall g\in \Delta(48)$. All the automorphisms of $\Delta(48)$ can be generated from the generators $u_1$, $u_2$ and inner automorphisms
\begin{equation}
\mathfrak{g}=\mathrm{conj}^{k}(a)\mathrm{conj}^{m}(c)\mathrm{conj}^{n}(d)u^{\mu}_1u^{\nu}_2,\quad \mathfrak{g}\in\mathrm{Aut(\Delta(48))}\,,
\end{equation}
where $k=0, 1, 2$, $m, n, \mu=0, 1, 2, 3$ and $\nu=0, 1$, $\mathrm{conj}(h)$ denotes the group conjugation with an element $h$, i.e. $\mathrm{conj(h)}: g\rightarrow hgh^{-1}$.

The outer automorphism $u_1$ acts on the irreducible representations of $\Delta(48)$ as
\begin{equation}
\label{eq:u_1_representation_trans}\mathbf{1^{\prime}}\stackrel{u_1}{\longleftrightarrow}\mathbf{1^{\prime\prime}},\quad \mathbf{3}\stackrel{u_1}{\longrightarrow}\mathbf{3^{\prime}}\stackrel{u_1}{\longrightarrow}\mathbf{\overline{3}}\stackrel{u_1}{\longrightarrow}\mathbf{\overline{3}^{\prime}}\stackrel{u_1}{\longrightarrow}\mathbf{3}, \quad \mathbf{\widetilde{3}}\stackrel{u_1}{\longrightarrow}\mathbf{\widetilde{3}}\,,
\end{equation}
where $\mathbf{3}\stackrel{u_1}{\longrightarrow}\mathbf{3^{\prime}}$ is to be read as $\rho_{\mathbf{3^{\prime}}}=\rho_{\mathbf{3}}\circ u_1$ etc. The outer automorphism $u_2$ acts as
\begin{equation}
\label{eq:u_2_representation_trans}\mathbf{1^{\prime}}\stackrel{u_2}{\longrightarrow}\mathbf{1^{\prime}},\quad \mathbf{1^{\prime\prime}}\stackrel{u_2}{\longrightarrow}\mathbf{1^{\prime\prime}},\quad \mathbf{3}\stackrel{u_2}{\longleftrightarrow}\mathbf{3^{\prime}},\quad \mathbf{\overline{3}}\stackrel{u_2}{\longleftrightarrow}\mathbf{\overline{3}^{\prime}},\quad \mathbf{\widetilde{3}}\stackrel{u_2}{\longrightarrow}\mathbf{\widetilde{3}}\,.
\end{equation}
A nontrivial CP transformation is in fact a representation of the corresponding automorphism in the sense of Eq.~\eqref{eq:consistence_equation}. For the one-dimensional representations, it can be easily fixed. Furthermore, Eq.~\eqref{eq:u_1_representation_trans} and Eq.~\eqref{eq:u_2_representation_trans} imply that the CP transformations for the automorphism $u_{1,2}$ can be defined consistently only if all the complex triplet representations $\mathbf{3}$, $\mathbf{\overline{3}}$, $\mathbf{3}^{\prime}$ and $\mathbf{\overline{3}^{\prime}}$ are present. Hence the generalized CP transformations $X(u_{1,2})$ corresponding to the generators $u_1$ and $u_2$ of the outer automorphism group act on the vector
\begin{equation}
\label{eq:vector_fields_phi}\Phi=\left(\begin{array}{c}
\phi_{\mathbf{3}}\\
\phi^{*}_{\mathbf{3}}\\
\phi_{\mathbf{3^{\prime}}}\\
\phi^{*}_{\mathbf{3^{\prime}}}
\end{array}\right)\,,
\end{equation}
and they are uniquely determined by the following consistency equations up to an overall phase,
{\small\begin{equation}
\left\{\begin{array}{l}
X(u_1)\rho^{*}(a)X^{-1}(u_1)=\rho(u_{1}(a))=\rho(a^2)\\
X(u_1)\rho^{*}(c)X^{-1}(u_1)=\rho(u_{1}(c))=\rho(cd^2)
\end{array}\right.,~~\left\{\begin{array}{l}
X(u_2)\rho^{*}(a)X^{-1}(u_2)=\rho(u_{2}(a))=\rho(a)\\
X(u_2)\rho^{*}(c)X^{-1}(u_2)=\rho(u_{2}(c))=\rho(cd^2)
\end{array}\right.\,.
\end{equation}}
The solutions to these equations are
\begin{equation}
\label{eq:CP_u1_u2}X(u_1)=\left(\begin{array}{cccc}
0 & 0 & 0 &  P_{23}  \\
0 & 0 & P_{23} & 0 \\
\mathbbm{1}_3 &  0  &  0  & 0 \\
0  &  \mathbbm{1}_3 & 0 & 0
\end{array}\right),\qquad X(u_2)=\left(\begin{array}{cccc}
0 &  0  & 0  &  \mathbbm{1}_3\\
0 &  0  &  \mathbbm{1}_3  & 0  \\
0 & \mathbbm{1}_3  &  0  & 0 \\
\mathbbm{1}_3  & 0  & 0  & 0
\end{array}\right)\,,
\end{equation}
up to an overall phase, where $\mathbbm{1}_3$ denotes a $3\times3$ unit matrix, and $P_{23}$ is a permutation matrix
\begin{equation}
P_{23}=\left(\begin{array}{ccc}
1 & ~0~ & 0 \\
0 & ~0~ & 1 \\
0 & ~1~ & 0
\end{array}\right)\,.
\end{equation}
For the remaining three-dimensional representation $\mathbf{\widetilde{3}}$, we find the associated CP transformation matrices
\begin{equation}
X_\mathbf{\widetilde{3}}(u_1)=\mathbbm{1}_3,\qquad X_\mathbf{\widetilde{3}}(u_2)=P_{23}\,,
\end{equation}
which represent the automorphism $u_{1,2}$ via $X_\mathbf{\widetilde{3}}(u_i)\rho^{*}_\mathbf{\widetilde{3}}(g)X^{-1}_\mathbf{\widetilde{3}}(u_i)=\rho_\mathbf{\widetilde{3}}(u_i(g))$ with $i=1, 2$. \\[-0.1in]

Moreover, given two automorphisms $\mathfrak{g_1}$ and $\mathfrak{g_2}$ of $\Delta(48)$ and the associated CP transformations $X(\mathfrak{g_1})$ and $X(\mathfrak{g_2})$, the product $\mathfrak{g_2}\mathfrak{g_1}$ is also an automorphism, and the corresponding generalized CP transformation matrix is given by~\cite{Holthausen:2012dk}\footnote{Since we have $\rho\left(\mathfrak{g_2}\left(\mathfrak{g_1}\left(g\right)\right)\right)=X(\mathfrak{g_2})\rho^{*}\left(\mathfrak{g_1}\left(g\right)\right)X^{-1}(\mathfrak{g_2})
=X(\mathfrak{g_2})W\rho\left(\mathfrak{g_1}\left(g\right)\right)W^{-1}X^{-1}(\mathfrak{g_2})=X(\mathfrak{g_2})WX(\mathfrak{g_1})\rho^{*}\left(g\right)X^{-1}(\mathfrak{g_1})W^{-1}X^{-1}(\mathfrak{g_2})$, therefore the CP transformation for the automorphism product $\mathfrak{g_2}\mathfrak{g_1}$ is $X(\mathfrak{g_2}\mathfrak{g_1})=X(\mathfrak{g_2})WX(\mathfrak{g_1})$. }
\begin{equation}
\label{eq:CP_product}X\left(\mathfrak{g_2}\mathfrak{g_1}\right)=X(\mathfrak{g_2})WX(\mathfrak{g_1})\,,
\end{equation}
where $W$ exchanges the complex conjugate components of the vector $\Phi$ with $\Phi^{*}=W\Phi$ which implies $\rho(g)=W\rho^{*}(g)W^{-1}$. For the reducible  $\Phi\sim\mathbf{3}\oplus\mathbf{\overline{3}}\oplus\mathbf{3^{\prime}}\oplus\mathbf{\overline{3}^{\prime}}$ shown in Eq.~\eqref{eq:vector_fields_phi}, the $W$ matrix is of the form
\begin{equation}
W=\left(\begin{array}{cccc}
0 &  \mathbbm{1}_3  &  0  &  0 \\
\mathbbm{1}_3   &  0  & 0  &  0 \\
0 & 0 &  0  & \mathbbm{1}_3 \\
0 &  0 & \mathbbm{1}_3 & 0
\end{array}\right)\,.
\end{equation}
Furthermore, the CP transformation for the inner automorphism $\mathrm{conj}(h)$ is given by
\begin{equation}
\label{eq:CP_innaut}X\left(\mathrm{conj}(h)\right)=\rho(h)W\,.
\end{equation}
As a result, all the generalized CP transformation matrix $X(\mathfrak{g})$ with $\mathfrak{g}\in\mathrm{Aut}\left(\Delta(48)\right)$ can be straightforwardly obtained with the help of the general relations in Eq.~\eqref{eq:CP_product} and Eq.~\eqref{eq:CP_innaut}. A complete classification of possible CP transformations which can be consistently implemented within $\Delta(48)$ family symmetry is achieved.

The 8 outer automorphisms generated by $u_1$ and $u_2$ lead to different CP transformations and should have distinct physical implications. It is remarkable to note that most of the generalized CP transformations (for example $X(u_1)$ and $X(u_2)$ in Eq.~\eqref{eq:CP_u1_u2}) only exchange different representations rather than map each irreducible representation into its complex conjugate representation as conventional CP. This comment is generally true for the generalized CP framework in particular the case that the order of the outer automorphism group is large than 2. In the present work, we minimally extend the $\Delta(48)$ family symmetry to include only those nontrivial CP transformations which map one irreducible representation into its complex conjugate, the corresponding outer automorphism should be of order 2, and we find that there are three such kinds of outer automorphisms: $\mathfrak{h_1}=u^2_1$, $\mathfrak{h_2}=u_1u_2$ and $\mathfrak{h_3}=u^3_1u_2$. For $\mathfrak{h_1}=u^2_1$, its actions on the $\Delta(48)$ generators $a$ and $c$ are
\begin{equation}
a\stackrel{\mathfrak{h_1}}{\longrightarrow}a,\qquad c\stackrel{\mathfrak{h_1}}{\longrightarrow}c^3\,.
\end{equation}
It interchanges $\mathbf{3}$($\mathbf{3^{\prime}}$) with $\mathbf{\overline{3}}$($\mathbf{\overline{3}^{\prime}}$), and it is represented by
\begin{equation}
X_{\mathbf{3}(\mathbf{\overline{3}})}(\mathfrak{h_1})=X_{\mathbf{3^{\prime}}(\mathbf{\overline{3}^{\prime}})}(\mathfrak{h_1})=P_{23}\,.
\end{equation}
For $\mathfrak{h_2}=u_1u_2$, the generators $a$ and $c$ transform as
\begin{equation}
a\stackrel{\mathfrak{h_2}}{\longrightarrow}a^2,\qquad c\stackrel{\mathfrak{h_2}}{\longrightarrow}c^3\,.
\end{equation}
$\mathfrak{h_2}$ maps $\mathbf{3^{\prime}}$ and $\mathbf{\overline{3}^{\prime}}$ into each other, and the corresponding CP transformation is given by
\begin{equation}
X_{\mathbf{3^{\prime}}(\mathbf{\overline{3}^{\prime}})}(\mathfrak{h_2})=\mathbbm{1}_3\,.
\end{equation}
Finally for $\mathfrak{h_3}=u^3_1u_2$, it acts on $a$ and $c$ as
\begin{equation}
a\stackrel{\mathfrak{h_3}}{\longrightarrow}a^2,\qquad c\stackrel{\mathfrak{h_3}}{\longrightarrow}c\,.
\end{equation}
The irreducible representations $\mathbf{3}$ and $\mathbf{\overline{3}}$ are exchanged under $\mathfrak{h_3}$, and the associated CP transformation is
\begin{equation}
X_{\mathbf{3}(\mathbf{\overline{3}})}(\mathfrak{h_3})=\mathbbm{1}_3\,.
\end{equation}

\section{\label{sec:model_independent_analysis}Model-independent analysis of lepton mixing patterns within $\Delta(48)\rtimes H_{\rm{CP}}$ }

\subsection{The basic framework}
\cleqn

\begin{figure}[t!]
\begin{center}
\includegraphics[width=0.38\textwidth]{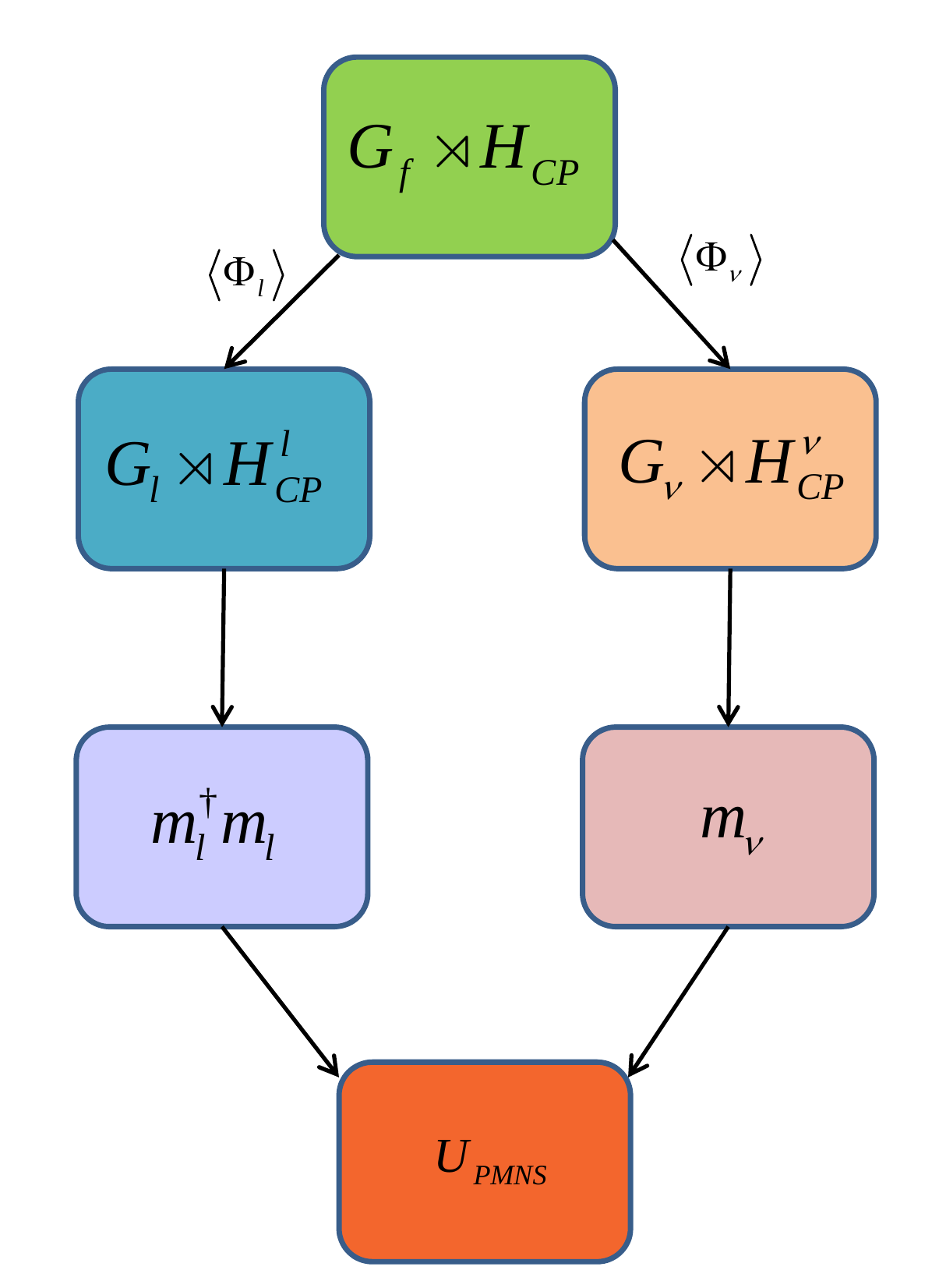} \caption{\label{fig:procedure_cartoon} Leptonic flavor mixing from the mismatch of the remnant symmetries in the neutrino and the charged lepton sectors of the theory. When $G_\nu$ is chosen to be $Z_2$, $G_\nu\rtimes H^\nu_{\rm{CP}}$ can be reduced to $Z_2\times H^\nu_{\rm{CP}}$.}
\end{center}
\end{figure}

First of all, we briefly review the setup which we will use to predict the mixing matrices from remnant symmetries. The basic formalism has already been stated clearly in Ref.~\cite{Ding:2013hpa}. In the paradigm of the family symmetry $G_f$ combined with the generalised CP symmetry $H_{\rm{CP}}$, both $G_f$ and $H_{\rm{CP}}$ are generally broken into some remnant symmetries in the neutrino and the  charged lepton sectors respectively at leading order (LO), and the mismatch between the residual symmetries generates the Pontecorvo-Maki-Nakagawa-Sakata (PMNS) mixing matrix. In this approach, remnant symmetries of the symmetry group $G_f\rtimes H_{\rm{CP}}$ are assumed and we do not consider how to dynamically achieve the remnant symmetries here. Notice that there are generally more than one vacuum alignment preserving the remnant symmetry, yet they all lead to the same PMNS matrix. A concrete model realizing the general results would be built in section~\ref{sec:model_construction}.

As illustrated in Fig.~\ref{fig:procedure_cartoon}, the residual family symmetries in the neutrino and charged lepton sectors are denoted as $G_{\nu}$ and $G_l$, respectively and the residual CP symmetries are represented by $H^{\nu}_{\rm{CP}}$ and $H^{l}_{\rm{CP}}$, respectively. The misalignment between $G_{\nu}\rtimes H^{\nu}_{\rm{CP}}$ and $G_{l}\rtimes H^{l}_{\rm{CP}}$ leads to specific forms for the PMNS matrix and the resulting lepton mixing parameters. Without loss of generality, the three generations of the left-handed lepton doublets $l$ are assumed to be embedded into a faithful three-dimensional irreducible representation $\mathbf{3}$ of $G_f$. The invariance under the residual family symmetries $G_{\nu}$ and $G_l$ constrains the neutrino mass matrix $m_{\nu}$ and the charged lepton mass matrix $m_{l}$ as follows
\begin{eqnarray}
\nonumber&&\rho^{T}_{\mathbf{3}}(g_{\nu_i})m_{\nu}\rho_{\mathbf{3}}(g_{\nu_
i})=m_{\nu}, \qquad~ g_{\nu_i}\in G_{\nu},\\
\label{eq:inv_family}&&\rho^{\dagger}_{\mathbf{3}}(g_{l_i})m^{\dagger}_{l}m_{l}\rho_{\mathbf{3}}(g_{l_i})=m^{\dagger}_{l}m_{l},\quad
g_{l_i}\in G_l\,.
\end{eqnarray}
where the charged lepton mass matrix $m_l$ is given in the convention in which the left-handed (right-handed) fields are on the right-hand (left-hand) side of $m_l$, $\rho_{\mathbf{3}}(g)$ is the representation matrix for the group element $g$ in the representation $\mathbf{3}$. Meanwhile, the neutrino and the charged lepton mass matrices are also constrained by residual CP symmetries as
\begin{eqnarray}
\nonumber&&X^{T}_{\nu\mathbf{3}}m_{\nu}X_{\nu\mathbf{3}}=m^{*}_{\nu}, \qquad~~
\quad X_{\nu\mathbf{3}}\in H^{\nu}_{\rm{CP}},\\
\label{eq:inv_CP}&&X^{\dagger}_{l\mathbf{3}}m^{\dagger}_{l}m_{l}X_{l\mathbf{3}}=(m^{\dagger}_{l}m_{l})^{*},\quad
X_{l\mathbf{3}}\in H^{l}_{\rm{CP}}\,.
\end{eqnarray}
where $X_{\nu\mathbf{3}}$ and $X_{l\mathbf{3}}$ denote the remnant CP symmetries in the neutrino and the charged lepton sectors, respectively. Since there are both remnant family and CP symmetries, the corresponding consistency equation similar to Eq.~\eqref{eq:consistence_equation} has to be satisfied. That is to say the elements
$X_{\nu}$ of $H^{\nu}_{\rm{CP}}$ and $X_{l}$ of $H^{l}_{\rm{CP}}$ should fulfill the following conditions:
\begin{eqnarray}
\nonumber&&X_{\nu}\rho^{*}(g_{\nu_i})X^{-1}_{\nu}=\rho(g_{\nu_j}),\qquad
g_{\nu_i},g_{\nu_j}\in G_{\nu},\\
\label{eq:consistency_remnant}&&X_{l}\rho^{*}(g_{l_i})X^{-1}_{l}=\rho_(g_{l_j}),\qquad~~\,
g_{l_i},g_{l_j}\in G_{l}\,.
\end{eqnarray}
From the invariance conditions of Eqs.~\eqref{eq:inv_family} and \eqref{eq:inv_CP}, we can reconstruct the mass matrices $m_{\nu}$ and
$m_l$, and predict the lepton mixing matrix
$U_{\rm{PMNS}}$. 

What's more, if the remnant family symmetries are chosen to be $G^{\prime}_{\nu}$ and $G^{\prime}_{l}$ which are conjugate to $G_{\nu}$ and $G_{l}$ under the element $h\in G_f$ as $G^{\prime}_{\nu}=hG_{\nu}h^{-1}$, $G^{\prime}_{l}=hG_{l}h^{-1}$, the same result for $U_{\rm{PMNS}}$ would be obtained~\cite{Ding:2013bpa}. As a consequence, we only need to analyze the independent pairs of $G_{\nu}$ and $G_{l}$ which are not related by group conjugation.

\subsection{Remnant symmetries of $\Delta(48)\rtimes H_{\rm{CP}}$ }

We require that the theory respect the full symmetry $\Delta(48)\rtimes H_{\rm{CP}}$ at high energy. Since most CP transformations defined by outer automorphisms interchange different irreducible representations rather than map one representation into its complex conjugate (i.e. transforming a particle to its anti-particle), in the present work we would like to implement the minimal set of CP transformations which map one irreducible representation into its complex conjugation instead of the whole generalised CP transformations associated with the outer automorphism group of $\Delta(48)$\footnote{In Ref.~\cite{Chen:2014tpa}, the authors claimed that only these generalized CP transformations sending each representation to its complex conjugate are physical otherwise one could introduce a special subset of representations. Hence the physical generalised CP symmetry is imposed in the present work.}. Namely $H_{\rm{CP}}$ is chosen to be the generalized CP transformations given by the outer automorphism $\mathfrak{h_1}$, $\mathfrak{h_2}$ or $\mathfrak{h_3}$ up to inner automorphisms in the present work. Note that each outer automorphism corresponds to 48 automorphism elements once the inner automorphism is taken into account.

In the following section, we will perform a model-independent study of admissible lepton mixing within $\Delta(48)\rtimes H_{\rm{CP}}$ by a scan of all the possible remnant symmetries $G^{\nu}_{\rm{CP}}\cong G_{\nu}\rtimes H^{\nu}_{\rm{CP}}$ and $G^{l}_{\rm{CP}}\cong G_{l}\rtimes H^{l}_{\rm{CP}}$, as shown in Fig.~\ref{fig:procedure_cartoon}. The three generations of the lepton doublet fields are assigned to a triplet $\mathbf{3}$ of $\Delta(48)$, the results for embedding into other faithful triplets $\mathbf{\overline{3}}$, $\mathbf{3^{\prime}}$ and $\mathbf{\overline{3}^{\prime}}$ are also presented thereafter. The residual family symmetry $G_{l}$ in the charged lepton sector is chosen to be an abelian cyclic subgroup $G_{l}\cong Z_m$ with $m\geq3$\footnote{For the case $G_{l}=Z_2$ whose eigenvalues are completely or partially degenerate, the three generations of charged
leptons can not be distinguished by this $G_{l}$ at low energy. As a consequence, the lepton mixing matrix can not be fixed uniquely in this case.}. Moreover, we assume that the light neutrinos are Majorana particles such that the group $G_{\nu}$ must be restricted to the Klein subgroup $K_4$ or a $Z_2$ subgroup of $\Delta(48)$.

From Appendix~\ref{sec:appendix_A_group_theory}, we see that the cyclic subgroups of $\Delta(48)$ can only be $Z_2$, $Z_3$ or $Z_4$. Firstly, we consider the case of $G_{\nu}=K_4$ and $G_{l}=Z_3$, where $Z_3$ denotes any $Z_3$ subgroup $Z^{(x,y)}_3(x,y=0,1,2,3)$, we find a unique mixing pattern:
\begin{equation}
||U_{\rm{PMNS}}||=\frac{1}{\sqrt{3}}\left(\begin{array}{ccc}
1  &  1  &  1 \\
1  &  1  &  1 \\
1  &  1  &  1
\end{array}\right)\,,
\end{equation}
which implies maximal solar and atmospheric mixing angles, and the reactor angle is predicted to fulfill $\sin^2\theta_{13}=1/3$. Obviously this scenario is not compatible with the current experimental data~\cite{Tortola:2012te,GonzalezGarcia:2012sz,Capozzi:2013csa}. Then, we consider the case of $G_{\nu}=K_4$ and $G_{l}=Z_4$, the resulting PMNS matrix is a trivial unit matrix up to permutations of rows and columns. This case is not phenomenologically viable as well. Consequently, we proceed to degrade the remnant symmetry group $G_{\nu}$ from $K_4$ to $Z_2$, and $G_l$ is still $Z_3$ or $Z_4$. In this scenario, the lepton mixing angles and CP phases are predicted in terms of a single real parameter. For the case $G_{\nu}=Z_2$ and $G_{l}=Z_4$, one column of the PMNS matrix is determined to be $\left(1,0,0\right)^{T}$ up to permutations. The experimental data can not be accommodated. Finally, we concentrate on the residual family symmetries $G_{\nu}=Z_2$ and $G_{l}=Z_3$. There are $3\times16=48$ possible combinations, but we find that all of them are conjugate to each other, as shown in Eq.~\eqref{eq:z2_conjugation} and Eq.~\eqref{eq:z3_conjugation}. As a result, it is sufficient to consider only $G_{\nu}=Z^{c^2}_2=\{1,c^2\}$ and $G_{l}=Z^{(0,0)}_3=\{1,a, a^2\}$ without loss of generality.

We begin the study of constraints on the charged lepton sector. The invariance under the residual flavor symmetry $G_{l}=Z^{(0,0)}_3=\{1,a, a^2\}$ requires that
\begin{equation}
\rho^{\dagger}_{\mathbf{3}}(a)m^{\dagger}_{l}m_{l}\rho_{\mathbf{3}}(a)=m^{\dagger}_{l}m_{l}\,.
\end{equation}
In the chosen basis where the representation matrix of the element $a$ is diagonal for different irreducible representations, as shown in Table~\ref{tab:rep_matrix}, we can straightforwardly obtain that the hermitian combination $m^{\dagger}_{l}m_{l}$ is diagonal. The phasing and permutation freedom of the column vectors can be used to bring it into the form $\text{diag}(m^2_{e},m^2_{\mu},m^2_{\tau})$, where $m_e$, $m_{\mu}$ and $m_{\tau}$ represent the electron, muon and tau masses, respectively. Hence the unitary matrix $U_l$, which diagonalizes $m^{\dagger}_{l}m_{l}$ as $U^{\dagger}_{l}m^{\dagger}_{l}m_{l}U_{l}=\text{diag}(m^2_{e},m^2_{\mu},m^2_{\tau})$, is determined to be
\begin{equation}
\label{eq:Ul}U_{l}=P_{l}K_{l}\,,
\end{equation}
where $K_{l}$ is a diagonal phase matrix, $P_{l}$ is a permutation matrix which can be one of the following six values:
\begin{eqnarray}
\nonumber&& P_{123}=\left(\begin{array}{ccc}
1  &  0  &  0 \\
0  &  1  &  0 \\
0  &  0  &  1
\end{array}\right),\qquad P_{231}=\left(\begin{array}{ccc}
0  &  1  &  0 \\
0  &  0  &  1  \\
1  &  0  &  0
\end{array}\right),\qquad P_{312}=\left(\begin{array}{ccc}
0   &  0  &  1 \\
1   &  0  &  0\\
0   &  1  &  0
\end{array}\right)\,,\\
\label{eq:permutation_matrix}&& P_{213}=\left(\begin{array}{ccc}
0  &  1  &  0  \\
1  &  0  &  0  \\
0  &  0  &  1
\end{array}\right),\qquad P_{321}=\left(\begin{array}{ccc}
0   &  0  &  1  \\
0   &  1  &  0  \\
1   &  0  &  0
\end{array}\right),\qquad P_{132}=\left(\begin{array}{ccc}
1   &   0  &   0 \\
0   &   0  &   1 \\
0   &   1  &   0
\end{array}\right)\,.
\end{eqnarray}
Since charged leptons are Dirac particles, the diagonal phase matrix $K_l$ can be rotated away by redefining the phases of right-handed charged leptons and thus $K_l$ is unphysical.

The underlying symmetry $\Delta(48)\rtimes H_{\rm{CP}}$ is broken down to $G^{\nu}_{\rm{CP}}\cong Z^{c^2}_2\times H^{\nu}_{\rm{CP}}$ in the neutrino sector. Notice that the semidirect product between the flavor and generalized CP symmetries is reduced to direct product, and this is generally true for the remnant flavor symmetry being $Z_2$~\cite{Ding:2013hpa,Ding:2013bpa}. The remnant CP symmetry $H^{\nu}_{\rm{CP}}$ has to be compatible with the remnant $Z^{c^2}_2=\left\{1,c^2\right\}$ family symmetry. This is to say the automorphism $\mathfrak{g}$ associated with $H^{\nu}_{\rm{CP}}$ should map the element $c^2$ into $c^2$, i.e. $\mathfrak{g}(c^2)=c^2$. This leads to
\begin{equation}
\mathfrak{g}=\mathrm{conj}^m(c)\mathrm{conj}^n(d)\mathfrak{h_i},\quad i=1,2,3\,,
\end{equation}
with $m,n=0, 1, 2, 3$. Therefore only 16 of the 48 generalized CP transformations are consistent with the residual $Z^{c^2}_2$ flavor symmetry no matter which one of the generalized CP symmetries $\mathfrak{h_i}(i=1,2,3)$ is imposed on the theory. The corresponding CP transformation matrix takes the form
\begin{equation}
X(\mathfrak{g})=X\left(\mathrm{conj}^m(c)\right)WX\left(\mathrm{conj}^n(d)\right)WX(\mathfrak{h_i})=\rho^m(c)\rho^n(d)X(\mathfrak{h_i})=\rho(c^md^n)X(\mathfrak{h_i})\,.
\end{equation}
It is straightforward to check that the consistency equation is really satisfied, i.e.
\begin{equation}
X(\mathfrak{g})\rho^{*}(c^2)X^{-1}(\mathfrak{g})=\rho(c^2)\,.
\end{equation}
Hence nontrivial CP transformations of $H^{\nu}_{\rm{CP}}$ could be the following unitary matrices:
\begin{equation}
X_{\nu}=\rho(c^md^n)X(\mathfrak{h_i}), \qquad m,n=0,1,2,3\,.
\end{equation}
We can construct the light neutrino mass matrix $m_{\nu}$
by demanding that it respects both the residual flavour symmetry $Z^{c^2}_2$ and the generalised CP symmetry $H^{\nu}_{\rm{CP}}$:
\begin{eqnarray}
\label{eq:neutrino_remnant_flavor}&&\rho^{T}_{\mathbf{3}}(c^2)m_{\nu}\rho_{\mathbf{3}}(c^2)=m_{\nu}\,,\\
\label{eq:neutrino_remnant_CP}&&X^{T}_{\nu\mathbf{3}}m_{\nu}X_{\nu\mathbf{3}}=m^{*}_{\nu}\,,
\end{eqnarray}
The most general neutrino mass matrix $m_{\nu}$ which satisfies Eq.~\eqref{eq:neutrino_remnant_flavor} can be expressed as
\begin{equation}
\label{eq:numm_inv_rem_fla}m_{\nu}=\alpha
\left(\begin{array}{ccc}
 2 & -1 & -1 \\
 -1 & 2 & -1 \\
 -1 & -1 & 2
\end{array}\right)+\beta
\left(\begin{array}{ccc}
 1 & 0 & 0 \\
 0 & 0 & 1 \\
 0 & 1 & 0
\end{array}\right)+\gamma
\left(\begin{array}{ccc}
 0 & 1 & 1 \\
 1 & 1 & 0 \\
 1 & 0 & 1
\end{array}\right)+\epsilon
\left(\begin{array}{ccc}
 0 & 1 & -1 \\
 1 & -1 & 0 \\
 -1 & 0 & 1
\end{array}\right)\,,
\end{equation}
where $\alpha$, $\beta$, $\gamma$ and $\epsilon$ are complex parameters, and they are further constrained by the remnant CP invariant condition in Eq.~\eqref{eq:neutrino_remnant_CP}. The different constraints on $\alpha$, $\beta$, $\gamma$ and $\epsilon$ for all the consistent residual CP symmetries are summarized in Table~\ref{tab:constraints_by_GCP}. The scenarios (e.g.,  $X_{\nu}=\rho(c)X(\mathfrak{h_3})$) which predict degenerate light neutrino masses are not viable, and hence are not included in this table. It is interesting to note that the invariance under the residual CP transformation $X_{\nu}=\rho(c^md^n)X(\mathfrak{h_i})$ reads
\begin{equation}
\left[\rho(c^md^n)X(\mathfrak{h_i})\right]^T m_\nu\left[\rho(c^md^n)X(\mathfrak{h_i})\right] = m^{*}_{\nu}\,.
\end{equation}
Taking into account the invariance condition of Eq.~\eqref{eq:neutrino_remnant_flavor} under $Z^{c^2}_2$, we have
\begin{equation}
\left[\rho(c^{m+2}d^n)X(\mathfrak{h_i})\right]^T m_\nu\left[\rho(c^{m+2}d^n)X(\mathfrak{h_i})\right] = m^{*}_{\nu}\,.
\end{equation}
This implies that $X_{\nu}=\rho(c^md^n)X(\mathfrak{h_i})$ and $X_{\nu}=\rho(c^{m+2}d^n)X(\mathfrak{h_i})$ impose the same constraint on the light neutrino mass matrix.

%%%%%%%%%%%%%%%%%%%%%%%%%%%%%%%%%%%%%%%%%%%%%%%%%%%%%%%%%%%%%%%%%%%%%%%%%%%%%%%%%%%%%%%%%%%%%%%%%%%%%%%%%%%%%%%%%%%%%%%%%%%%%%%%%%%%%%%%%%%%%%%%%%%

\begin{table}[hptb]
\addtolength{\tabcolsep}{-2pt}
\begin{center}
\begin{tabular}{|c|c|c|c|c|}
\hline\hline
%%%%%%%%%%%%%%%%%%%%%

 & \multirow{2}{*}{~$l$~ }  &\multirow{2}{*}{$H_{\rm{CP}}^{\nu}$ }& {Constraints on} & \multirow{2}{*}{$\Omega$}  \\

  &   &   &   $\alpha$, $\beta$, $\gamma$, $\epsilon$  &  \\ \hline

%%%%%%%%%%%%%%%%%%%%%
\multirow{4}{*}{I} & $\mathbf{3}$ & $X(\mathfrak{h_1}),\rho(c^2)X_{\mathbf{r}}(\mathfrak{h_1})$ &   \multirow{4}{*}{$\begin{array}{c}
 \alpha_{\mathrm{Im}}=\beta_{\mathrm{Im}}=0,\\
 \gamma_{\mathrm{Im}}=\epsilon_{\mathrm{Re}}=0
 \end{array}$}  & \multirow{4}{*}{$\left(\begin{array}{ccc}
 1  & 0 & 0  \\
 0  & 1  & 0  \\
 0  & 0  & i
\end{array}\right)$}  \\ \cdashline{2-3}

   & $\mathbf{3^{\prime}}$ &$X(\mathfrak{h_1}), \rho(c^2)X(\mathfrak{h_1})$ &   &  \\

  & $\mathbf{\overline{3}}$ &$X(\mathfrak{h_1}),
\rho(c^2)X(\mathfrak{h_1})$ &   &  \\

  & $\mathbf{\overline{3}^{\prime}}$ &$X(\mathfrak{h_1}),
\rho(c^2)X(\mathfrak{h_1})$ &   &  \\\hline

%%%%%%%%%%%%%%%%%%%%%%%%%%%%%%%%%%%%%%%%%%%%%%%%%%%%%%%%%%%%%%%%%%%%%%%%%%%%%%%%%%%%%%%%%%%%%%%%%%%%%%%%%%%%%%%%%%%%%%%%%% %%                                                       II                                                                 %%%%%%%%%%%%%%%%%%%%%%%%%%%%%%%%%%%%%%%%%%%%%%%%%%%%%%%%%%%%%%%%%%%%%%%%%%%%%%%%%%%%%%%%%%%%%%%%%%%%%%%%%%%%%%%%%%%%%%%%%%

\multirow{4}{*}{II}  &   $\mathbf{3}$ &  $\rho(d)X(\mathfrak{h_1}),
 \rho(c^2d)X(\mathfrak{h_1})
 $  &   \multirow{4}{*}{$\begin{array}{c}
 \alpha_{\mathrm{Im}}=-\alpha_{\mathrm{Re}}, \\
 \beta_{\mathrm{Re}}=-2\gamma_{\mathrm{Re}},\\
 \beta_{\mathrm{Im}}=\gamma_{\mathrm{Im}}+\sqrt{3}\;\epsilon_{\mathrm{Re}}, \\ \epsilon_{\mathrm{Im}}=-\sqrt{3}\;\gamma_{\mathrm{Re}}
 \end{array}$} & \multirow{4}{*}{$
\left(\begin{array}{ccc}
 e^{\frac{5 i \pi }{8}} \sin \frac{\pi
   }{8} & 0 & e^{\frac{i \pi }{8}} \cos
   \frac{\pi }{8} \\
 0 & e^{\frac{i \pi }{4}} & 0 \\
 -e^{\frac{5 i \pi }{8}} \cos \frac{\pi
   }{8} & 0 & e^{\frac{i \pi }{8}} \sin
   \frac{\pi }{8}
\end{array}\right)$ }  \\ \cdashline{2-3}

 & $\mathbf{3^{\prime}}$  &  $\rho(d^3)X(\mathfrak{h_1}),  \rho(c^2d^3)X(\mathfrak{h_1}),$ &   &  \\

  & $\mathbf{\overline{3}}$ &  $\rho(cd^3)X(\mathfrak{h_1}),
\rho(c^3d^3)X(\mathfrak{h_1})$ &   &  \\

   & $\mathbf{\overline{3}^{\prime}}$ &  $\rho(cd^3)X(\mathfrak{h_1}),
\rho(c^3d^3)X(\mathfrak{h_1})$ &   &  \\ \hline

%%%%%%%%%%%%%%%%%%%%%%%%%%%%%%%%%%%%%%%%%%%%%%%%%%%%%%%%%%%%%%%%%%%%%%%%%%%%%%%%%%%%%%%%%%%%%%%%%%%%%%%%%%%%%%%%%%%%%%%%%% %%                                                       III                                                                 %%%%%%%%%%%%%%%%%%%%%%%%%%%%%%%%%%%%%%%%%%%%%%%%%%%%%%%%%%%%%%%%%%%%%%%%%%%%%%%%%%%%%%%%%%%%%%%%%%%%%%%%%%%%%%%%%%%%%%%%%%

\multirow{4}{*}{III} &  $\mathbf{3}$ &  $\rho(d^2)X(\mathfrak{h_1}),
 \rho(c^2d^2)X(\mathfrak{h_1})$  &  \multirow{4}{*}{$\begin{array}{c}
 \alpha_{\mathrm{Re}}=\beta_{\mathrm{Im}}=0,\\
 \gamma_{\mathrm{Im}}=\epsilon_{\mathrm{Re}}=0
 \end{array}$}  &   \multirow{4}{*}{$\frac{1}{\sqrt{2}}
\left(\begin{array}{ccc}
 e^{\frac{i \pi }{4}} & 0 & -e^{-\frac{i \pi
   }{4}} \\
 0 & \sqrt{2} & 0 \\
 -e^{\frac{i \pi }{4}} & 0 & -e^{-\frac{i \pi
   }{4}}
\end{array}\right)$}  \\ \cdashline{2-3}

  & $\mathbf{3^{\prime}}$  &  $\rho(d^2)X(\mathfrak{h_1}),
  \rho(c^2d^2)X(\mathfrak{h_1})$ &   &  \\

 & $\mathbf{\overline{3}}$ &  $\rho(d^2)X(\mathfrak{h_1}),
\rho(c^2d^2)X(\mathfrak{h_1})$ &   &  \\

 & $\mathbf{\overline{3}^{\prime}}$ &  $\rho(d^2)X(\mathfrak{h_1}),
\rho(c^2d^2)X(\mathfrak{h_1})$ &   &  \\ \hline

%%%%%%%%%%%%%%%%%%%%%%%%%%%%%%%%%%%%%%%%%%%%%%%%%%%%%%%%%%%%%%%%%%%%%%%%%%%%%%%%%%%%%%%%%%%%%%%%%%%%%%%%%%%%%%%%%%%%%%%%%% %%                                                       IV                                                                %%%%%%%%%%%%%%%%%%%%%%%%%%%%%%%%%%%%%%%%%%%%%%%%%%%%%%%%%%%%%%%%%%%%%%%%%%%%%%%%%%%%%%%%%%%%%%%%%%%%%%%%%%%%%%%%%%%%%%%%%%

\multirow{4}{*}{IV} & $\mathbf{3}$ &  $\rho(d^3)X(\mathfrak{h_1}),
 \rho(c^2d^3)X(\mathfrak{h_1})$  &  \multirow{4}{*}{$\begin{array}{c}
  \alpha_{\mathrm{Im}}=\alpha_{\mathrm{Re}},\\
  \beta_{\mathrm{Re}}=-2\gamma_{\mathrm{Re}},\\
  \beta_{\mathrm{Im}}=\gamma_{\mathrm{Im}}+\sqrt{3}\;\epsilon_{\mathrm{Re}}, \\ \epsilon_{\mathrm{Im}}=-\sqrt{3}\;\gamma_{\mathrm{Re}}
  \end{array}$}   &    \multirow{4}{*}{$
\left(\begin{array}{ccc}
 -e^{\frac{3 i \pi }{8}} \cos \frac{\pi
   }{8} & 0 & e^{-\frac{i \pi }{8}} \sin
   \frac{\pi }{8} \\
 0 & e^{-\frac{i \pi }{4}} & 0 \\
 e^{\frac{3 i \pi }{8}} \sin \frac{\pi
   }{8} & 0 & e^{-\frac{i \pi }{8}} \cos
   \frac{\pi }{8}
\end{array}\right)$ }  \\ \cdashline{2-3}

 & $\mathbf{3^{\prime}}$  &  $ \rho(d)X(\mathfrak{h_1}),
  \rho(c^2d)X(\mathfrak{h_1}) $ &   &  \\

   & $\mathbf{\overline{3}}$ &  $\rho(cd)X(\mathfrak{h_1}),
  \rho(c^3d)X(\mathfrak{h_1})  $ &   &  \\

 & $\mathbf{\overline{3}^{\prime}}$ &  $\rho(cd)X(\mathfrak{h_1}),
\rho(c^3d)X(\mathfrak{h_1})  $ &   &  \\  \hline

%%%%%%%%%%%%%%%%%%%%%%%%%%%%%%%%%%%%%%%%%%%%%%%%%%%%%%%%%%%%%%%%%%%%%%%%%%%%%%%%%%%%%%%%%%%%%%%%%%%%%%%%%%%%%%%%%%%%%%%%%% %%                                                       V                                                                 %%%%%%%%%%%%%%%%%%%%%%%%%%%%%%%%%%%%%%%%%%%%%%%%%%%%%%%%%%%%%%%%%%%%%%%%%%%%%%%%%%%%%%%%%%%%%%%%%%%%%%%%%%%%%%%%%%%%%%%%%%

\multirow{4}{*}{V} & $\mathbf{3}$  &  $\rho(c)X(\mathfrak{h_1}),
 \rho(c^3)X(\mathfrak{h_1})$  &  \multirow{4}{*}{$\begin{array}{c}
  \alpha_{\mathrm{Im}}=0, \\
  \beta_{\mathrm{Re}}=\gamma_{\mathrm{Re}},\\
  \beta_{\mathrm{Im}}=-2\gamma_{\mathrm{Im}}, \\
  \epsilon_{\mathrm{Im}}=0
  \end{array}$}  &   \multirow{4}{*}{$\frac{1}{\sqrt{2}}
\left(\begin{array}{ccc}
 1 & 0 & i \\
 0 & ~\sqrt{2}~~ & 0 \\
 -1 & 0 & i
\end{array}\right)$}  \\ \cdashline{2-3}

  & $\mathbf{3^{\prime}}$  &  $ \rho(cd^2)X(\mathfrak{h_1}),
  \rho(c^3d^2)X(\mathfrak{h_1})$ &   &  \\

  & $\mathbf{\overline{3}}$ &  $\rho(c)X(\mathfrak{h_1}),
\rho(c^3)X(\mathfrak{h_1}) $ &   &  \\

   & $\mathbf{\overline{3}^{\prime}}$ &  $\rho(cd^2)X(\mathfrak{h_1}),
\rho(c^3d^2)X(\mathfrak{h_1}) $ &   &  \\\hline

%%%%%%%%%%%%%%%%%%%%%%%%%%%%%%%%%%%%%%%%%%%%%%%%%%%%%%%%%%%%%%%%%%%%%%%%%%%%%%%%%%%%%%%%%%%%%%%%%%%%%%%%%%%%%%%%%%%%%%%%%% %%                                                       VI                                                                 %%%%%%%%%%%%%%%%%%%%%%%%%%%%%%%%%%%%%%%%%%%%%%%%%%%%%%%%%%%%%%%%%%%%%%%%%%%%%%%%%%%%%%%%%%%%%%%%%%%%%%%%%%%%%%%%%%%%%%%%%%

\multirow{4}{*}{VI}  & $\mathbf{3}$ &  $\rho(cd)X(\mathfrak{h_1}),
 \rho(c^3d)X(\mathfrak{h_1})$  &   \multirow{4}{*}{$\begin{array}{c}
  \alpha_{\mathrm{Im}}=-\alpha_{\mathrm{Re}},\\
  \beta_{\mathrm{Re}}=-2\gamma_{\mathrm{Re}},\\
  \beta_{\mathrm{Im}}=\gamma_{\mathrm{Im}}-\sqrt{3}\;\epsilon_{\mathrm{Re}}, \\ \epsilon_{\mathrm{Im}}=\sqrt{3}\;\gamma_{\mathrm{Re}}
  \end{array}$}  & \multirow{4}{*}{$
\left(\begin{array}{ccc}
 e^{\frac{5 i \pi }{8}} \cos
   \frac{\pi }{8} & 0 &
   e^{\frac{i \pi }{8}} \sin
   \frac{\pi }{8} \\
 0 & e^{\frac{i \pi }{4}} & 0 \\
 -e^{\frac{5 i \pi }{8}} \sin
   \frac{\pi }{8} & 0 &
   e^{\frac{i \pi }{8}} \cos
   \frac{\pi }{8}
\end{array}\right)$}  \\\cdashline{2-3}

  & $\mathbf{3^{\prime}}$  & $\rho(cd)X(\mathfrak{h_1}),
  \rho(c^3d)X(\mathfrak{h_1})$ &   &  \\

  & $\mathbf{\overline{3}}$ & $\rho(d^3)X(\mathfrak{h_1}),
\rho(c^2d^3)X(\mathfrak{h_1})$ &   &  \\

 & $\mathbf{\overline{3}^{\prime}}$  & $\rho(d)X(\mathfrak{h_1}),
\rho(c^2d)X(\mathfrak{h_1})$ &   &  \\\hline

%%%%%%%%%%%%%%%%%%%%%%%%%%%%%%%%%%%%%%%%%%%%%%%%%%%%%%%%%%%%%%%%%%%%%%%%%%%%%%%%%%%%%%%%%%%%%%%%%%%%%%%%%%%%%%%%%%%%%%%%%% %%                                                       VII                                                                 %%%%%%%%%%%%%%%%%%%%%%%%%%%%%%%%%%%%%%%%%%%%%%%%%%%%%%%%%%%%%%%%%%%%%%%%%%%%%%%%%%%%%%%%%%%%%%%%%%%%%%%%%%%%%%%%%%%%%%%%%%

\multirow{4}{*}{VII} & $\mathbf{3}$  &   $\rho(cd^2)X_{\mathbf{r}}(\mathfrak{h_1}),
 \rho(c^3d^2)X_{\mathbf{r}}(\mathfrak{h_1})$   &  \multirow{4}{*}{$\begin{array}{c}
\alpha_{\mathrm{Re}}=0, \\
\beta_{\mathrm{Re}}=\gamma_{\mathrm{Re}}, \\
\beta_{\mathrm{Im}}=-2\gamma_{\mathrm{Im}}, \\
\epsilon_{\mathrm{Im}}=0
\end{array}$}  &   \multirow{4}{*}{$
\left(\begin{array}{ccc}
 e^{\frac{i \pi }{4}} & 0 & 0 \\
 0 & 1 & 0 \\
 0 & 0 & e^{\frac{3 i \pi }{4}}
\end{array}\right)$}  \\\cdashline{2-3}

  & $\mathbf{3^{\prime}}$  & $\rho(c)X(\mathfrak{h_1}),
  \rho(c^3)X(\mathfrak{h_1})$ &   &  \\

  & $\mathbf{\overline{3}}$ & $\rho(cd^2)X(\mathfrak{h_1}),
\rho(c^3d^2)X(\mathfrak{h_1})$ &   &  \\

  & $\mathbf{\overline{3}^{\prime}}$ & $\rho(c)X(\mathfrak{h_1}),
\rho(c^3)X(\mathfrak{h_1})$ &   &  \\\hline

%%%%%%%%%%%%%%%%%%%%%%%%%%%%%%%%%%%%%%%%%%%%%%%%%%%%%%%%%%%%%%%%%%%%%%%%%%%%%%%%%%%%%%%%%%%%%%%%%%%%%%%%%%%%%%%%%%%%%%%%%% %%                                                       VIII                                                                 %%%%%%%%%%%%%%%%%%%%%%%%%%%%%%%%%%%%%%%%%%%%%%%%%%%%%%%%%%%%%%%%%%%%%%%%%%%%%%%%%%%%%%%%%%%%%%%%%%%%%%%%%%%%%%%%%%%%%%%%%%

\multirow{4}{*}{VIII}  &  $\mathbf{3}$ &   $\rho(cd^3)X(\mathfrak{h_1}),
 \rho(c^3d^3)X(\mathfrak{h_1})$   &   \multirow{4}{*}{$\begin{array}{c}
  \alpha_{\mathrm{Im}}=\alpha_{\mathrm{Re}}, \\
  \beta_{\mathrm{Re}}=-2 \gamma_{\mathrm{Re}},\\
  \beta_{\mathrm{Im}}=\gamma_{\mathrm{Im}}-\sqrt{3}\;\epsilon_{\mathrm{Re}}, \\ \epsilon_{\mathrm{Im}}=\sqrt{3}\;\gamma_{\mathrm{Re}}
  \end{array}$}   &    \multirow{4}{*}{$
\left(\begin{array}{ccc}
 -e^{\frac{3 i \pi }{8}} \sin \frac{\pi
   }{8} & 0 & e^{-\frac{i \pi }{8}} \cos
   \frac{\pi }{8} \\
 0 & e^{-\frac{i \pi }{4}} & 0 \\
 e^{\frac{3 i \pi }{8}} \cos \frac{\pi
   }{8} & 0 & e^{-\frac{i \pi }{8}} \sin
   \frac{\pi }{8}
\end{array}\right)$}  \\\cdashline{2-3}

  & $\mathbf{3^{\prime}}$  & $\rho(cd^3)X(\mathfrak{h_1}),
  \rho(c^3d^3)X(\mathfrak{h_1})$ &   &  \\

 & $\mathbf{\overline{3}}$ & $\rho(d)X(\mathfrak{h_1}),
\rho(c^2d)X(\mathfrak{h_1})$ &   &  \\

  & $\mathbf{\overline{3}^{\prime}}$ & $\rho(d^3)X(\mathfrak{h_1}),
\rho(c^2d^3)X(\mathfrak{h_1})$ &   &  \\\hline

%%%%%%%%%%%%%%%%%%%%%%%%%%%%%%%%%%%%%%%%%%%%%%%%%%%%%%%%%%%%%%%%%%%%%%%%%%%%%%%%%%%%%%%%%%%%%%%%%%%%%%%%%%%%%%%%%%%%%%%%%% %%                                                       IX                                                                 %%%%%%%%%%%%%%%%%%%%%%%%%%%%%%%%%%%%%%%%%%%%%%%%%%%%%%%%%%%%%%%%%%%%%%%%%%%%%%%%%%%%%%%%%%%%%%%%%%%%%%%%%%%%%%%%%%%%%%%%%%

\multirow{4}{*}{IX} & $\mathbf{3}$  &  $X(\mathfrak{h_3}),
 \rho(c^2)X(\mathfrak{h_3})$  &   \multirow{4}{*}{$\begin{array}{c}
 \alpha_{\mathrm{Im}}=\beta_{\mathrm{Im}}=0, \\
 \gamma_{\mathrm{Im}}=\epsilon_{\mathrm{Im}}=0
 \end{array}$}  &   \multirow{4}{*}{$\left(\begin{array}{ccc}
 1 & 0 & 0 \\
 0 &  1 & 0 \\
 0 & 0 & 1\end{array}\right)$}   \\\cdashline{2-3}

   & $\mathbf{3^{\prime}}$  & $  X(\mathfrak{h_2}),
  \rho(c^2)X(\mathfrak{h_2})$ &   &  \\

 & $\mathbf{\overline{3}}$  & $X(\mathfrak{h_3}),
\rho(c^2)X(\mathfrak{h_3})$ &   &  \\

 & $\mathbf{\overline{3}^{\prime}}$  & $X(\mathfrak{h_2}),
\rho(c^2)X(\mathfrak{h_2})$ &   &  \\\hline

%%%%%%%%%%%%%%%%%%%%%%%%%%%%%%%%%%%%%%%%%%%%%%%%%%%%%%%%%%%%%%%%%%%%%%%%%%%%%%%%%%%%%%%%%%%%%%%%%%%%%%%%%%%%%%%%%%%%%%%%%% %%                                                       X                                                                 %%%%%%%%%%%%%%%%%%%%%%%%%%%%%%%%%%%%%%%%%%%%%%%%%%%%%%%%%%%%%%%%%%%%%%%%%%%%%%%%%%%%%%%%%%%%%%%%%%%%%%%%%%%%%%%%%%%%%%%%%%

\multirow{4}{*}{X} & $\mathbf{3}$ & $\rho(cd^2)X(\mathfrak{h_3}),
\rho(c^3d^2)X(\mathfrak{h_3})$  &  \multirow{4}{*}{$\begin{array}{c}
\alpha_{\mathrm{Re}}=0, \\
\beta_{\mathrm{Re}}=\gamma_{\mathrm{Re}},\\
\epsilon_{\mathrm{Re}}=0, \\
\beta_{\mathrm{Im}}=-2\gamma_{\mathrm{Im}}
\end{array}$}  & \multirow{4}{*}{$
\left(\begin{array}{ccc}
 e^{\frac{i \pi }{4}} & 0 & 0 \\
 0 & 1 & 0 \\
 0 & 0 & e^{\frac{i \pi }{4}}
\end{array}\right)$}   \\\cdashline{2-3}

  & $\mathbf{3^{\prime}}$  & $\rho(c)X(\mathfrak{h_2}),
  \rho(c^3)X(\mathfrak{h_2}),  $ &   &  \\

  & $\mathbf{\overline{3}}$ & $\rho(cd^2)X(\mathfrak{h_3}),
\rho(c^3d^2)X(\mathfrak{h_3})  $ &   &  \\

 & $\mathbf{\overline{3}^{\prime}}$  & $\rho(c)X(\mathfrak{h_2}),
\rho(c^3)X(\mathfrak{h_2})  $ &   &  \\ \hline\hline

\end{tabular}
\caption{\label{tab:constraints_by_GCP}The generalised CP transformations consistent with a residual $Z^{c^2}_2$ family symmetry in the neutrino sector and the resulting constraints on the parameters $\alpha$, $\beta$, $\gamma$ and $\epsilon$ in the neutrino mass matrix of Eq.~\eqref{eq:numm_inv_rem_fla}. The second column stands for the $\Delta(48)$ triplet to which the lepton doublet $l$ is assigned. The subscripts ``\text{Re}'' and ``\text{Im}'' denote the real and imaginary parts, respectively.}
\end{center}
\end{table}

%%%%%%%%%%%%%%%%%%%%%%%%%%%%%%%%%%%%%%%%%%%%%%%%%%%%%%%%%%%%%%%%%%%%%%%%%%%%%%%%%%%%%%%%%%%%%%%%%%%%%%%%%%%%%%%%%%%%%%%%%%%%%%%%%%%%%%%%%%%%%%%%%%%

Performing a tri-bimaximal transformation $U_{\rm{TB}}$~\cite{TBM} on the neutrino mass matrix $m_{\nu}$ of Eq.~\eqref{eq:numm_inv_rem_fla}, we obtain
\begin{equation}
m^{\prime}_{\nu}=U^{T}_{\rm{TB}}m_{\nu}U_{\rm{TB}}=
\left(\begin{array}{ccc}
 3 \alpha +\beta -\gamma  & 0 & -\sqrt{3} \epsilon  \\
 0 & \beta +2 \gamma  & 0 \\
 -\sqrt{3}\epsilon  & 0 & 3 \alpha -\beta +\gamma
\end{array}\right)\,,
\end{equation}
where
\begin{equation}
\label{TB} U_{\rm{TB}}=\left(\begin{array}{ccc}
 \sqrt{\frac{2}{3}} & ~~\frac{1}{\sqrt{3}}~~
   & 0 \\
 -\frac{1}{\sqrt{6}} & ~~\frac{1}{\sqrt{3}}~~
   & -\frac{1}{\sqrt{2}} \\
 -\frac{1}{\sqrt{6}} & ~~\frac{1}{\sqrt{3}}~~
   & \frac{1}{\sqrt{2}}
\end{array}\right)\,.
\end{equation}
Furthermore, $m^{\prime}_{\nu}$ can be diagonalized  by a unitary matrix $U^{\prime}_{\nu}$ as
\begin{equation}
\label{eq:Unup}U^{\prime T}_{\nu}m^{\prime}_{\nu}U^{\prime}_{\nu}=\text{diag}(m_1,m_2,m_3)\,.
\end{equation}
For all the cases listed in Table~\ref{tab:constraints_by_GCP}, we find that $U^{\prime}_{\nu}$ can be factorized into the form
\begin{equation}
U^{\prime}_{\nu}=\Omega R(\theta)P_{\nu}K_{\nu}\,,
\end{equation}
where $\Omega$ is a constant unitary matrix to make $\Omega^{T}m^{\prime}_{\nu}\Omega$ real, and the explicit forms of $\Omega$ for different remnant CP symmetries are summarized in Table~\ref{tab:constraints_by_GCP}. $R(\theta)$ is a rotation matrix with
\begin{equation}
R(\theta)=\left(\begin{array}{ccc}
\cos\theta  &  ~~0~~   &   \sin\theta \\
0  &  ~~1~~   &  0  \\
-\sin\theta  &   ~~0~~   &  \cos\theta
\end{array}\right)\,.
\end{equation}
Being analogous to the charged lepton sector, $P_{\nu}$ is a permutation matrix, and it can also take the six possible forms shown in Eq.~\eqref{eq:permutation_matrix}. Because light neutrino masses are unconstrained in the present framework, the contribution from the neutrino sector to the lepton mixing is fixed up to permutations of columns. This is exactly the reason why $P_{\nu}$ appears in Eq.~\eqref{eq:Unup}. $K_{\nu}$ is a diagonal matrix with non-vanishing entries being $\pm1$ or $\pm i$ to ensure the neutrino masses $m_{1,2,3}$ are positive.
Combining the result for the charged lepton unitary matrix $U_{l}$ in Eq.~\eqref{eq:Ul}, we find that the PMNS matrix is predicted to be
\begin{equation}
U_{\rm{PMNS}}=K^{\dagger}_{l}P^{\dagger}_{l}U_{\rm{TB}}\Omega R(\theta)P_{\nu}K_{\nu}\,,
\label{eq:PMNSfactor}
\end{equation}
Notice that the phase matrix $K_{l}$ can always be absorbed into the charged lepton fields, and therefore would be omitted henceforth. Since the contribution from the matrix $K_{\nu}$ is only possibly shifting the Majorana phases by $\pi$, the factor $K_{\nu}$ will be neglected as well in the following of this section. It is necessary to emphasize again that the lepton mixing matrix $U_{\rm{PMNS}}$ is only determined up to permutations of rows and columns here (i.e. $P_{l}$ and $P_{\nu}$), because both the charged lepton masses and the neutrino masses are not constrained within the current framework. In the PDG convention~\cite{pdg}, the PMNS matrix is cast into the form
\begin{equation}
U_{\rm{PMNS}}=V\,\text{diag}(1,e^{i\frac{\alpha_{21}}{2}},e^{i\frac{\alpha_{31}}{2}}),
\label{eq:pmns_pdg}
\end{equation}
with
\begin{equation}
V=\left(\begin{array}{ccc}
c_{12}c_{13}  &   s_{12}c_{13}   &   s_{13}e^{-i\delta_{\rm{CP}}}  \\
-s_{12}c_{23}-c_{12}s_{23}s_{13}e^{i\delta_{\rm{CP}}}   &  c_{12}c_{23}-s_{12}s_{23}s_{13}e^{i\delta_{\rm{CP}}}  &  s_{23}c_{13}  \\
s_{12}s_{23}-c_{12}c_{23}s_{13}e^{i\delta_{\rm{CP}}}   & -c_{12}s_{23}-s_{12}c_{23}s_{13}e^{i\delta_{\rm{CP}}}  &  c_{23}c_{13}
\end{array}\right).
\end{equation}
where $c_{ij}=\cos\theta_{ij}$ and $s_{ij}=\sin\theta_{ij}$, $\delta_{\rm{CP}}$ denotes the Dirac CP phase, and $\alpha_{21}$ and $\alpha_{31}$ are the Majorana CP phases if neutrinos are Majorana particles.

\subsection{Possible mixing patterns}
We find 10 cases corresponding to different remnant symmetries listed in Table~\ref{tab:constraints_by_GCP}. Phenomenological predictions for lepton mixing parameters have been reported in our previous work~\cite{Ding:2013nsa}, where the results are just presented without details. In the following, we will demonstrate how to derive these interesting results, and what's more, predictions for the PMNS matrix and light neutrino masses are listed for each case. These results are also useful for the phenomenological analysis of the constructed model in section~\ref{sec:model_construction}. For each symmetry breaking pattern, all possible permutation matrices $P_{l}$ and $P_{\nu}$ would be considered. Among all possible permutations, the one which could be compatible with the observed lepton mixing angles with $\theta_{23}$ in the first octant\footnote{$\sin^2\theta_{23}$ has two best-fit values $\sin^2\theta_{23}=0.413$ and $\sin^2\theta_{23}=0.594$~\cite{GonzalezGarcia:2012sz}. The octant of the atmospheric mixing angle $\theta_{23}$ has not been fixed so far.} will be shown in the following. By switching the second and the third rows, $\theta_{23}$ in the second octant can be accommodated. Now we begin to study the ten cases one by one.
\begin{description}[labelindent=-0.5em, leftmargin=0.3em]

\item[~~I.~]{$H_{\rm{CP}}^{\nu}=\left\{X(\mathfrak{h_1}),\rho(c^2)X(\mathfrak{h_1})\right\}$}

In this case, we have $\alpha_{\mathrm{Im}}=\beta_{\mathrm{Im}}= \gamma_{\mathrm{Im}}=\epsilon_{\mathrm{Re}}=0$, where the subscripts $``\text{Re}"$ and $``\text{Im}"$ denote the real and imaginary parts, respectively. The unitary matrix $\Omega$ that we choose is
    \begin{equation}
    \Omega=\left(\begin{array}{ccc}
 1  & ~0~ & 0  \\
 0  & ~1~  & 0  \\
 0  & ~0~  & i
\end{array}\right)\,.
    \end{equation}
Taking into account the different values of permutations $P_l$ and $P_{\nu}$ in Eq.~\eqref{eq:PMNSfactor}, we obtain the PMNS matrix which is 
compatible with experimental data as follows
\begin{equation}
U_{\rm{PMNS}}=U_{\rm{TB}}\Omega R(\theta)=\frac{1}{\sqrt{6}}
\left(\begin{array}{ccc}
 2 \cos\theta & ~~\sqrt{2}~~ &  2\sin\theta
   \\
 -\cos\theta+i \sqrt{3} \sin\theta &
   ~~\sqrt{2}~~ & -\sin\theta-i \sqrt{3} \cos\theta \\
 -\cos\theta-i \sqrt{3} \sin\theta &
   ~~\sqrt{2}~~ & -\sin\theta+i\sqrt{3} \cos\theta
\end{array}\right)\,,
\label{eq:PMNScaseI}
\end{equation}
where the phase matrices $K_{l}$ and $K_{\nu}$ have been omitted, and for simplicity they would be omitted as well in the following cases. The rotation angle $\theta$ is determined by
\begin{equation}
\tan2\theta=-\frac{\epsilon_{\mathrm{Im}}}{\sqrt{3}\;\alpha_{\mathrm{Re}}}\,.
\end{equation}
The corresponding predictions for the lepton mixing parameters are presented in Table~\ref{tab:caseI}. Note that which quadrants all the CP phases lie in cannot be determined in the present framework, and thus we only list the absolute values of tangents or sines of the CP phases. The lepton mixing angles in particularly the sizable $\theta_{13}$ compatible with the experimental observations can be achieved. For the value of $\theta=0.184$, we have $\sin^2\theta_{13}\simeq0.0222$, $\sin^2\theta_{12}\simeq0.341$ and $\sin^2\theta_{23}=1/2$, which are in excellent agreement with the present data~\cite{Tortola:2012te,GonzalezGarcia:2012sz,Capozzi:2013csa}. Moreover, the Dirac CP violation is predicted to be maximal $\delta_{\rm{CP}}=\pm\pi/2$, while the Majorana CP phases $\alpha_{21}$ and $\alpha_{31}$ are trivial. Finally mass eigenvalues of light neutrinos are
\begin{eqnarray}
\nonumber&&m_1=\left|\beta_{\mathrm{Re}}-\gamma_{\mathrm{Re}}+ \mathrm{sign}\left(\alpha_{\mathrm{Re}}\cos2\theta\right) \sqrt{3\epsilon^2_{\mathrm{Im}}+9\alpha^2_{\mathrm{Re}}}\right|,\\
\nonumber&&m_2=\left|\beta_{\mathrm{Re}}+2\gamma_{\mathrm{Re}}\right|,\\
&&m_3=\left|\beta_{\mathrm{Re}}-\gamma_{\mathrm{Re}}- \mathrm{sign}\left(\alpha_{\mathrm{Re}}\cos2\theta\right) \sqrt{3\epsilon^2_{\mathrm{Im}}+9\alpha^2_{\mathrm{Re}}}\right|\,,
\end{eqnarray}
where four parameters $\alpha_{\mathrm{Re}}$, $\beta_{\mathrm{Re}}$, $\gamma_{\mathrm{Re}}$ and $\epsilon_{\mathrm{Im}}$ are involved in these expressions. Therefore, the measured neutrino mass-squared differences $\Delta m^2_{ij}$ (for $ij=21,31,32$) can be easily accommodated. No constraints on light neutrino masses are imposed in the present context, and hence the neutrino mass spectrum can be either normal ordering (NO) or inverted ordering (IO) in this case.

\item[~~II.~]{$H_{\rm{CP}}^{\nu}=\left\{\rho(d)X(\mathfrak{h_1}),
\rho(c^2d)X(\mathfrak{h_1})\right\}$}

The residual CP symmetry constrains the parameters as: $\alpha_{\mathrm{Im}}=-\alpha_{\mathrm{Re}}$, $\beta_{\mathrm{Re}}=-2\gamma_{\mathrm{Re}}$,
$\beta_{\mathrm{Im}}=\gamma_{\mathrm{Im}}+\sqrt{3}\;\epsilon_{\mathrm{Re}}$ and  $\epsilon_{\mathrm{Im}}=-\sqrt{3}\;\gamma_{\mathrm{Re}}$.
The unitary transformation $\Omega$ is expressed as
\begin{equation}
\Omega=
\left(\begin{array}{ccc}
 e^{\frac{5 i\pi}{8}} \sin \frac{\pi}{8} & ~0~ & e^{\frac{i\pi}{8}} \cos
   \frac{\pi}{8} \\
 0 & ~e^{\frac{i\pi}{4}}~ & 0 \\
 -e^{\frac{5 i\pi}{8}} \cos \frac{\pi
   }{8} & ~0~ & e^{\frac{i\pi}{8}}\sin\frac{\pi}{8}
\end{array}\right)\,.
\end{equation}
Using the freedom of permutations of rows and columns, the phenomenologically interesting PMNS matrix reads
{\small
\begin{eqnarray}
\nonumber&&U_{\rm{PMNS}}=P_{321}U_{\rm{TB}}\Omega R(\theta)\\
\nonumber&&=\frac{1}{\sqrt{3}}
\left(\begin{array}{ccc}
-e^{\frac{3 i\pi}{8}}
   \cos \left(\theta +\frac{\pi}{24}\right)-e^{\frac{7 i\pi}{8}} \cos \left(\theta-\frac{\pi
   }{24}\right) & ~~e^{\frac{i\pi}{4}}~ & -e^{\frac{3 i\pi}{8}} \sin
   \left(\theta +\frac{\pi}{24}\right)-e^{\frac{7 i\pi}{8}}
   \sin \left(\theta-\frac{\pi}{24}\right) \\
 e^{\frac{3 i\pi}{8}}
   \sin \left(\theta +\frac{5\pi}{24}\right)-e^{\frac{7 i\pi}{8}} \sin \left(\theta-\frac{5\pi}{24}\right) &
   ~~e^{\frac{i\pi}{4}}~ & -e^{\frac{3 i\pi}{8}}\cos
   \left(\theta +\frac{5\pi}{24}\right)+e^{\frac{7 i\pi}{8}}
   \cos \left(\theta-\frac{5\pi}{24}\right) \\
 -\cos \left(\theta-\frac{\pi}{4}\right)+e^{\frac{i\pi}{4}} \cos \left(\theta+\frac{\pi}{4}\right) & ~~e^{\frac{i\pi}{4}}~ &
   \cos \left(\theta+\frac{\pi}{4}\right)+e^{\frac{i\pi}{4}} \cos \left(\theta-\frac{\pi}{4} \right)
\end{array}\right)\,,\\
&&\label{eq:PMNS_II}
\end{eqnarray}}
\!\!where the angle $\theta$ is 
\begin{equation}
\tan2\theta=\frac{\sqrt{3}\;\gamma_{\mathrm{Re}}-\epsilon_{\mathrm{Re}}}{\sqrt{6}\;\alpha_{\mathrm{Re}}}\,.
\end{equation}
All the mixing angles and CP-violating phases are expressed in terms of the parameter $\theta$, as shown in Table~\ref{tab:caseII_VI_and_IV_VIII}. Light neutrino masses in this case are shown as
\begin{eqnarray}
\nonumber&&\hskip-0.2in m_1=\left|3\gamma_{\mathrm{Re}}+\sqrt{3}\;\epsilon_{\mathrm{Re}}+
\mathrm{sign}\left(\alpha_{\mathrm{Re}}\cos2\theta\right)\sqrt{18\alpha^2_{\mathrm{Re}}+\left(3\gamma_{\mathrm{Re}}-\sqrt{3}\;\epsilon_{\mathrm{Re}}\right)^2}\right|,\\
\nonumber&&\hskip-0.2in m_2=\left|3\gamma_{\mathrm{Im}}+\sqrt{3}\;\epsilon_{\mathrm{Re}}\right|,\\
\label{eq:nmass_II}&&\hskip-0.2in m_3=\left|3\gamma_{\mathrm{Re}}+\sqrt{3}\;\epsilon_{\mathrm{Re}}-
\mathrm{sign}\left(\alpha_{\mathrm{Re}}\cos2\theta\right)\sqrt{18\alpha^2_{\mathrm{Re}}+\left(3\gamma_{\mathrm{Re}}-\sqrt{3}\;\epsilon_{\mathrm{Re}}\right)^2}\right|\,.
\end{eqnarray}
Following analogous arguments as before, we see that light neutrino masses remain unconstrained in this case as well.

\item[~~III.~]{$\left\{H_{\rm{CP}}^{\nu}=\rho(d^2)X(\mathfrak{h_1}),\rho(c^2d^2)X(\mathfrak{h_1})\right\}$}

The parameters $\alpha$, $\beta$, $\gamma$ and $\epsilon$ are constrained to satisfy $\alpha_{\mathrm{Re}}=\beta_{\mathrm{Im}}=\gamma_{\mathrm{Im}}=\epsilon_{\mathrm{Re}}=0$. We find the unitary matrix
\begin{equation}
\Omega=\frac{1}{\sqrt{2}}
\left(\begin{array}{ccc}
 e^{\frac{i \pi }{4}} & ~0~ & -e^{-\frac{i \pi
   }{4}} \\
 0 & ~\sqrt{2}~ & 0 \\
 -e^{\frac{i \pi }{4}} & ~0~ & -e^{-\frac{i \pi
   }{4}}
\end{array}\right)\,.
\end{equation}
The PMNS matrix takes the form
\begin{equation}
 U_{\rm{PMNS}}=P_{312}U_{\rm{TB}}\Omega R(\theta)=\frac{1}{2\sqrt{3}}
\left(\begin{array}{ccc}
 -e^{\frac{i\pi }{4}}\left(e^{-i \theta }+\sqrt{3} e^{i \theta }\right) &
   ~~2~~ & e^{-\frac{i\pi }{4}}\left(e^{-i\theta}-\sqrt{3} e^{i\theta}\right) \\
  2 e^{-i\left(\theta-\frac{\pi }{4} \right)} & ~~2~~ &  -2 e^{-i \left(\theta +\frac{\pi
   }{4}\right)} \\
 -e^{\frac{i\pi }{4}}\left(e^{-i \theta }-\sqrt{3} e^{i \theta }\right) &
   ~~2~~ & e^{-\frac{i\pi }{4}}\left(e^{-i \theta }+\sqrt{3} e^{i\theta}\right)
\end{array}\right)\,,
\end{equation}
with
\begin{equation}
\tan2\theta=\frac{\gamma_{\mathrm{Re}}-\beta_{\mathrm{Re}}}{3\alpha_{\mathrm{Im}}}\,.
\end{equation}
Predictions for lepton mixing angles and CP-violating phases are collected in Table~\ref{tab:caseII_VI_and_IV_VIII}. Light neutrino masses are given by
\begin{eqnarray}
\nonumber&&m_1=\left|\sqrt{3}\;\epsilon_{\mathrm{Im}}+\mathrm{sign}\left(\alpha_{\mathrm{Im}}\cos2\theta\right)\sqrt{9 \alpha^2_{\mathrm{Im}}+(\beta_{\mathrm{Re}}-\gamma_{\mathrm{Re}})^2}\right|,\\
\nonumber&&m_2=\left|\beta_{\mathrm{Re}}+2\gamma_{\mathrm{Re}}\right|,\\
&&m_3=\left|\sqrt{3}\;\epsilon_{\mathrm{Im}}-\mathrm{sign}\left(\alpha_{\mathrm{Im}}\cos2\theta\right)\sqrt{9\alpha^2_{\mathrm{Im}}+(\beta_{\mathrm{Re}}-\gamma_{\mathrm{Re}})^2}\right|\,.
\end{eqnarray}

\item[~~IV.~]{$H_{\rm{CP}}^{\nu}=\left\{\rho(d^3)X(\mathfrak{h_1}), \rho(c^2d^3)X(\mathfrak{h_1})\right\}$}

This residual CP symmetry implies $\alpha_{\mathrm{Im}}=\alpha_{\mathrm{Re}}$,  $\beta_{\mathrm{Re}}=-2\gamma_{\mathrm{Re}}$, $\beta_{\mathrm{Im}}=\gamma_{\mathrm{Im}}+\sqrt{3}\;\epsilon_{\mathrm{Re}}$ and $\epsilon_{\mathrm{Im}}=-\sqrt{3}\;\gamma_{\mathrm{Re}}$.
The unitary transformation $\Omega$ takes the form
\begin{equation}
\label{eq:omega_IV} \Omega=
\left(\begin{array}{ccc}
 -e^{\frac{3 i\pi}{8}} \cos \frac{\pi
   }{8} & ~0~ & e^{-\frac{i\pi}{8}} \sin
   \frac{\pi}{8} \\
 0 & ~e^{-\frac{i\pi}{4}}~ & 0 \\
 e^{\frac{3 i\pi}{8}} \sin \frac{\pi
   }{8} & ~0~ & e^{-\frac{i\pi}{8}}\cos\frac{\pi}{8}
\end{array}\right)\,.
\end{equation}
The PMNS matrix takes the following form
{\small
\begin{eqnarray}
\nonumber&&U_{\rm{PMNS}}=P_{231}U_{\rm{TB}}\Omega R(\theta)P_{321}\\
\nonumber&&=\frac{1}{\sqrt{3}}
\left(\begin{array}{ccc}
 -e^{-\frac{3i\pi}{8}}
   \cos \left(\theta-\frac{\pi}{24}\right)-e^{\frac{i\pi}{8}} \cos \left(\theta+\frac{\pi}{24}\right) &
   ~~e^{-\frac{i\pi}{4}}~ & e^{-\frac{3i\pi}{8}}
   \sin \left(\theta-\frac{\pi}{24}\right)+e^{\frac{i\pi}{8}} \sin
   \left(\theta +\frac{\pi}{24}\right) \\
 -e^{-\frac{3i\pi}{8}} \sin \left(\theta-\frac{5 \pi
   }{24}\right)+e^{\frac{i\pi}{8}}
   \sin \left(\theta +\frac{5\pi}{24}\right) &
   e^{-\frac{i\pi}{4}} & -e^{-\frac{3i\pi}{8}}
   \cos \left(\theta-\frac{5\pi}{24}
   \right)+e^{\frac{i\pi}{8}} \cos
   \left(\theta +\frac{5\pi}{24}\right) \\
-i\cos \left(\theta-\frac{\pi}{4}\right)+e^{\frac{i\pi}{4}}\cos\left(\theta+\frac{\pi }{4}\right) &
   e^{-\frac{i\pi}{4}} & -i\cos\left(\theta+\frac{\pi}{4}\right)-e^{\frac{i\pi}{4}}\cos\left(\theta-\frac{\pi}{4}\right)
\end{array}\right)\,,\\
\label{eq:PMNS_IV}&&
\end{eqnarray}
}
with
\begin{equation}
\tan2\theta=\frac{\sqrt{3}\;\gamma_{\mathrm{Re}}+\epsilon_{\mathrm{Re}}}{\sqrt{6}\;\alpha_{\mathrm{Re}}}\,.
\end{equation}
Up to a factor of $\pi$ for the Majorana phase $\alpha_{21}$, the PMNS matrix in this case is the complex conjugate of the PMNS matrix of Eq.~\eqref{eq:PMNS_II} in case II. Given that the quadrants of the CP phases are unpredictable, we obatin the same mixing parameters as those of case II, as shown in Table~\ref{tab:caseII_VI_and_IV_VIII}. However, light neutrino masses take different values as follows
\begin{eqnarray}
\nonumber&&\hskip-0.2in
m_1=\left|3\gamma_{\mathrm{Re}}-\sqrt{3}\;\epsilon_{\mathrm{Re}}+\mathrm{sign}\left(\alpha_{\mathrm{Re}}\cos2\theta\right)\sqrt{18\alpha^2_{\mathrm{Re}}+\left(3\gamma_{\mathrm{Re}}+\sqrt{3}\;\epsilon_{\mathrm{Re}}\right)^2}\right|,\\
\nonumber&&\hskip-0.2in
m_2=\left|3\gamma_{\mathrm{Im}}+\sqrt{3}\;\epsilon_{\mathrm{Re}}\right|,\\
\label{eq:nmass_IV}&&\hskip-0.2in
m_3=\left|3\gamma_{\mathrm{Re}}-\sqrt{3}\;\epsilon_{\mathrm{Re}}-\mathrm{sign}\left(\alpha_{\mathrm{Re}}\cos2\theta\right)\sqrt{18\alpha^2_{\mathrm{Re}}+\left(3\gamma_{\mathrm{Re}}+\sqrt{3}\;\epsilon_{\mathrm{Re}}\right)^2}\right|\,.
\end{eqnarray}

\item[~~V.~]{$H_{\rm{CP}}^{\nu}=\left\{\rho(c)X(\mathfrak{h_1}), \rho(c^3)X(\mathfrak{h_1})\right\}$}

Invariance under this remnant CP leads to $\alpha_{\mathrm{Im}}=\epsilon_{\mathrm{Im}}=0$, $\beta_{\mathrm{Re}}=\gamma_{\mathrm{Re}}$ and $\beta_{\mathrm{Im}}=-2\gamma_{\mathrm{Im}}$. The unitary transformation $\Omega$ turns out to be
\begin{equation}
    \Omega=\frac{1}{\sqrt{2}}
\left(\begin{array}{ccc}
1 & ~0~ & i \\
 0 & ~\sqrt{2}~~ & 0 \\
 -1 & ~0~ & i
\end{array}\right)\,.
\end{equation}
The ``best'' PMNS matrix reads
\begin{equation}
U_{\rm{PMNS}}=P_{312}U_{\rm{TB}}\Omega R(\theta)=\frac{1}{2\sqrt{3}}
\left(\begin{array}{ccc}
 -e^{-i\theta}-\sqrt{3} e^{i \theta } &
   ~~2~~ & -i e^{-i \theta }+i \sqrt{3} e^{i
   \theta } \\
 2 e^{-i \theta } & ~~2~~ & 2 i e^{-i\theta} \\
 -e^{-i \theta }+\sqrt{3} e^{i \theta } &
   ~~2~~ & -i e^{-i \theta }-i \sqrt{3} e^{i\theta }
\end{array}\right)\,,
\end{equation}
where the rotation angle $\theta$ is
\begin{equation}
\tan2\theta=-\frac{\gamma_{\mathrm{Im}}}{\alpha_{\mathrm{Re}}}\,.
\end{equation}
The results for the mixing parameters are the same as those of case III except the Majorana phase $\alpha_{21}$, as shown in Table~\ref{tab:caseII_VI_and_IV_VIII}. The light neutrino masses are given by
\begin{eqnarray}
\nonumber&&m_1=\left|\sqrt{3}\;\epsilon_{\mathrm{Re}}+3\mathrm{sign}\left(\alpha_{\mathrm{Re}}\cos2\theta\right)
   \sqrt{\alpha^2_{\mathrm{Re}}+\gamma^2_{\mathrm{Im}}}\right|,\\
\nonumber&&m_2=3\left|\gamma_{\mathrm{Re}}\right|,\\
&&m_3=\left|\sqrt{3}\;\epsilon_{\mathrm{Re}}-3\mathrm{sign}\left(\alpha_{\mathrm{Re}}\cos2\theta\right)
   \sqrt{\alpha^2_{\mathrm{Re}}+\gamma^2_{\mathrm{Im}}}\right|\,.
\end{eqnarray}

\item[~~VI.~]{$H_{\rm{CP}}^{\nu}=\left\{\rho(cd)X(\mathfrak{h_1}), \rho(c^3d)X(\mathfrak{h_1})\right\}$}

The constraints on the parameters $\alpha$, $\beta$, $\gamma$ and $\epsilon$ are $\alpha_{\mathrm{Im}}=-\alpha_{\mathrm{Re}}$, $\beta_{\mathrm{Re}}=-2\gamma_{\mathrm{Re}}$, $\beta_{\mathrm{Im}}=\gamma_{\mathrm{Im}}-\sqrt{3}\;\epsilon_{\mathrm{Re}}$ and $\epsilon_{\mathrm{Im}}=\sqrt{3}\;\gamma_{\mathrm{Re}}$. The unitary transformation $\Omega$ is
\begin{equation}
\label{eq:omega_VI}\Omega=
\left(\begin{array}{ccc}
 e^{\frac{5 i\pi}{8}}\cos\frac{\pi}{8} & ~0~ &   e^{\frac{i\pi}{8}}\sin\frac{\pi}{8} \\
 0 & ~e^{\frac{i\pi}{4}}~ & 0 \\
 -e^{\frac{5 i\pi}{8}}\sin\frac{\pi}{8} & ~0~ &  e^{\frac{i\pi}{8}}\cos\frac{\pi}{8}
\end{array}\right)\,.
\end{equation}
The phenomenologically interesting PMNS matrix takes the form
{\small
\begin{eqnarray}
\nonumber&&U_{\rm{PMNS}}=P_{231}U_{\rm{TB}}\Omega R(\theta)P_{321}\\
\nonumber&&=\frac{1}{\sqrt{3}}\left(\begin{array}{ccc}
-e^{\frac{3i\pi}{8}} \cos \left(\theta-\frac{\pi
   }{24} \right)+e^{\frac{7 i\pi}{8}} \cos \left(\theta
   +\frac{\pi}{24}\right) & ~~e^{\frac{i\pi}{4}}~ & e^{\frac{3 i\pi}{8}} \sin\left(\theta-\frac{\pi}{24}\right)-e^{\frac{7 i\pi}{8}}\sin\left(\theta +\frac{\pi}{24}\right)
   \\
 -e^{\frac{3 i\pi}{8}}\sin \left(\theta-\frac{5\pi}{24}
   \right)-e^{\frac{7 i\pi}{8}}\sin\left(\theta +\frac{5\pi}{24}\right)
   & ~~e^{\frac{i\pi}{4}}~ & -e^{\frac{3 i \pi
   }{8}} \cos \left(\theta-\frac{5 \pi
   }{24} \right)-e^{\frac{7i\pi}{8}} \cos \left(\theta +\frac{5
\pi}{24}\right) \\
 i \cos \left(\theta-\frac{\pi}{4}\right)-e^{\frac{3 i\pi}{4}} \cos  \left(\theta +\frac{\pi}{4}\right) &
   ~~e^{\frac{i\pi}{4}}~ & i \cos\left(\theta +\frac{\pi}{4}\right)+e^{\frac{3 i
\pi}{4}} \cos \left(\theta-\frac{\pi}{4} \right)
\end{array}\right)\,.\\
\label{eq:PMNS_VI}&&
\end{eqnarray}
}
where
\begin{equation}
\tan2\theta=\frac{\sqrt{3}\;\gamma_{\mathrm{Re}}+\epsilon_{\mathrm{Re}}}{\sqrt{6}\;\alpha_{\mathrm{Re}}}\,.
\end{equation}
We see that the mixing matrix in Eq.~\eqref{eq:PMNS_VI} is identical to the complex conjugate of the corresponding one of case IV. Predictions for mixing parameters are listed in Table~\ref{tab:caseII_VI_and_IV_VIII}, and they are the same as those of cases II and IV. Light neutrino masses are determined to be
\begin{eqnarray}
\nonumber&&m_1=\left|3\gamma_{\mathrm{Re}}-\sqrt{3}\;\epsilon_{\mathrm{Re}}+\mathrm{sign}\left(\alpha_{\mathrm{Re}}\cos2\theta\right)\sqrt{18\alpha^2_{\mathrm{Re}}+\left(3\gamma_{\mathrm{Re}}+\sqrt{3}\;\epsilon_{\mathrm{Re}}\right)^2}\right|,\\
\nonumber&&m_2=\left|3\gamma_{\mathrm{Im}}-\sqrt{3}\;\epsilon_{\mathrm{Re}}\right|,\\
\label{eq:nmass_VI}&&m_3=\left|3\gamma_{\mathrm{Re}}-\sqrt{3}\;\epsilon_{\mathrm{Re}}-\mathrm{sign}\left(\alpha_{\mathrm{Re}}\cos2\theta\right)\sqrt{18\alpha^2_{\mathrm{Re}}+\left(3\gamma_{\mathrm{Re}}+\sqrt{3}\;\epsilon_{\mathrm{Re}}\right)^2}\right|\,.
\end{eqnarray}

\item[~~VII.~]{$H_{\rm{CP}}^{\nu}=\left\{\rho(cd)X(\mathfrak{h_1}), \rho(c^3d)X(\mathfrak{h_1})\right\}$}

We find the relations $\alpha_{\mathrm{Re}}=\epsilon_{\mathrm{Im}}=0$, $\beta_{\mathrm{Re}}=\gamma_{\mathrm{Re}}$ and $\beta_{\mathrm{Im}}=-2\gamma_{\mathrm{Im}}$ satisfied, which leads to
\begin{equation}
\Omega=\left(\begin{array}{ccc}
 e^{\frac{i\pi}{4}} & ~0~ & 0 \\
 0 & ~1~ & 0 \\
 0 & ~0~ & e^{\frac{3 i\pi}{4}}
\end{array}\right)\,.
\end{equation}
and the PMNS matrix
\begin{equation}
U_{\rm{PMNS}}=U_{\rm{TB}}\Omega R(\theta)=\frac{1}{\sqrt{6}}
\left(\begin{array}{ccc}
 2 e^{\frac{i\pi}{4}} \cos \theta &
   ~~\sqrt{2}~~ & 2e^{\frac{i\pi}{4}}\sin\theta \\
 -e^{\frac{i\pi}{4}}\cos\theta +\sqrt{3}\;e^{\frac{3 i\pi}{4}} \sin
   \theta &  ~~\sqrt{2}~~ & -e^{\frac{i\pi}{4}}\sin\theta-\sqrt{3}\;e^{\frac{3 i\pi}{4}} \cos\theta
   \\
 -e^{\frac{i\pi}{4}} \cos\theta-\sqrt{3}\;e^{\frac{3 i\pi}{4}} \sin  \theta & ~~\sqrt{2}~~ & -e^{\frac{i\pi}{4}}\sin\theta+\sqrt{3}\; e^{\frac{3i\pi}{4}}\cos\theta
\end{array}\right)\,,
\end{equation}
with
\begin{equation}
\tan2\theta=\frac{\epsilon_{\mathrm{Re}}}{\sqrt{3}\;\alpha_{\mathrm{Im}}}\,.
\end{equation}
The mixing parameters turn out to be the same as those of case I except that the Majorana phase $\alpha_{21}$ is maximal instead of zero in this scenario, as can be found from Table~\ref{tab:caseII_VI_and_IV_VIII}. The light neutrino masses are
\begin{eqnarray}
\nonumber&&m_1=\left|3 \gamma_{\mathrm{Im}}-\mathrm{sign}\left(\alpha_{\mathrm{Im}}\cos2\theta\right)\sqrt{9
   \alpha^2_{\mathrm{Im}}+3\epsilon^2_{\mathrm{Re}}}\right|,\\
\nonumber&&m_2=3\left|\gamma_{\mathrm{Re}}\right|,\\
&&m_3=\left|3 \gamma_{\mathrm{Im}}+\mathrm{sign}\left(\alpha_{\mathrm{Im}}\cos2\theta\right)\sqrt{9
   \alpha^2_{\mathrm{Im}}+3\epsilon^2_{\mathrm{Re}}}\right|\,.
\end{eqnarray}

\item[~~VIII.~]{$H_{\rm{CP}}^{\nu}=\left\{\rho(cd^3)X(\mathfrak{h_1}), \rho(c^3d^3)X(\mathfrak{h_1})\right\}$}

In this case, the residual CP symmetry constrains the parameters as: $\alpha_{\mathrm{Im}}=\alpha_{\mathrm{Re}}$, $\beta_{\mathrm{Re}}=-2\gamma_{\mathrm{Re}}$, $\beta_{\mathrm{Im}}=\gamma_{\mathrm{Im}}-\sqrt{3}\;\epsilon_{\mathrm{Re}}$ and $\epsilon_{\mathrm{Im}}=\sqrt{3}\;\gamma_{\mathrm{Re}}$. The unitary transformation $\Omega$ is
\begin{equation}
\Omega=\left(\begin{array}{ccc}
 -e^{\frac{3 i\pi}{8}} \sin \frac{\pi}{8} & ~0~ & e^{-\frac{i\pi}{8}} \cos\frac{\pi}{8} \\
 0 & ~e^{-\frac{i\pi}{4}}~ & 0 \\
 e^{\frac{3 i\pi}{8}}\cos\frac{\pi}{8} & ~0~ & e^{-\frac{i\pi}{8}}\sin\frac{\pi}{8}
\end{array}\right)\,,
\end{equation}
and the desired PMNS matrix is given by
{\small
\begin{eqnarray}
\nonumber&&U_{\rm{PMNS}}=P_{321}U_{\rm{TB}}\Omega R(\theta)\\
\nonumber&&=\frac{1}{\sqrt{3}}
\left(\begin{array}{ccc}
-e^{-\frac{3i\pi}{8}} \cos
   \left(\theta+\frac{\pi}{24}\right)+e^{\frac{i\pi}{8}}\cos \left(\theta-\frac{\pi}{24}\right) &
   ~~e^{-\frac{i\pi}{4}}~ & -e^{-\frac{3i\pi}{8}} \sin \left(\theta   +\frac{\pi}{24}\right)+e^{\frac{i\pi}{8}}\sin \left(\theta-\frac{\pi}{24}\right) \\
 e^{-\frac{3i\pi}{8}}\sin\left(\theta+\frac{5\pi}{24}\right)+e^{\frac{i\pi}{8}} \sin \left(\theta-\frac{5\pi}{24} \right) &
   ~~e^{-\frac{i\pi}{4}}~ & -e^{-\frac{3i\pi}{8}} \cos\left(\theta+   \frac{5\pi}{24}\right)-e^{\frac{i\pi}{8}} \cos \left(\theta-\frac{5\pi}{24}\right) \\
 -\cos\left(\theta-\frac{\pi }{4}\right)+e^{-\frac{i\pi }{4}}\cos \left(\theta +\frac{\pi}{4}\right) &
 ~e^{-\frac{i\pi}{4}}~ & \cos \left(\theta +\frac{\pi}{4}\right) +e^{-\frac{i\pi}{4}}\cos \left(\theta-\frac{\pi}{4}\right)
\end{array}\right)\,,\\
\label{eq:PMNS_VIII}&&
\end{eqnarray}
}
where
\begin{equation}
\tan2\theta=\frac{\sqrt{3}\;\gamma_{\mathrm{Re}}-\epsilon_{\mathrm{Re}}}{\sqrt{6}\;\alpha_{\mathrm{Re}}}\,.
\end{equation}
Notice that the PMNS matrix in Eq.~\eqref{eq:PMNS_VIII} is precisely the complex conjugate of the corresponding one of case II. Predictions for mixing parameters are presented in Table~\ref{tab:caseII_VI_and_IV_VIII}, and they are exactly the same as those of cases II, IV and VI. We have the light neutrino masses
\begin{eqnarray}
\nonumber&&m_1=\left|3\gamma_{\mathrm{Re}}+\sqrt{3}\;\epsilon_{\mathrm{Re}}+\mathrm{sign}\left(\alpha_{\mathrm{Re}}\cos2\theta\right)\sqrt{18\alpha^2_{\mathrm{Re}}+\left(3\gamma_{\mathrm{Re}}-\sqrt{3}\;\epsilon_{\mathrm{Re}}\right)^2}\right|,\\
\nonumber&&m_2=\left|3\gamma_{\mathrm{Im}}-\sqrt{3}\;\epsilon_{\mathrm{Re}}\right|,\\
\label{eq:nmass_VIII}&&m_3=\left|3\gamma_{\mathrm{Re}}+\sqrt{3}\;\epsilon_{\mathrm{Re}}-\mathrm{sign}\left(\alpha_{\mathrm{Re}}\cos2\theta\right)\sqrt{18\alpha^2_{\mathrm{Re}}+\left(3\gamma_{\mathrm{Re}}-\sqrt{3}\;\epsilon_{\mathrm{Re}}\right)^2}\right|\,.
\end{eqnarray}

\item[~~IX.~]{$H_{\rm{CP}}^{\nu}=\left\{X(\mathfrak{h_3}), \rho(c^2)X(\mathfrak{h_3})\right\}$}

The neutrino mass matrix $m_{\nu}$ is constrained to be real such that $\alpha_{\mathrm{Im}}=\beta_{\mathrm{Im}}=\gamma_{\mathrm{Im}}=\epsilon_{\mathrm{Im}}=0$ arises. The unitary transformation $\Omega$ is trivial $\Omega=\mathbbm{1}_3$ in this case, and
we can straightforwardly obtain the corresponding PMNS matrix as follows
\begin{equation}
U_{\rm{PMNS}}=P_{132}U_{\rm{TB}}\Omega R(\theta)=\frac{1}{\sqrt{6}}
\left(\begin{array}{ccc}
 2\cos\theta & ~~\sqrt{2}~~ &  2\sin\theta   \\
 -\cos\theta-\sqrt{3}\;\sin\theta &    ~~\sqrt{2}~~ & -\sin\theta+\sqrt{3}\;\cos\theta \\
 -\cos\theta+\sqrt{3}\;\sin\theta &   ~~\sqrt{2}~~ & -\sin\theta-\sqrt{3}\;\cos\theta
\end{array}\right)\,,
\end{equation}
with the angle $\theta$
\begin{equation}
\tan2\theta=\frac{\sqrt{3}\;\epsilon_{\mathrm{Re}}}{\beta_{\mathrm{Re}}-\gamma_{\mathrm{Re}}}\,.
\end{equation}
Expressions for lepton mixing parameters are shown in Table~\ref{tab:caseII_VI_and_IV_VIII}, both Dirac and Majorana CP phases are predicted to be trivial. The deviation of the atmospheric mixing from maximal mixing is expected. The light neutrino masses are determined to be
\begin{eqnarray}
\nonumber&&m_1=\left|3\alpha_{\mathrm{Re}}+\mathrm{sign}\left(\left(\beta_{\mathrm{Re}}-\gamma_{\mathrm{Re}}\right)\cos2\theta\right)\sqrt{\left(\beta_{\mathrm{Re}}-\gamma_{\mathrm{Re}}\right)^2+3\epsilon^2_{\mathrm{Re}}}\right|,\\
\nonumber&&m_2=\left|\beta_{\mathrm{Re}}+2\gamma_{\mathrm{Re}}\right|,\\
&&m_3=\left|3\alpha_{\mathrm{Re}}-\mathrm{sign}\left(\left(\beta_{\mathrm{Re}}-\gamma_{\mathrm{Re}}\right)\cos2\theta\right)\sqrt{\left(\beta_{\mathrm{Re}}-\gamma_{\mathrm{Re}}\right)^2+3\epsilon^2_{\mathrm{Re}}}\right|\,.
\end{eqnarray}

\item[~~X.~]{$H_{\rm{CP}}^{\nu}=\left\{\rho(cd^2)X(\mathfrak{h_3}), \rho(c^3d^2)X(\mathfrak{h_3})\right\}$}

The residual CP symmetry leads to $\alpha_{\mathrm{Re}}=\epsilon_{\mathrm{Re}}=0$, $\beta_{\mathrm{Re}}=\gamma_{\mathrm{Re}}$ and $\beta_{\mathrm{Im}}=-2\gamma_{\mathrm{Im}}$.
In this case, the unitary transformation $\Omega$ is of the form
\begin{equation}
\Omega=\left(\begin{array}{ccc}
 e^{\frac{i\pi}{4}} & ~~0~~ & 0 \\
 0 & ~~1~~ & 0 \\
 0 & ~~0~~ & e^{\frac{i\pi}{4}}
\end{array}\right)\,.
\end{equation}
The resulting ``best'' PMNS matrix
\begin{equation}
U_{\rm{PMNS}}=P_{132}U_{\rm{TB}}\Omega R(\theta)=\frac{1}{\sqrt{6}}
\left(\begin{array}{ccc}
 2 e^{\frac{i \pi }{4}} \cos \theta & ~~\sqrt{2}~~
   & 2 e^{\frac{i \pi }{4}} \sin \theta \\
 e^{\frac{i \pi }{4}} \left(-\cos\theta-\sqrt{3}\;\sin\theta\right) & ~~\sqrt{2}~~ &    e^{\frac{i\pi}{4}}\left(-\sin\theta+\sqrt{3}\;\cos\theta\right) \\
 e^{\frac{i \pi }{4}} \left(-\cos\theta+\sqrt{3}\;\sin\theta\right) & ~~\sqrt{2}~~ & e^{\frac{i\pi}{4}}\left(-\sin\theta-\sqrt{3}\;\cos\theta\right)
\end{array}\right)\,,
\end{equation}
is obtained, where
\begin{equation}
\tan2\theta=-\frac{\epsilon_{\mathrm{Im}}}{\sqrt{3}\;\gamma_{\mathrm{Im}}}\,.
\end{equation}
Predictions for mixing parameters are the same as those of case IX aside from the Majorana phase $\alpha_{21}$. Finally we present analytical results for light neutrino masses
\begin{eqnarray}
\nonumber&&m_1=\left|3\alpha_{\mathrm{Im}}-\mathrm{sign}\left(\gamma_{\mathrm{Im}}\cos2\theta\right)\sqrt{9
   \gamma^2_{\mathrm{Im}}+3\epsilon^2_{\mathrm{Im}}}\right|,\\
   \nonumber&&m_2=3\left|\gamma_{\mathrm{Re}}\right|,\\
&&m_3=\left|3\alpha_{\mathrm{Im}}+\mathrm{sign}\left(\gamma_{\mathrm{Im}}\cos2\theta\right)\sqrt{9
   \gamma^2_{\mathrm{Im}}+3\epsilon^2_{\mathrm{Im}}}\right|\,.
\end{eqnarray}

\end{description}

%%%%%%%%%%%%%%%%%%%%%%%%%%%%%%%%%%%%%%%%%%%%%%%%%%%%%%%%%%%%%%%%%%%%%%%%%%%%%%%%%%%%%%%%%%%%%%%%%%%%%%%%%%%%%%%%%%%%%%%%%%%%%%%%%%%%%%%%%%%%%%%%%%
\begin{table}[t!]
\begin{center}
\begin{tabular}{| c | c | c | c| c | c| c| }
\hline\hline
%%%%%%%%%%%%%%%%%%%%%
  &  ~~\text{\tt I}~~  &    ~\text{\tt VII}~   & ~\text{\tt IX}~ &  \text{\tt X}  & \text{\tt III}  &   \text{\tt V} \\  \hline

$\sin^2\theta_{13}$  & \multicolumn{4}{c|}{$\frac{2}{3}\sin^2\theta$}   & \multicolumn{2}{c|}{$\frac{1}{3}-\frac{\cos2\theta}{2\sqrt{3}}$ }      \\  \hline

$\sin^2\theta_{12}$  &  \multicolumn{4}{c|}{$\frac{1}{2+\cos2\theta}$}  & \multicolumn{2}{c|}{$\frac{2}{4+\sqrt{3}\cos2\theta}$ }  \\  \hline

\multirow{2}{*}{$\sin^2\theta_{23}$}  &  \multicolumn{2}{c|}{\multirow{2}{*}{$\frac{1}{2}$}} &  \multicolumn{2}{c|}{$\frac{1}{2}-\frac{\sqrt{3}\sin2\theta}{4+2\cos 2\theta}$} & \multicolumn{2}{c|}{$\frac{2}{4+\sqrt{3}\cos2\theta}$ }  \\\cdashline{4-7}

 & \multicolumn{2}{c|}{} & \multicolumn{2}{c|}{$\frac{1}{2}+\frac{\sqrt{3}\sin2\theta}{4+2\cos2\theta}$}   & \multicolumn{2}{c|}{$\frac{2+\sqrt{3}\cos2\theta}{4+\sqrt{3}\cos2\theta}$}  \\\hline

$\left|J_{\text{CP}}\right|$  & \multicolumn{2}{c|}{$\frac{1}{6\sqrt{3}}\left|\sin2\theta\right|$} &  \multicolumn{2}{c|}{0} &   \multicolumn{2}{c|}{$\frac{1}{6\sqrt{3}}\left|\sin2\theta\right|$}  \\\hline

$\left|\tan\delta_{\text{CP}}\right|$  & \multicolumn{2}{c|}{$+\infty$}  &  \multicolumn{2}{c|}{0}  & \multicolumn{2}{c|}{$\big|\frac{4+\sqrt{3}\cos2\theta}{1+\sqrt{3}\cos2\theta}\tan2\theta\big|$}     \\   \hline

$\left|\tan\alpha_{21}\right|$  & 0  &  $+\infty$  &   0 &   $+\infty$ & $\big|\frac{\sqrt{3}+2\cos2\theta}{\sin2\theta}\big|$ &   $\big|\frac{\sin2\theta}{\sqrt{3}+2\cos2\theta}\big|$   \\   \hline

$\left|\tan\alpha^{\prime}_{31}\right|$  &  \multicolumn{4}{c|}{0}  & \multicolumn{2}{c|}{$\big|\frac{4\sqrt{3}\sin2\theta}{1-3\cos4\theta}\big|$}  \\ \hline\hline

\multicolumn{7}{|c|}{\tt Best Fits}  \\\hline\hline

$\theta_{\mathrm{bf}}$  &    \multicolumn{2}{c|}{0.184}  & \multicolumn{2}{c|}{0.182}  &   \multicolumn{2}{c|}{0}  \\\hline

$\chi^2_{\mathrm{min}}(\theta_{23}<\pi/4)$  &  \multicolumn{2}{c|}{14.527}  &  \multicolumn{2}{c|}{9.548}  &   \multicolumn{2}{c|}{110.741}  \\\cdashline{1-7}

$\chi^2_{\mathrm{min}}(\theta_{23}>\pi/4)$  &  \multicolumn{2}{c|}{27.254}  &   \multicolumn{2}{c|}{9.303}  &  \multicolumn{2}{c|}{111.559} \\\hline

$\sin^2\theta_{13}$  & \multicolumn{2}{c|}{0.0222}  & \multicolumn{2}{c|}{0.0218} & \multicolumn{2}{c|}{0.0447}     \\  \hline

$\sin^2\theta_{12}$  &  \multicolumn{2}{c|}{0.341} &  \multicolumn{2}{c|}{0.341} &  \multicolumn{2}{c|}{0.349} \\  \hline

\multirow{2}{*}{$\sin^2\theta_{23}$}  &  \multicolumn{2}{c|}{\multirow{2}{*}{0.5}} &  \multicolumn{2}{c|}{0.395} & \multicolumn{2}{c|}{0.349}  \\\cdashline{4-7}

 & \multicolumn{2}{c|}{} & \multicolumn{2}{c|}{0.605}   &  \multicolumn{2}{c|}{0.651} \\\hline

$\left|J_{\text{CP}}\right|$  & \multicolumn{2}{c|}{0.0346} &  \multicolumn{2}{c|}{0} & \multicolumn{2}{c|}{0}   \\ \hline

$\left|\sin\delta_{\text{CP}}\right|$  & \multicolumn{2}{c|}{1}  &  \multicolumn{2}{c|}{0}  & \multicolumn{2}{c|}{0}   \\   \hline

$\left|\sin\alpha_{21}\right|$  & 0  &  1 &   0 &   1  & 1  &   0  \\   \hline

$\left|\sin\alpha^{\prime}_{31}\right|$  &  \multicolumn{4}{c|}{0}  & \multicolumn{2}{c|}{0} \\ \hline\hline

\end{tabular}
\caption{\label{tab:caseI}Predictions for mixing parameters in cases I, III, V, VII, IX and X, where $``+\infty"$ for $\left|\tan\delta_{\text{CP}}\right|$, $\left|\tan\alpha_{21}\right|$ and $\left|\tan\alpha^{\prime}_{31}\right|$ denotes that the absolute value of the corresponding CP phase is $\pi/2$.  The Majorana phase $\alpha^{\prime}_{31}$ has been redefined to include the Dirac CP phase by $\alpha^{\prime}_{31}=\alpha_{31}-2\delta_{\text{CP}}$. Predictions for exchanging the second and the third rows of the PMNS matrix are shown below the dashed line.}
\end{center}
\end{table}

%%%%%%%%%%%%%%%%%%%%%%%%%%%%%%%%%%%%%%%%%%%%%%%%%%%%%%%%%%%%%%%%%%%%%%%%%%%%%%%%%%%%%%%%%%%%%%%%%%%%%%%%%%%%%%%%%%%%%%%%%%%%%%%%%%%%%%%%%%%%%%%%%%%

\begin{table}[t!]
\begin{center}
%\resizebox{\textwidth}{!}{
\begin{tabular}{|c|c|}
\hline\hline
%%%%%%%%%%%%%%%%%%%%%
 & ~~\text{\tt II, IV, VI and VIII}~~  \\  \hline

$\sin^2\theta_{13}$  & $\frac{1}{3}-\frac{1+\sqrt{3}}{6\sqrt{2}}\cos2\theta$  \\   \hline

$\sin^2\theta_{12}$  &  $\frac{2 \sqrt{2}}{4\sqrt{2}+\left(1+\sqrt{3}\right) \cos2 \theta}$     \\  \hline

\multirow{2}{*}{$\sin^2\theta_{23}$}  &   $\frac{2\sqrt{2}-\left(1-\sqrt{3}\right)\cos2\theta}{4\sqrt{2}+\left(1+\sqrt{3}\right) \cos2\theta}$    \\ \cdashline{2-2}

  &   $\frac{2\sqrt{2}+2\cos2\theta}{4 \sqrt{2}+\left(1+\sqrt{3}\right)\cos2\theta}$  \\  \hline

$\left|J_{\text{CP}}\right|$  &  $\frac{1}{6\sqrt{3}}\left|\sin2\theta\right|$   \\ \hline

$\left|\tan\delta_{\text{CP}}\right|$ &  $\big|\frac{4\sqrt{2}+\left(1+\sqrt{3}\right) \cos2\theta}{1-\sqrt{3}-\sqrt{2}\cos2\theta}\tan2\theta\big|$ \\ \hline

$\left|\tan\alpha_{21}\right|$  &
$\big|\frac{1+\sqrt{3}+2\sqrt{2}\cos2\theta+\left(1-\sqrt{3}\right) \sin2\theta}{1+\sqrt{3}+2\sqrt{2}\cos2\theta-\left(1-\sqrt{3}\right) \sin2\theta }\big|$   \\ \hline

$\left|\tan\alpha^{\prime}_{31}\right|$  &  $\big|\frac{4 \sin2\theta}{2-3\sqrt{3}+\left(2+\sqrt{3}\right)\cos4\theta}\big|$     \\\hline\hline

\multicolumn{2}{|c|}{\tt Best Fits}  \\\hline\hline

$\theta_{\mathrm{bf}}$  &    $\pm0.130$    \\\hline

$\chi^2_{\mathrm{min}}(\theta_{23}<\pi/4)$  &  9.124  \\\cdashline{1-2}

$\chi^2_{\mathrm{min}}(\theta_{23}>\pi/4)$  &  9.838  \\\hline

$\sin^2\theta_{13}$  &   0.0222  \\   \hline

$\sin^2\theta_{12}$  &  0.341     \\  \hline

\multirow{2}{*}{$\sin^2\theta_{23}$}  &  0.426    \\ \cdashline{2-2}

  &   0.574  \\  \hline

$\left|J_{\text{CP}}\right|$  &  0.0248   \\ \hline

$\left|\sin\delta_{\text{CP}}\right|$ &  0.725 \\ \hline

$\left|\sin\alpha_{21}\right|$  &  0.682 \text{or}   0.731  \\\hline

$\left|\sin\alpha^{\prime}_{31}\right|$  &  0.9992 \\\hline\hline

\end{tabular}
\caption{\label{tab:caseII_VI_and_IV_VIII}Predictions for mixing parameters in cases II, IV, VI and VIII. The Majorana phase $\alpha^{\prime}_{31}$ has been redefined to include the Dirac CP phase by $\alpha^{\prime}_{31}=\alpha_{31}-2\delta_{\text{CP}}$. The predictions for exchanging the second and the third rows of the PMNS matrix are shown below the dashed line.}
\end{center}
\end{table}

\begin{table}[hptb]
\begin{center}
\begin{tabular}{|c|c|c|c|}
\hline\hline

{\tt Parameters}  &   $\sin^2\theta_{12}$   &   $\sin^2\theta_{23}$   &   $\sin^2\theta_{13}$ \\\hline

{\tt Best-Fit $\pm$ $1\sigma$ values }   &   $0.302^{+0.013}_{-0.012}$  &  $0.413^{+0.037}_{-0.025}\oplus0.594^{+0.021}_{-0.022}$  &  $0.0227^{+0.0023}_{-0.0024}$  \\\hline

\tt $3\sigma$ values    &   $0.267\to0.344$  &  $0.342\to0.667$  &  $0.0156\to0.0299$  \\\hline\hline

\end{tabular}
\caption{\label{tab:fitting}The best-fit, $1\sigma$ and $3\sigma$ values of neutrino mixing parameters are taken from Ref.~\cite{GonzalezGarcia:2012sz}. }
\end{center}
\end{table}

For all the ten cases discussed above, we see that the residual $Z^{c^2}_2$ flavor symmetry enforces the second column of the resulting PMNS matrix being proportional to $(1,1,1)^{T}/\sqrt{3}$. As a consequence, we have the relations
\begin{equation}
\sin^2\theta_{12}=\frac{1}{3\cos^2\theta_{13}}\,.
\end{equation}
The measured value of the reactor mixing angle $\sin^2\theta_{13}\simeq0.0227$ leads to $\sin^2\theta_{12}\simeq0.341$, which are compatible with the experimentally allowed regions~\cite{Tortola:2012te,GonzalezGarcia:2012sz,Capozzi:2013csa}.
It is remarkable that all the mixing parameters depend on only one free parameter $\theta$ with period $\pi$ in the present context, and in particular the CP phases are nontrivial function of $\theta$ except cases I, VII, IX and X.

In order to measure quantitatively whether and how well the different mixing schemes listed in Table~\ref{tab:caseI} and Table~\ref{tab:caseII_VI_and_IV_VIII} can explain the current experimental data, we perform a $\chi^2$ analysis and the $\chi^2$ function is constructed as usual
\begin{equation}
\chi^2(\theta)=\sum_{ij=12,13,23}\frac{\left[\sin^2\theta_{ij}(\theta)-\left(\sin^2\theta_{ij}\right)^{\rm{ex}}\right]^2}{\sigma^2_{ij}}\,,
\end{equation}
where $\sin^2\theta_{ij}$ depending on the parameter $\theta$ is the theoretical prediction of the present work, as collected in Table~\ref{tab:caseI} and Table~\ref{tab:caseII_VI_and_IV_VIII}. $\left(\sin^2\theta_{ij}\right)^{\rm{ex}}$ represents the experimentally measured value of the mixing angle, and $\sigma_{ij}$ is the corresponding $1\sigma$ error. Their values are taken from Ref.~\cite{GonzalezGarcia:2012sz} and are summarized in Table~\ref{tab:fitting}. Since $\sin^2\theta_{23}$ now has two best-fit values $\sin^2\theta_{23}=0.413$ and $\sin^2\theta_{23}=0.594$~\cite{GonzalezGarcia:2012sz}, two $\chi^2$ functions associated with these two central values are constructed.

For each remnant CP symmetry, all the possible permutations of rows and columns of the PMNS matrix are studied, and the corresponding minimal values of $\chi^2$ are calculated. Then the arrangement of the PMNS matrix with the smallest $\chi^2$ minimal value is chosen, and the analytical expressions for the mixing parameters and the best-fit results are presented in Table~\ref{tab:caseI} and Table~\ref{tab:caseII_VI_and_IV_VIII}. We find that the PMNS matrix with the smallest $\chi^2$ for the central value $\sin^2\theta_{23}=0.594$ is related to the one with the smallest $\chi^2$ for $\sin^2\theta_{23}=0.413$ by exchanging the second and the third rows, and the corresponding results for mixing parameters and their best-fit values are shown below the dashed line. Because the sign of $\tan\alpha_{21}$ and $\tan\alpha^{\prime}_{31}$ depends on the CP parity of the neutrino states which is contained in the matrix $K_{\nu}$, we present the absolute values $\left|\tan\alpha_{21}\right|$ and $\left|\tan\alpha^{\prime}_{31}\right|$ in Table~\ref{tab:caseI} and Table~\ref{tab:caseII_VI_and_IV_VIII}.
We see that, first of all, the measured three lepton mixing angles can be accommodated very well for certain values of the parameter $\theta$ except cases III and V, which predict $\sin^2\theta_{12}=\sin^2\theta_{23}$ or $\sin^2\theta_{12}=1-\sin^2\theta_{23}$. Furthermore, since mixing angles as well as CP phases are predicted in terms of a single parameter $\theta$, different mixing parameters are strongly correlated with each other.

\begin{figure}[t!]
\begin{center}
\includegraphics[width=1\textwidth]{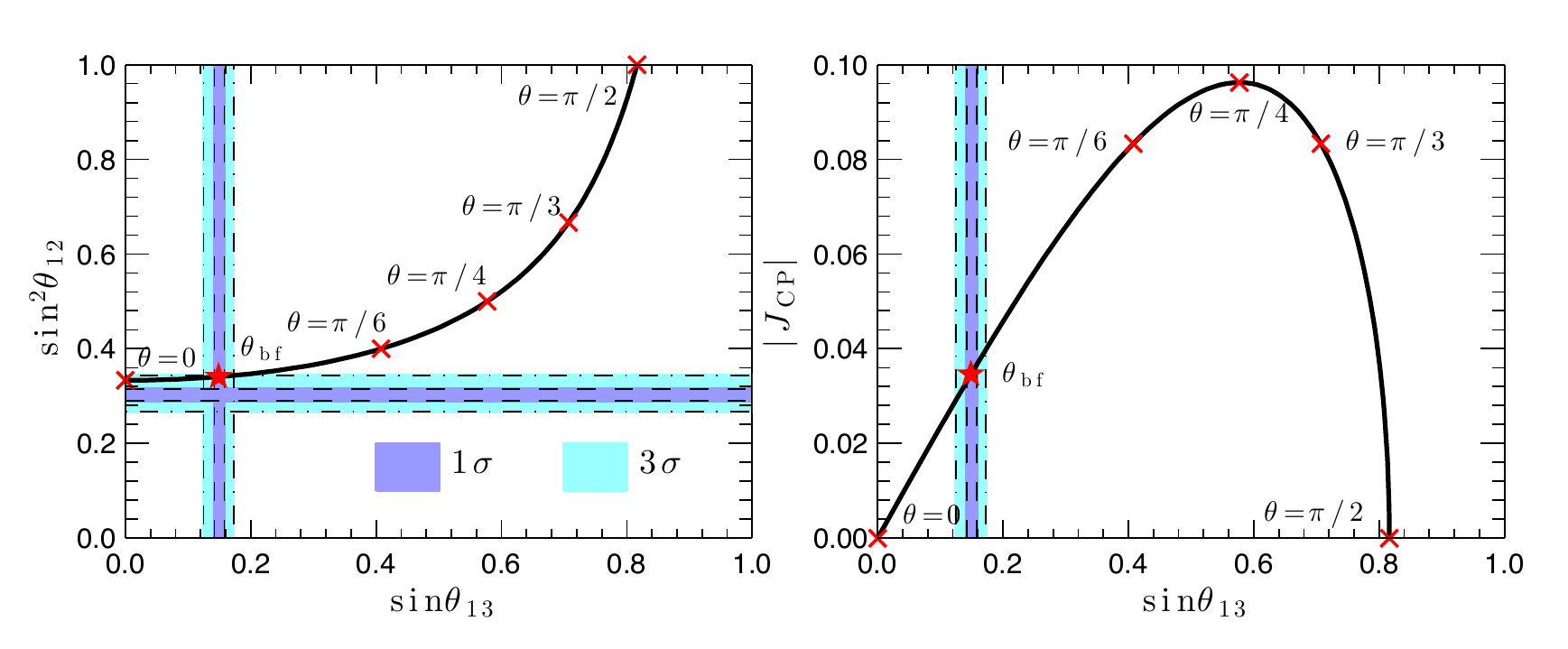} \caption{\label{fig:caseI} Relations of mixing angles ($\sin\theta_{13}$, $\sin^2\theta_{12}$) and the Jarlskog invariant $J_{\rm CP}$ for cases I and VII. The other mixing parameters $\theta_{23}$, $\delta_{\rm{CP}}$, $\alpha_{21}$ and $\alpha^{\prime}_{31}$ are not shown here since they take constant values. We mark the best-fit value $\theta_{\rm bf}$ of the parameter $\theta$ with a red star, and also mark $\theta=0,\pi/6,\pi/4,\pi/3,\pi/2$ with a cross on the curve. The $1\sigma$ and $3\sigma$ ranges for the mixing angles are taken from Table~\ref{tab:fitting}.}
\end{center}
\end{figure}

\begin{figure}[t!]
\begin{center}
\includegraphics[width=1\textwidth]{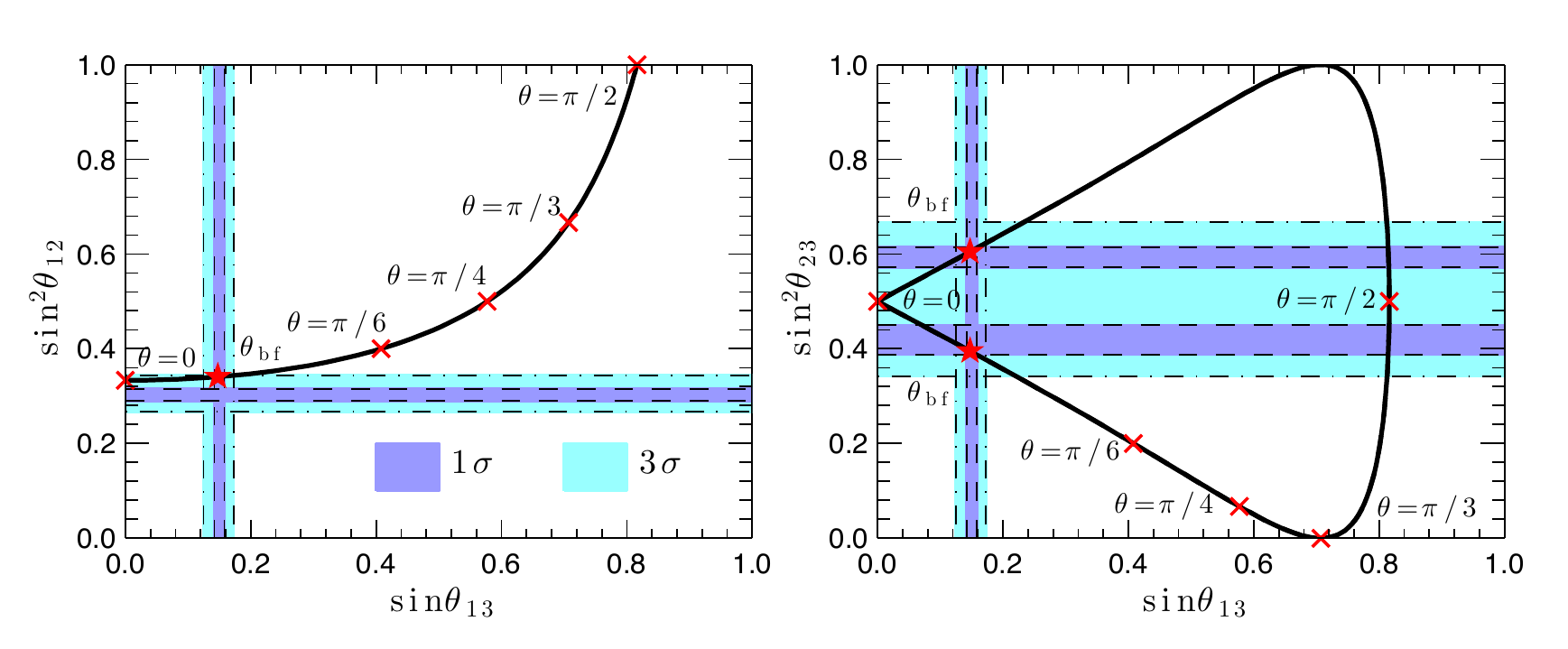} \caption{\label{fig:caseIX} Relations of mixing angles ($\sin\theta_{13}$, $\sin^2\theta_{12}$, $\sin^2\theta_{23}$) for cases IX and X. CP-violating phases are not shown here since they are predicted to be constant. We mark the best-fit value $\theta_{\rm bf}$ of the parameter $\theta$ with a red star, and also mark $\theta=0,\pi/6,\pi/4,\pi/3,\pi/2$ with a cross on the curve. The $1\sigma$ and $3\sigma$ ranges for the mixing angles are taken from Table~\ref{tab:fitting}.}
\end{center}
\end{figure}

\begin{figure}[t!]
\begin{center}
\includegraphics[width=1\textwidth]{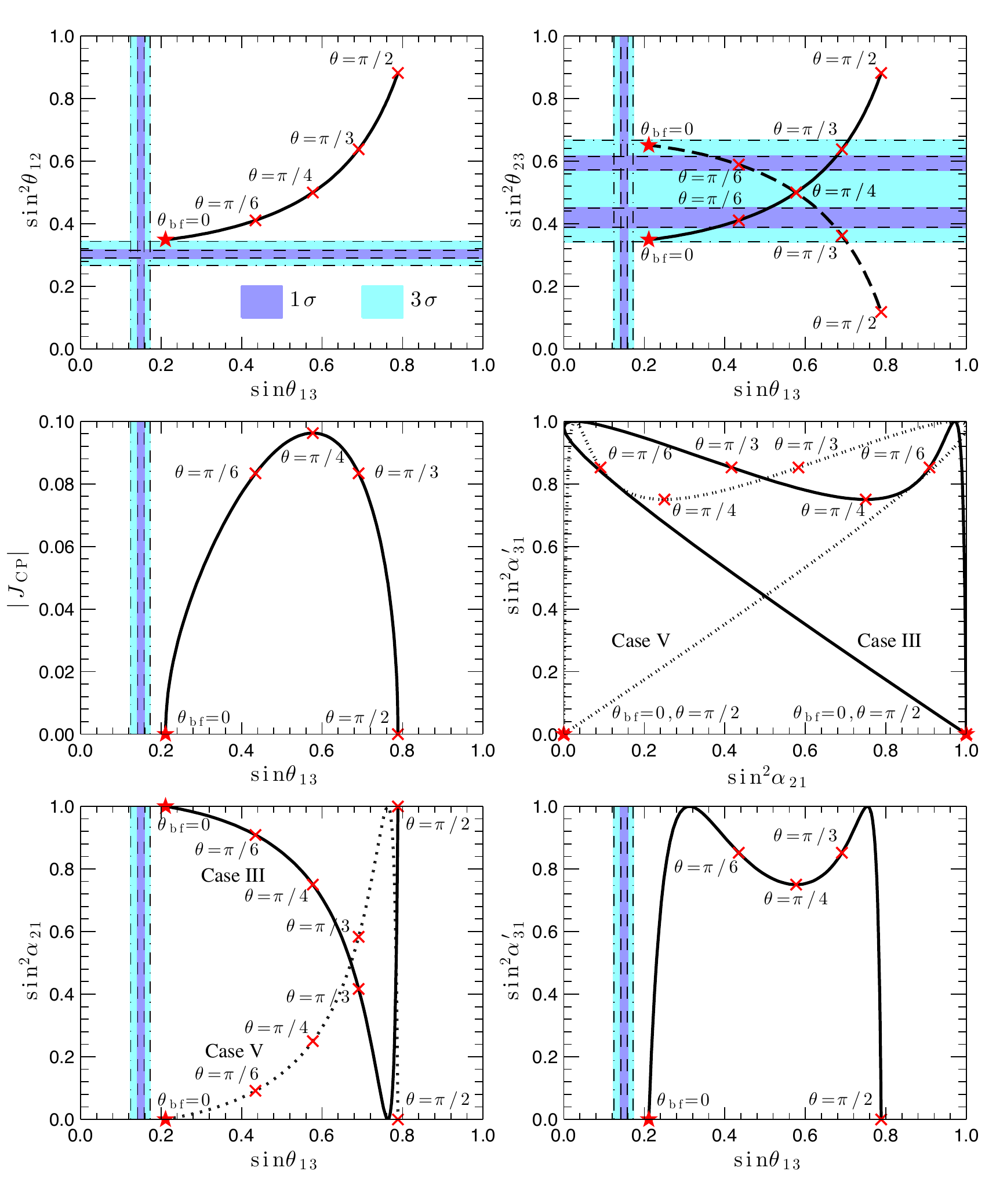} \caption{\label{fig:caseIII} Correlations among mixing angles ($\sin\theta_{13}$, $\sin^2\theta_{12}$, $\sin^2\theta_{23}$) and CP parameters ($J_{\rm{CP}}$, $\sin^2\alpha_{21}$, $\sin^2\alpha'_{31}$) for cases III and V. On the top right panel, the solution for $\sin^2\theta_{23}$ in the first octant is shown in a solid line, and the solution for $\sin^2\theta_{23}$ in the second octant is shown in a dashed line. The results for $\sin^2\alpha_{21}$ in case III and case V are shown in solid and dotted lines, respectively. We mark the best-fit value $\theta_{\rm bf}=0$ with a red star, and mark $\theta=\pi/6,\pi/4,\pi/3,\pi/2$ with a cross on the curve. The $1\sigma$ and $3\sigma$ ranges for the mixing angles are taken from Table~\ref{tab:fitting}.}
\end{center}
\end{figure}

\begin{figure}[t!]
\begin{center}
\includegraphics[width=1\textwidth]{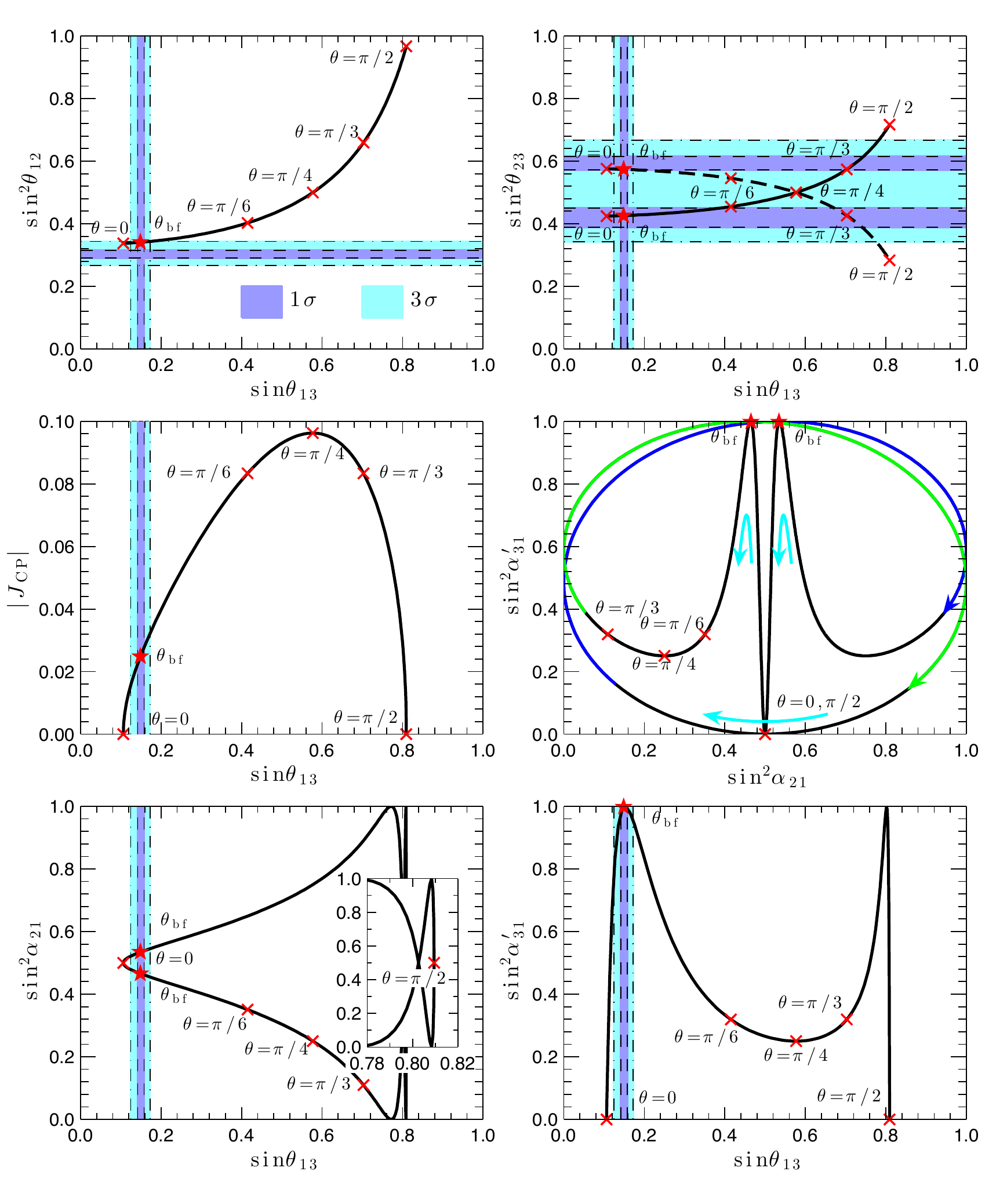} \caption{\label{fig:caseII} Correlations between mixing angles ($\sin\theta_{13}$, $\sin^2\theta_{12}$, $\sin^2\theta_{23}$) and CP parameters ($J_{\rm{CP}}$, $\sin^2\alpha_{21}$, $\sin^2\alpha^{\prime}_{31}$) for cases II, IV, VI and VIII. On the top right panel, the solution for $\sin^2\theta_{23}$ in the first octant is shown in a solid line, and the solution for $\sin^2\theta_{23}$ in the second octant is shown in a dashed line. We mark the best-fit value $\theta_{\rm bf}$ of the parameter $\theta$ with a red star, and mark $\theta=0,\pi/6,\pi/4,\pi/3,\pi/2$ with a cross on the red curve. The $1\sigma$ and $3\sigma$ ranges for the mixing angles are taken from Table~\ref{tab:fitting}.}
\end{center}
\end{figure}

The correlations among mixing parameters for each cases are displayed in Figs.~\ref{fig:caseI}, \ref{fig:caseIX}, \ref{fig:caseIII} and \ref{fig:caseII}. Some comments based on these figures are presented in the following.

\begin{itemize}
\item{In case I and case VII, the solar and reactor mixing angles are related by $3\sin^2\theta_{12}\cos^2\theta_{13}$ $=1$, and the atmospheric neutrino mixing is predicted to be maximal $\theta_{23}=\pi/4$. Hence excellent agreement with the experimental data can be achieved, as shown in Fig.~\ref{fig:caseI}. Moreover, the Dirac CP is maximally violated $\delta_{\rm{CP}}=\pm\pi/2$, while the Majorana CP phases $\sin\alpha_{21}=\sin\alpha_{31}=0$ in case I and  $\cos\alpha_{21}=\sin\alpha_{31}=0$ in case VII. Comparing with previous work on $S_4$ family symmetry combined with the generalised CP~\cite{Ding:2013hpa}, we see that the mixing pattern in case I can also be achieved within the context of $S_4$ while the pattern in case VII can not be obtained. }
    
\item{In case IX and case X, $\theta_{12}$ and $\theta_{13}$ are predicted to be of the same form as those of case I, whereas the atmospheric mixing angle $\theta_{23}$ deviates from maximal mixing. The following correlations are found:
    \begin{eqnarray}
     3\sin^2\theta_{12}\cos^2\theta_{13}=1,\qquad \sin^2\theta_{23}=\frac{1}{2}\pm\frac{\sin\theta_{13}\sqrt{2-3\sin^2\theta_{13}}}{2\cos^2\theta_{13}}\,,
    \end{eqnarray}
which are plotted in Fig.~\ref{fig:caseIX}. For the best-fit value of the reactor angle $\theta_{13}=8.66^{\circ}$~\cite{GonzalezGarcia:2012sz}, the other two mixing angles are determined to be $\theta_{12}\simeq35.73^{\circ}$, $\theta_{23}\simeq38.82^{\circ}$ or $\theta_{23}\simeq51.18^{\circ}$, which are compatible with the preferred values from global fits. There is no CP violation except that the Majorana phase $\alpha_{21}$ is maximal $\alpha_{21}=\pm\pi/2$ in case X. The mixing texture of case IX is also admissible in $S_4$ family symmetry with generalised CP~\cite{Ding:2013hpa} although the one in case X is not achievable.}

\item{Predictions for cases III and V are shown in Fig.~\ref{fig:caseIII}. All the lepton mixing parameters nontrivially depend on $\theta$. Especially, we have $\theta_{23}=\theta_{12}$ or $\theta_{23}=90^\circ-\theta_{12}$. The best-fit value of $\theta$ is $\theta_{\text{bf}}=0$ for which $\sin^2\theta_{13}$ is minimized with $\sin^2\theta_{13}\left(\theta_{\text{bf}}\right)=\left(2-\sqrt{3}\right)/6$.
    Hence in this scenario $\theta_{13}$ has a lower bound $\theta_{13}\geq12.2^{\circ}$ which is beyond the $3\sigma$ range of the experimental data. Furthermore, we have $\sin^2\theta_{12}\left(\theta_{\text{bf}}\right)=\left(8-2\sqrt{3}\right)/13$, $\sin^2\theta_{23}\left(\theta_{\text{bf}}\right)=\left(8-2\sqrt{3}\right)/13$ or $\sin^2\theta_{23}\left(\theta_{\text{bf}}\right)=\left(5+2\sqrt{3}\right)/13$, therefore the predicted mixing pattern reduces to Toorop-Feruglio-Hagedorn mixing~\cite{Toorop:2011jn,Ding:2012xx,King:2012in} for $\theta_{\text{bf}}=0$. Note that imposing $S_4$ family symmetry and generalised CP can lead to the mixing pattern of case III but not that of case V~\cite{Ding:2013hpa}.}
\item{Cases II, IV, VI, and VIII are most interesting. The resulting mixing pattern are firstly proposed in Ref. \cite{Ding:2013nsa} (denoted as pattern D there). The mixing parameters are predicted to be of the same form in all the four cases, and the correlations among them are shown in Fig.~\ref{fig:caseII}. The following relations among the mixing angles are satisfied:}
\begin{eqnarray}
\nonumber&&3\sin^2\theta_{12}\cos^2\theta_{13}=1,\\
\nonumber&&\sin^2\theta_{23}=2-\sqrt{3}-(3-2\sqrt{3})\sin^2\theta_{12},\qquad \theta_{23}<\pi/4,\\
&&\sin^2\theta_{23}=\sqrt{3}-1+(3-2\sqrt{3})\sin^2\theta_{12},\qquad \theta_{23}>\pi/4\,.
\end{eqnarray}
Notice that this mixing pattern can accommodate the three lepton mixing angles very well. For the best-fit value $\theta_{\text{bf}}\simeq0.130$, we have $\sin^2\theta_{13}\left(\theta_{\text{bf}}\right)\simeq0.0222$, $\sin^2\theta_{12}\left(\theta_{\text{bf}}\right)\simeq0.341$, $\sin^2\theta_{23}\left(\theta_{\text{bf}}\right)\simeq0.426$ or $\sin^2\theta_{23}\left(\theta_{\text{bf}}\right)\simeq0.574$ which are in the experimentally favored ranges~\cite{Tortola:2012te,GonzalezGarcia:2012sz,Capozzi:2013csa}. For the CP phases, the best-fit values are $\left|\sin\delta_{\rm{CP}}\left(\theta_{\text{bf}}\right)\right|=0.725$, $\left|\sin\alpha^{\prime}_{31}\left(\theta_{\text{bf}}\right)\right|=0.999$, $\left|\sin\alpha_{21}\left(\theta_{\text{bf}}\right)\right|=0.682$ or
$\left|\sin\alpha_{21}\left(\theta_{\text{bf}}\right)\right|=0.731$. Hence $\alpha^{\prime}_{31}$ is approximately maximal with $\alpha^{\prime}_{31}\simeq\pm\pi/2$ while $\delta_{\rm{CP}}$ and $\alpha_{21}$ don't take simple values $0$, $\pi$ or $\pm\pi/2$. This is a distinguishing new feature of $\Delta(48)$ family symmetry compared with widely discussed $A_4$, $S_4$ family symmetries. Moreover, we note that the predicted value of the Dirac CP phase $\delta_{\rm{CP}}$ is compatible with the current $1\sigma$ preferred range $0.9\pi\leq\delta_{\rm{CP}}\leq2.0\pi$ from global fits~\cite{GonzalezGarcia:2012sz}. Future dedicated long-baseline neutrino oscillation experiments such as LBNE~\cite{Adams:2013qkq} and Hyper-Kamiokande~\cite{Abe:2011ts} can measure the Dirac phase with a certain precision such that this mixing pattern could be tested. Although this mixing pattern cannot be obtained from $A_4$ and $S_4$ family symmetry, it can be realized in $\Delta(96)$ family symmetry with generalised CP~\cite{Ding:2014ssa}. However, the associated remnant symmetries within $\Delta(48)$ and $\Delta(96)$ frameworks are different. From the model building perspective, the $\Delta(48)$ family symmetry should be preferred over $\Delta(96)$ family symmetry to produce this mixing pattern since the group structure of $\Delta(48)$ is  simpler than $\Delta(96)$.  
\end{itemize}

\section{\label{sec:model_construction}Model with $\Delta(48)$ and generalised CP symmetries}
%%%%%%%%%%%%%%%%%%%%%%%%%%%%

\cleqn

\begin{table} [t!]
\begin{center}
\begin{tabular}{|c||c|c|c|c|c|c||c|c|c|c|c|c|c|c|c|}
\hline\hline

{\tt Field}    &     $l$    &      $e^{c}$   &     $\mu^{c}$    &    $\tau^c$  &     $\nu^{c}$  &   $h_{u,d}$   &    $\phi_{l}$ & $\varphi_l$   &  $\rho_{l}$   &  $\varphi$  &  $\phi$  &  $\xi$   &   $\rho$   &   $\chi$   &  $\sigma$  \\ \hline

&   &     &    &    &    &    &  &    &   &    &    &     &     &       \\ [-0.16in]

$\Delta(48)$  & $\mathbf{3}$  &    $\mathbf{1}$   &    $\mathbf{1}$     &    $\mathbf{1}$   &   $\mathbf{\overline{3}}$  &  $\mathbf{1}$   &  $\mathbf{\overline{3}}$    &  $\mathbf{\overline{3}{}^{\prime}}$  & $\mathbf{3}^{\prime}$  & $\mathbf{\overline{3}}$   &   $\mathbf{\widetilde{3}}$  &   $\mathbf{1}$  &  $\mathbf{1}$   &   $\mathbf{3}^{\prime}$  &  $\mathbf{1}$   \\  \hline

$Z_2$         &   1   &   1   &   1   &  1   &  1    &   1
&  1   &   1  &  1  &  $-1$   &  1  &  1 &  $-1$  & $-1$ &  1 \\ \hline

$Z_5$         &   1   &  $\omega^3_5$   &   1  &  $\omega^2_5$    &  1 &  1   &   $\omega^3_5$  &  $\omega^2_5$   &  $\omega_5$   &  1  &  1   &    1 &  1   &  1  &  1  \\ \hline

$Z_6$         &   1   &  $\omega^5_6$   &   $\omega_6$  &  $\omega^3_6$  &  1  &  1   &    $\omega^3_6$  &  $\omega^2_6$   &  $\omega_6$   &  $\omega^2_6$    &  $\omega^4_6$     &    $\omega^4_6$    &  $\omega^4_6$    &  $\omega^4_6$  &  $\omega^2_6$  \\ \hline

$U(1)_R$      &    1  &   1  &   1  &  1  &   1    &  0   &   0  &  0  &  0   &   0  &  0  &  0 &  0   &   0  &  0 \\\hline\hline
\end{tabular}
\caption{\label{tab:matter_flavon_new_model}Matter and flavon fields and their transformation properties under the family symmetry $\Delta(48) \times Z_2\times Z_5\times Z_6$ and $U(1)_R$, where $l=\left(l_{\tau}, l_{\mu}, l_{e}\right)$ is the left-handed lepton doublet fields with $\omega_5=e^{2i\pi/5}$ and $\omega_6=e^{i\pi/3}$.}
\end{center}
\end{table}
%%%%%%%%%%%%%%%%%%%%%%%%%%%%
Inspired by the general analysis in the previous section, we will construct a dynamical model in this section, where the new mixing pattern in Table~\ref{tab:caseII_VI_and_IV_VIII} is realized. The model is based on $\Delta(48)\rtimes H_{\rm{CP}}$, which is supplemented by the auxiliary symmetry $Z_2\times Z_5\times Z_6$. $Z_5\times Z_6$ distinguishes the flavons in the neutrino sector from those entering into the charged lepton sector, and $Z_2$ further distinguishes the flavons $\varphi$, $\rho$, $\chi$ from $\phi$, $\zeta$, $\sigma$. The three generations of left-handed lepton doublets $l$ and the right-handed neutrinos $\nu^{c}$ are embedded into $\Delta(48)$ triplets $\mathbf{3}$ and $\mathbf{\overline{3}}$ respectively, while the right-handed charged leptons $e^{c}$, $\mu^{c}$ and $\tau^{c}$ are all invariant under $\Delta(48)$. The field content and their transformation properties are shown in Table~\ref{tab:matter_flavon_new_model}.

\subsection{Basic structure}
Our model is formulated within the framework of supersymmetry, and the neutrino masses are generated via type-I seesaw mechanism. The superpotential relevant to the charged lepton and the neutrino masses are
\begin{equation}
w=w^{\text{eff}}_l+w^{\text{eff}}_\nu \,,
\end{equation}
with
\begin{eqnarray}
\nonumber w^{\text{eff}}_{l}=&& \frac{y_{\tau}}{\Lambda}\left(l\phi_l\right)_{\mathbf{1}}\tau^{c}h_d+\frac{y_{\mu}}{\Lambda^2}\left(l\left(\phi_l\varphi_l\right)_{\mathbf{\overline{3}}}\right)_{\mathbf{1}}\mu^{c}h_d
+\frac{y_{e_1}}{\Lambda^3}\left(l\left(\rho_l\left(\phi_l\phi_l\right)_{\mathbf{\widetilde{3}}}\right)_{\mathbf{\overline{3}}}\right)_{\mathbf{1}}e^{c}h_d\\
\label{eq:w_charged_lepton}&&+\frac{y_{e_2}}{\Lambda^3}\left(l\left(\phi_l\left(\varphi_l\varphi_l\right)_{\mathbf{3}{}^{\prime}_S}\right)_{\mathbf{\overline{3}}}\right)_{\mathbf{1}}e^{c}h_d\,,\\
\nonumber w^{\text{eff}}_{\nu}=&& y\left(l\nu^c\right)_{\mathbf{1}}h_u+\frac{x_{1}}{\Lambda}\left(\left(\nu^c\nu^c\right)_{\mathbf{3}_S}\varphi\right)_{\mathbf{1}}\rho
+\frac{x_2}{\Lambda}\left(\left(\nu^c\nu^c\right)_{\mathbf{3}_S}\left(\varphi\chi\right)_{\mathbf{\overline{3}}}\right)_{\mathbf{1}}\\
\label{eq:w_neutrino}&&+\frac{x_{3}}{\Lambda}\left(\left(\nu^c\nu^c\right)_{\mathbf{\widetilde{3}}}\left(\varphi\chi\right)_{\mathbf{\widetilde{3}}}\right)_{\mathbf{1}}
+\frac{x_{4}}{\Lambda}\left(\left(\nu^c\nu^c\right)_{\mathbf{\widetilde{3}}}\phi\right)_{\mathbf{1}}\sigma\,,
\end{eqnarray}
where all the couplings are constrained to be real by the generalised CP symmetry. As we shall show in section~\ref{subsec:vacuum_alignment}, at LO the flavons develop the following VEV configuration:
\begin{eqnarray}
\nonumber&\langle\phi_l\rangle=\left(\begin{array}{c}
1\\
0\\
0
\end{array}\right)v_{\phi_l},\quad \langle\varphi_l\rangle=\left(\begin{array}{c}
0\\
1\\
0
\end{array}\right)v_{\varphi_l},\quad
 \langle\rho_l\rangle=\left(\begin{array}{c}
0\\
1\\
0
\end{array}\right)v_{\rho_l},\\
\nonumber&\langle\varphi\rangle=\left(\begin{array}{c}
1\\
1\\
1
\end{array}\right)v_{\varphi}, \quad~~
\langle\phi\rangle=\left(\begin{array}{c}
1\\
1\\
1
\end{array}\right)v_{\phi}, \quad~\,
\langle\chi\rangle=\left(\begin{array}{c}
1\\
1\\
1
\end{array}\right)v_{\chi}, \\
\label{eq:vev}&\hskip-0.4in\langle\xi\rangle=v_{\xi}, \qquad\qquad~\,
\langle\rho\rangle=v_{\rho},\qquad\qquad~
\langle\sigma\rangle=v_{\sigma}\,.
\end{eqnarray}
With this vacuum configuration, we find that the charged lepton mass matrix is diagonal with
\begin{equation}
m_{\tau}=y_{\tau}\frac{v_{\phi_l}}{\Lambda}v_d,\quad m_{\mu}=\omega^2y_{\mu}\frac{v_{\phi_l}v_{\varphi_l}}{\Lambda^2}v_d,\quad m_e=y_{e_1}\frac{v_{\rho_l}v^2_{\phi_l}}{\Lambda^3}v_d+2\omega^2y_{e_2}\frac{v_{\phi_l}v^2_{\varphi_l}}{\Lambda^3}v_d\,,
\end{equation}
where $v_d=\langle h_d\rangle$. It is remarkable that the mass hierarchy among charged leptons can be correctly reproduced for
\begin{equation}
v_{\phi_l}\sim\lambda^2\Lambda,\qquad v_{\varphi_l}\sim\lambda^2\Lambda\,,
\end{equation}
where $\lambda\simeq0.23$ is the Cabibbo angle. As a result, at LO there is no contribution to the lepton mixing matrix from the charged lepton sector. For the neutrino sector, we can straightforwardly read out the Dirac and the right-handed neutrino Majorana matrices as
\begin{eqnarray}
\nonumber&m_{D}=yv_{u}
\left(\begin{array}{ccc}
0 &  0  &  1 \\
0 &  1  &  0 \\
1 &  0  &  0
\end{array}\right),\\
&m_M=x_1\frac{v_{\varphi}v_{\rho}}{\Lambda}
\left(\begin{array}{ccc}
2  & -1 & -1 \\
-1 & 2  & -1  \\
-1 & -1 & 2
\end{array}\right)+3x_3\frac{v_{\varphi}v_{\chi}}{\Lambda}
\left(\begin{array}{ccc}
1   &   \omega   &   \omega^2  \\
\omega   &  \omega^2  &  1  \\
\omega^2   &   1   &   \omega
\end{array}\right)+x_4\frac{v_{\phi}v_{\sigma}}{\Lambda}
\left(\begin{array}{ccc}
1   &   1   &   1  \\
1   &  1  &  1  \\
1  &   1   &   1
\end{array}\right)\,.
\end{eqnarray}
Note that the $x_2$ term in Eq.~\eqref{eq:w_neutrino} does not contribute to the $m_{M}$. The reason is that the contraction $\left(\varphi\chi\right)_{\mathbf{\overline{3}}}$ vanishes for the alignment in Eq.~\eqref{eq:vev}. The light neutrino mass matrix is given by the see-saw relation:
\begin{equation}
m_{\nu}=-m^{T}_Dm_{M}m_D\,.
\end{equation}
Therefore we have
\begin{eqnarray}
\nonumber \widetilde{m}_{\nu}&=&P^{T}_{321}m_{\nu}P_{321}\\
   \label{eq:mnu_model}    &=&\alpha\left(\begin{array}{ccc}
                          2  &  -1  &  -1 \\
                          -1 &   2  & -1 \\
                          -1 &   -1 & 2
                          \end{array}\right)
                       +\beta\left(\begin{array}{ccc}
                          1  &  0 &   0 \\
                          0  &  0 & 1  \\
                          0  &  1 &  0
                          \end{array}\right)
                         +\gamma\left(\begin{array}{ccc}
                          0  &  1  &  1 \\
                          1  &  1  &  0 \\
                          1  &  0  &  1
                          \end{array}\right)+\epsilon\left(\begin{array}{ccc}
                          0  &  1  &  -1  \\
                          1  &  -1 &  0  \\
                          -1 &  0  & 1
                          \end{array}\right)\,,
\end{eqnarray}
where
\begin{eqnarray}
\nonumber&&\alpha=-\frac{y^2v^2_{u}}{9}\frac{\Lambda}{x_1v_{\varphi}v_{\rho}},\qquad \beta=-\frac{y^2v^2_{u}}{9}\left[\frac{\Lambda}{x_4v_{\phi}v_{\sigma}}-\frac{3x_3v_{\chi}\Lambda}{x^2_1v_{\varphi}v^2_{\rho}}\right],\\
&&\gamma=-\frac{y^2v^2_{u}}{9}\left[\frac{\Lambda}{x_4v_{\phi}v_{\sigma}}+\frac{3x_3v_{\chi}\Lambda}{2x^2_1v_{\varphi}v^2_{\rho}}\right],\qquad \epsilon=i\frac{y^2v^2_{u}}{2\sqrt{3}}\frac{x_3v_{\chi}\Lambda}{x^2_1v_{\varphi}v^2_{\rho}}\,.
\end{eqnarray}
Notice that $\widetilde{m}_{\nu}$ is of the same form as the neutrino mass matrix in Eq.~\eqref{eq:numm_inv_rem_fla} which is the
most general Majorana neutrino mass matrix invariant under the action of the residual flavor symmetry $Z^{c^2}_2=\left\{1, c^2\right\}$. Imposing the generalised CP symmetry renders all coupling constant real, and CP is spontaneously broken by the complex flavon VEVs. The phase structure of the VEVs depends on the driving potential. Anticipating the results of the vacuum alignment in section~\ref{subsec:vacuum_alignment}, we see that the flavons can develop VEVs with phases
\begin{equation}
\frac{v_{\varphi}}{\left|v_{\varphi}\right|}=\pm e^{\pm i\frac{\pi}{4}},\qquad
\label{eq:VEV_phases}\frac{v_{\chi}}{\left|v_{\chi}\right|}=\pm e^{\pm i\frac{\pi}{4}},\qquad \frac{v_{\phi}}{\left|v_{\phi}\right|}=\pm i\,,
\end{equation}
where the signs depend on the undetermined real couplings of the driving potential. The VEVs of the $\Delta(48)$ singlet flavons $\xi$, $\rho$ and $\sigma$ are real. Since the overall sign of the VEVs can be absorbed by parameter redefinition, four scenarios arise in the present model.

\begin{description}[labelindent=-0.5em, leftmargin=0.3em]

\item[~~(a).~] {$\frac{v_{\varphi}}{\left|v_{\varphi}\right|}=\pm e^{i\frac{\pi}{4}}$, $\frac{v_{\chi}}{\left|v_{\chi}\right|}=\pm e^{i\frac{\pi}{4}}$}

Referring to the Appendix \ref{sec:appendix_B}, we can see that the vacuum of the flavons $\varphi$, $\chi$ and $\phi$ are invariant under $\rho(d)X(\mathfrak{h_1})$ and $\rho(c^2d)X(\mathfrak{h_1})$. Hence the
generalised CP symmetry is broken to $H^{\nu}_{\rm{CP}}=\left\{\rho(d)X(\mathfrak{h_1}), \rho(c^2d)X(\mathfrak{h_1})\right\}$ in the neutrino sector in this case. To facilitate the presentation, we define the parameters:
\begin{equation}
r=-\frac{y^2v^2_{u}}{9}\frac{\Lambda}{x_1v_{\varphi}v_{\rho}}e^{i\frac{\pi}{4}},\quad
s=-i\frac{y^2v^2_{u}}{9}\frac{\Lambda}{x_4v_{\phi}v_{\sigma}},\quad
t=-\frac{y^2v^2_{u}}{3}\frac{x_3v_{\chi}\Lambda}{x^2_1v_{\varphi}v^2_{\rho}}\,.
\end{equation}
Obviously all the three parameters $r$, $s$ and $t$ are real, and we have
\begin{equation}
\alpha=re^{-i\frac{\pi}{4}},\quad \beta=-t-is,\quad \gamma=\frac{1}{2}t-is,\quad \epsilon=-\frac{\sqrt{3}}{2}it\,.
\end{equation}
It is straightforward to check that the following relations are satisfied
\begin{eqnarray}
\nonumber&&\alpha_{\mathrm{Im}}=-\alpha_{\mathrm{Re}}=-\frac{r}{\sqrt{2}},\qquad\qquad~ \beta_{\mathrm{Re}}=-2\gamma_{\mathrm{Re}}=-t,\\
&&\beta_{\mathrm{Im}}=\gamma_{\mathrm{Im}}+\sqrt{3}\;\epsilon_{\mathrm{Re}}=-s,\qquad \epsilon_{\mathrm{Im}}=-\sqrt{3}\;\gamma_{\mathrm{Re}}=-\frac{\sqrt{3}}{2}t\,.
\end{eqnarray}
Hence this case is exactly case II of the general analysis in section~\ref{sec:model_independent_analysis}. As a result, the PMNS matrix is of the form shown in Eq.~\eqref{eq:PMNS_II}, and the predictions for the lepton mixing parameters can be found in Table~\ref{tab:caseII_VI_and_IV_VIII}. The formulae for the light neutrino masses in Eq.~\eqref{eq:nmass_II} are simplified to
\begin{eqnarray}
\nonumber&m_1=\frac{3}{2}\left|t+\mathrm{sign}\left(r\cos2\theta\right)\sqrt{t^2+4r^2}\right|,\\
&m_2=3\left|s\right|,\qquad m_3=\frac{3}{2}\left|t-\mathrm{sign}\left(r\cos2\theta\right)\sqrt{t^2+4r^2}\right|,
\end{eqnarray}
while the expression for $\theta$ is 
\begin{equation}
\tan2\theta=\frac{t}{2r}\,.
\end{equation}
Note that three real parameters $r$, $s$ and $t$ are involved in the light neutrino mass matrix. They can be determined by experimental data of $\theta_{13}$ and mass-squared differences. 8 solutions are found when the best-fit values of $\sin^2\theta_{13}=0.0227$, $\Delta m^2_{21}=7.50\times 10^{-5}~\text{eV}^2$ and $\Delta m^2_{31}=2.473\times 10^{-3}~\text{eV}^2$ ($\Delta m^2_{32}=-2.427\times 10^{-3}~\text{eV}^2$) for NO(IO)~\cite{GonzalezGarcia:2012sz} are taken.
%%%
The resulting predictions for the lepton mixing parameters and the light neutrino masses are summarized in Table~\ref{tab:model_predictions_II}. We see that the neutrino mass spectrum can be NO or IO in the present model. The sum of neutrino masses is $\sum_{i}m_{i}\simeq133.713\,\text{meV}$ for NO and $\sum_{i}m_{i}\simeq154.891\,\text{meV}$ for IO, which are well compatible with the latest results from the Planck satellite: $\sum_{i}m_i<0.23\,\text{eV}$(95\% CL; Planck+WMAP-pol+highL+BAO)~\cite{Ade:2013zuv}.

Besides mixing angles, CP-violating phases and light neutrino masses, this model yields definite predictions for the effective mass parameters $m_\beta$ in beta decay experiments and $m_{\beta\beta}$ in neutrinoless double-beta decay ($0\nu\beta\beta$) experiments, where $m_\beta$ and $m_{\beta\beta}$ are defined through
\begin{equation}
m_{\beta}=\Big[\sum_{i}\left|\left(U_{\rm{PMNS}}\right)_{e i}\right|^2m_i^2\Big]^{1/2}, \qquad
m_{\beta\beta}=\Big|\sum_{i}\left(U_{\rm{PMNS}}\right)_{ei}^2m_i\Big|\,,
\end{equation}
respectively. For the best-fit values of $\theta_{13}$ and $\Delta m^2_{ij}$ shown above, we find that $m_\beta$ is predicted to be $m_\beta\simeq36.842\,\text{meV}$ for NO and $m_\beta\simeq59.476 \,\text{meV}$ for IO. This is still below the expected sensitivity of the KATRIN experiment~\cite{Katrin}, whereas the IO case could be tested by the proposed next-generation experiments, such as Project 8 and PTOLEMY~\cite{mbeta_MH}. The effective mass $m_{\beta\beta}$ can be $32.285\,\text{meV}$, $17.962\,\text{meV}$ for NO or $55.125\,\text{meV}$, $27.041\,\text{meV}$ for IO, depending on which quadrant the Majorana phase $\alpha_{21}$ lies in. The current best upper bounds on $m_{\beta\beta}$ are given by EXO-200~\cite{Auger:2012ar} and KamLAND-Zen~\cite{Gando:2012zm}, with a combined result $m_{\beta\beta}<120-250\,\text{meV}$ ~\cite{Gando:2012zm}. With an uncertainty from nuclear physics, the next generation experiments EXO-1000~\cite{Danilov:2000pp},
CUORE~\cite{Arnaboldi:2002du},
GERDA III~\cite{Abt:2004yk},
KamLAND2-Zen~\cite{Shirai:2013jwa} {\it et. al.} are expected to push the $m_{\beta\beta}$ sensitivity to tens of meV, and thus have the potential to rule out our model.

\begin{table}[t!]
\begin{center}
\begin{tabular}{|c|c|c|c|c|} \hline\hline

& & & &\\[-0.18in]
$\left(\begin{array}{c}
r\\
s\\
t
\end{array}\right)$   &   $\pm\left(\begin{array}{c}
15.590\\
12.253\\
-8.502
\end{array}
\right)$   &    $\pm\left(\begin{array}{c}
-15.590\\
12.253\\
8.502
\end{array}
\right)$   &   $\pm\left(\begin{array}{c}
15.204 \\
20.113\\
8.292
\end{array}
\right)$   &    $\pm\left(\begin{array}{c}
15.204\\
-20.113\\
8.292
\end{array}
\right)$  \\ 
& & & &\\[-0.18in]
\hline

$\sin^2\theta_{12}$  &  \multicolumn{4}{c|}{0.341}    \\ \hline

$\theta_{12}/^{\circ}$ & \multicolumn{4}{c|}{35.734}   \\ \hline

$\sin^2\theta_{23}$  &   \multicolumn{4}{c|}{0.426}     \\ \hline

$\theta_{23}/^{\circ}$  &  \multicolumn{4}{c|}{40.759}   \\ \hline

$\sin\delta_{\rm{CP}}$   &  \multicolumn{2}{c|}{0.733}    &  \multicolumn{2}{c|}{$-0.733$}   \\ \hline

$\delta_{\rm{CP}}/^{\circ}$  &  \multicolumn{2}{c|}{47.166}   & \multicolumn{2}{c|}{312.834}  \\ \hline

$\sin\alpha_{21}$   &  0.732  &  $-0.732$   &  0.682  &   $-0.682$  \\ \hline

$\alpha_{21}/^{\circ}$  &  47.020  &  227.020  &  42.980  & 222.980 \\ \hline

$\sin\alpha_{31}$   &   \multicolumn{2}{c|}{0.0938}  &  \multicolumn{2}{c|}{$-0.0938$}     \\ \hline

$\alpha_{31}/^{\circ}$  &  \multicolumn{2}{c|}{5.383}   &  \multicolumn{2}{c|}{354.617}      \\ \hline

$\sin\alpha^{\prime}_{31}$   &  \multicolumn{2}{c|}{$-0.9998$}   &  \multicolumn{2}{c|}{0.9998}     \\ \hline

$\alpha^{\prime}_{31}/^{\circ}$  &  \multicolumn{2}{c|}{271.051}   & \multicolumn{2}{c|}{88.949}      \\ \hline  \hline

$m_1$   &    \multicolumn{2}{c|}{35.724}   &  \multicolumn{2}{c|}{59.714}   \\ \hline

$m_2$   &   \multicolumn{2}{c|}{36.759}    &  \multicolumn{2}{c|}{60.339}     \\ \hline

$m_3$  &   \multicolumn{2}{c|}{61.231}    &  \multicolumn{2}{c|}{34.839}     \\ \hline

$\text{mass order}$   & \multicolumn{2}{c|}{NO}  & \multicolumn{2}{c|}{IO}      \\ \hline

$m_{\beta}$  &  \multicolumn{2}{c|}{36.842}  &   \multicolumn{2}{c|}{59.476}  \\\hline

$m_{\beta\beta}$   &   32.285 &  17.962  &  55.125  &  27.041  \\ \hline\hline

\end{tabular}
\caption{\label{tab:model_predictions_II} Predictions for lepton mixing parameters, light neutrino masses and the effective masses $m_{\beta}$ of beta decay and $m_{\beta\beta}$ of the neutrinoless double-beta decay, where the unit of $r$, $s$, $t$ and mass is meV. The phases of the VEVs $v_{\varphi}$ and $v_{\chi}$ fulfill $v_{\varphi}/\left|v_{\varphi}\right|=\pm e^{i\frac{\pi}{4}}$ and $v_{\chi}/\left|v_{\chi}\right|=\pm e^{i\frac{\pi}{4}}$.}
\end{center}
\end{table}

\item[~~(b).~] {$\frac{v_{\varphi}}{\left|v_{\varphi}\right|}=\pm e^{-i\frac{\pi}{4}}$, $\frac{v_{\chi}}{\left|v_{\chi}\right|}=\pm e^{i\frac{\pi}{4}}$}

The remnant CP symmetry in the neutrino sector turns out to be $H^{\nu}_{\rm{CP}}=\left\{\rho(cd^3)X(\mathfrak{h_1}),\right.$
$\left.\rho(c^3d^3)X(\mathfrak{h_1})\right\}$ in this case. This corresponds to case VIII investigated in section~\ref{sec:model_independent_analysis}. Similar to the previous case, we introduce the following three real parameters:
\begin{equation}
r=-\frac{y^2v^2_{u}}{9}\frac{\Lambda}{x_1v_{\varphi}v_{\rho}}e^{-i\frac{\pi}{4}},\quad s=-i\frac{y^2v^2_{u}}{9}\frac{\Lambda}{x_4v_{\phi}v_{\sigma}},\quad
t=-i\frac{y^2v^2_{u}}{3}\frac{x_3v_{\chi}\Lambda}{x^2_1v_{\varphi}v^2_{\rho}}\,,
\end{equation}
which is related to the parameters $\alpha$, $\beta$, $\gamma$ and $\epsilon$ in Eq.~\eqref{eq:mnu_model} as
\begin{equation}
\alpha=re^{i\frac{\pi}{4}},\quad \beta=i\left(t-s\right),\quad \gamma=-i\left(s+\frac{t}{2}\right), \quad \epsilon=-\frac{\sqrt{3}}{2}t\,.
\end{equation}
As a consequence, the following relations are satisfied
\begin{eqnarray}
\nonumber&\alpha_{\mathrm{Im}}=\alpha_{\mathrm{Re}}=r/\sqrt{2},\qquad \beta_{\mathrm{Re}}=-2\gamma_{\mathrm{Re}}=0,\\
&\beta_{\mathrm{Im}}=\gamma_{\mathrm{Im}}-\sqrt{3}\;\epsilon_{\mathrm{Re}}=t-s,\qquad \epsilon_{\mathrm{Im}}=\sqrt{3}\;\gamma_{\mathrm{Re}}=0\,.
\end{eqnarray}
The constraints of case VIII are reproduced, as shown in Table~\ref{tab:constraints_by_GCP}. Therefore predictions for the PMNS matrix and light neutrino masses are given by Eq.~\eqref{eq:PMNS_VIII} and Eq.~\eqref{eq:nmass_VIII} respectively, and the resulting lepton mixing parameters are displayed in Table~\ref{tab:caseII_VI_and_IV_VIII}. Similar to the previous case, the light neutrino sector is also controlled by three real parameters $r$, $s$ and $t$, and hence the model is quite predictive. Imposing the measured values of $\theta_{13}$ and $\Delta m^2_{ij}$~\cite{GonzalezGarcia:2012sz}, predictions for lepton mixing parameters, neutrino masses, and the effective mass parameters $m_{\beta}$, $m_{\beta\beta}$ are presented in Table~\ref{tab:model_predictions_VIII}. Comparing with the results of scenario (b) in Table~\ref{tab:model_predictions_II}, we see that this scenario gives rise to nearly the same predictions for lepton mixing parameters and neutrino masses as scenario (b), except the Majorana phase $\alpha_{21}$. As a consequence, the predictions for $m_{\beta}$ coincide while $m_{\beta\beta}$ takes different values in the two scenarios.

\begin{table}[t!]
\begin{center}
\begin{tabular}{|c|c|c|c|c|} \hline\hline

& & & &\\[-0.18in]

$\left(\begin{array}{c}
r\\
s\\
t
\end{array}\right)$   &   $\pm\left(\begin{array}{c}
15.590\\
12.253\\
8.502
\end{array}
\right)$   &    $\pm\left(\begin{array}{c}
15.590 \\
-12.253 \\
8.502
\end{array}
\right)$   &   $\pm\left(\begin{array}{c}
15.204 \\
20.113  \\
-8.292
\end{array}
\right)$   &    $\pm\left(\begin{array}{c}
-15.204 \\
20.113 \\
8.292
\end{array}
\right)$   \\ 
& & & &\\[-0.18in]
\hline

$\sin^2\theta_{12}$  &  \multicolumn{4}{c|}{0.341}  \\ \hline

$\theta_{12}/^{\circ}$ &  \multicolumn{4}{c|}{35.734}    \\ \hline

$\sin^2\theta_{23}$  &  \multicolumn{4}{c|}{0.426}     \\ \hline

$\theta_{23}/^{\circ}$  &  \multicolumn{4}{c|}{40.759}  \\ \hline

$\sin\delta_{\rm{CP}}$   &   \multicolumn{2}{c|}{0.733}  &  \multicolumn{2}{c|}{$-0.733$}   \\ \hline

$\delta_{\rm{CP}}/^{\circ}$  &  \multicolumn{2}{c|}{47.166}   &  \multicolumn{2}{c|}{312.834}       \\ \hline

$\sin\alpha_{21}$   &  0.682   &  $-0.682$  &   0.732    & $-0.732$  \\ \hline

$\alpha_{21}/^{\circ}$  &  137.020   &  317.020  &  132.980  &  312.980  \\ \hline

$\sin\alpha_{31}$   &  \multicolumn{2}{c|}{0.0938}    &  \multicolumn{2}{c|}{$-0.0938$}     \\ \hline

$\alpha_{31}/^{\circ}$  &    \multicolumn{2}{c|}{5.383}   &   \multicolumn{2}{c|}{354.617}     \\ \hline

$\sin\alpha^{\prime}_{31}$   &  \multicolumn{2}{c|}{$-0.9998$}  &  \multicolumn{2}{c|}{0.9998}     \\ \hline

$\alpha^{\prime}_{31}/^{\circ}$  &  \multicolumn{2}{c|}{271.051}    &  \multicolumn{2}{c|}{88.949}    \\ \hline  \hline

$m_1$   &  \multicolumn{2}{c|}{35.724}     &  \multicolumn{2}{c|}{59.714}     \\ \hline

$m_2$   &   \multicolumn{2}{c|}{36.759}    &  \multicolumn{2}{c|}{60.339}    \\ \hline

$m_3$  &  \multicolumn{2}{c|}{61.231}    &   \multicolumn{2}{c|}{34.839}   \\ \hline

$\text{mass order}$   &  \multicolumn{2}{c|}{NO}    &  \multicolumn{2}{c|}{IO}    \\ \hline

$m_{\beta}$  &  \multicolumn{2}{c|}{36.842}  &  \multicolumn{2}{c|}{59.476}  \\ \hline

$m_{\beta\beta}$   &  15.696   &  33.445  &  29.211  &  54.006   \\ \hline\hline

\end{tabular}
\caption{\label{tab:model_predictions_VIII} Predictions for lepton mixing parameters, light neutrino masses and the effective masses $m_{\beta}$ of beta decay and $m_{\beta\beta}$ of the neutrinoless double-beta decay, where the unit of $a$, $b$, $c$ and mass is meV. The phases of the VEVs $v_{\varphi}$ and $v_{\chi}$ are $v_{\varphi}/\left|v_{\varphi}\right|=\pm e^{-i\frac{\pi}{4}}$ and $v_{\chi}/\left|v_{\chi}\right|=\pm e^{i\frac{\pi}{4}}$.}
\end{center}
\end{table}

  \item[~~(c).~] {$\frac{v_{\varphi}}{\left|v_{\varphi}\right|}=\pm e^{-i\frac{\pi}{4}}$, $\frac{v_{\chi}}{\left|v_{\chi}\right|}=\pm e^{-i\frac{\pi}{4}}$}

Comparing with the results of Appendix~\ref{sec:appendix_B}, we see that the VEVs of the flavons $\varphi$ and $\chi$ break the generalised CP symmetry to $H_{\rm{CP}}^{\nu}=\left\{\rho(cd^3)X(\mathfrak{h_1}), \rho(c^3d^3)X(\mathfrak{h_1})\right\}$ in the neutrino sector. Hence this case is identical to the case IV of the general analysis in section~\ref{sec:model_independent_analysis}. It is straightforward to check that the neutrino mass matrix $m^{\prime}_{\nu}$ is of the same form as the corresponding one of case IV. As a result, the lepton mixing matrix is
{\small
\begin{eqnarray*}
\nonumber&&\hskip-0.05in U_{\rm{PMNS}}=P_{321}U_{\rm{TB}}\Omega R\left(\theta\right)\\
\label{eq:PMNS_IV_model}&&=\frac{1}{\sqrt{3}}
\left(\begin{array}{ccc}
-e^{-\frac{3i\pi}{8}}
   \cos \left(\theta-\frac{5\pi}{24}
   \right)+e^{\frac{i\pi}{8}} \cos
   \left(\theta +\frac{5\pi}{24}\right)  &
   e^{-\frac{i\pi}{4}} &  -e^{-\frac{3i\pi}{8}} \sin \left(\theta-\frac{5 \pi
   }{24}\right)+e^{\frac{i\pi}{8}}\sin \left(\theta +\frac{5\pi}{24}\right)\\

e^{-\frac{3i\pi}{8}}
   \sin \left(\theta-\frac{\pi}{24}\right)+e^{\frac{i\pi}{8}} \sin
   \left(\theta +\frac{\pi}{24}\right) &
   ~~e^{-\frac{i\pi}{4}}~ &  -e^{-\frac{3i\pi}{8}}
   \cos \left(\theta-\frac{\pi}{24}\right)-e^{\frac{i\pi}{8}} \cos \left(\theta+\frac{\pi}{24}\right) \\

-i\cos\left(\theta+\frac{\pi}{4}\right)-e^{\frac{i\pi}{4}}\cos\left(\theta-\frac{\pi}{4}\right) &    e^{-\frac{i\pi}{4}} &  -i\cos \left(\theta-\frac{\pi}{4}\right)+e^{\frac{i\pi}{4}}\cos\left(\theta+\frac{\pi }{4}\right)
\end{array}\right)\,,
\end{eqnarray*}
}
\!\!where the unitary transformation $\Omega$ is of the form in Eq.~\eqref{eq:omega_IV}, and the angle $\theta$ satisfies $\tan2\theta=-t/(2r)$.
Therefore the reactor mixing angle $\theta_{13}$ is predicted to be\footnote{We can reorder the first and the third light neutrino masses such that the first and the third column of the above PMNS matrix is permutated. The resulting $\theta_{13}$ fulfills $\sin^2\theta_{13}=\frac{1}{3}+\frac{\sqrt{6}-\sqrt{2}}{12}\cos2\theta\geq\frac{1}{3}-\frac{\sqrt{6}-\sqrt{2}}{12}\simeq0.247$.}
\begin{equation}
\sin^2\theta_{13}=\frac{1}{3}+\frac{\sqrt{2}-\sqrt{6}}{12}\cos2\theta\geq\frac{1}{3}+\frac{\sqrt{2}-\sqrt{6}}{12}\simeq0.247.
\end{equation}
Obviously the observed value of $\theta_{13}$ cannot be produced in this scenario. As has been pointed out in section~\ref{sec:model_independent_analysis}, one has to permute the rows of $U_{\rm{PMNS}}$ as done in Eq.~\eqref{eq:PMNS_IV} in order to achieve agreement with the present data. This permutation corresponds to exchanging the three charged lepton masses. However, the charged lepton masses are predicted to be of different order of magnitude in the present model, so that this permutation is forbidden.

\item[~~(d).~] {$\frac{v_{\varphi}}{\left|v_{\varphi}\right|}=\pm e^{i\frac{\pi}{4}}$, $\frac{v_{\chi}}{\left|v_{\chi}\right|}=\pm e^{-i\frac{\pi}{4}}$}

The remnant CP symmetry in the neutrino sector is $H_{\rm{CP}}^{\nu}=\left\{\rho(cd)X(\mathfrak{h_1}),\rho(c^3d)X(\mathfrak{h_1})\right\}$. The constraints $\alpha_{\mathrm{Im}}=-\alpha_{\mathrm{Re}}$, $\beta_{\mathrm{Re}}=-2\gamma_{\mathrm{Re}}$, $\beta_{\mathrm{Im}}=\gamma_{\mathrm{Im}}-\sqrt{3}\;\epsilon_{\mathrm{Re}}$ and $\epsilon_{\mathrm{Im}}=\sqrt{3}\;\gamma_{\mathrm{Re}}$ are met for this phase structure. Hence this is exactly case VI discussed in section~\ref{sec:model_independent_analysis}. The PMNS matrix is of the form
{\small
\begin{eqnarray*}
\nonumber&&\hskip-0.05in U_{\rm{PMNS}}=P_{321}U_{\rm{TB}}\Omega\\
\label{eq:PMNS_VI_model}&&=\frac{1}{\sqrt{3}}
\left(\begin{array}{ccc}
-e^{\frac{3 i \pi
   }{8}} \cos \left(\theta-\frac{5 \pi
   }{24} \right)-e^{\frac{7i\pi}{8}} \cos \left(\theta +\frac{5
\pi}{24}\right)  & ~e^{\frac{i\pi}{4}}~ & -e^{\frac{3 i\pi}{8}} \sin
   \left(\theta-\frac{5\pi}{24}
   \right)-e^{\frac{7 i\pi}{8}} \sin
   \left(\theta +\frac{5\pi}{24}\right) \\

e^{\frac{3 i\pi}{8}} \sin
   \left(\theta-\frac{\pi}{24}
   \right)-e^{\frac{7 i\pi}{8}} \sin
   \left(\theta +\frac{\pi}{24}\right)
 & ~e^{\frac{i\pi}{4}}~ & -e^{\frac{3i\pi}{8}} \cos \left(\theta-\frac{\pi
   }{24} \right)+e^{\frac{7 i\pi}{8}}\cos\left(\theta+\frac{\pi}{24}\right)   \\

i \cos\left(\theta +\frac{\pi}{4}\right)+e^{\frac{3 i
\pi}{4}} \cos \left(\theta-\frac{\pi}{4} \right)  & ~e^{\frac{i\pi}{4}}~ &  i \cos \left(\theta-\frac{\pi}{4}\right)-e^{\frac{3 i\pi}{4}} \cos
   \left(\theta +\frac{\pi}{4}\right)
\end{array}\right)\,,
\end{eqnarray*}}
\!\!where the unitary matrix $\Omega$ is the one in Eq.~\eqref{eq:omega_VI} and the rotation angle $\theta$ is given by $\tan2\theta=-t/(2r)$. As a consequence, the reactor mixing angle is\footnote{If we exchange the first and the third columns of this PMNS matrix, $\theta_{13}$ would be given by $\sin^2\theta_{13}=\frac{1}{3}+\frac{\sqrt{6}-\sqrt{2}}{12}\cos2\theta\geq\frac{1}{3}-\frac{\sqrt{6}-\sqrt{2}}{12}$.}
\begin{equation}
\sin^2\theta_{13}=\frac{1}{3}-\frac{\sqrt{6}-\sqrt{2}}{12}\cos2\theta\geq\frac{1}{3}-\frac{\sqrt{6}-\sqrt{2}}{12}\simeq0.247\,,
\end{equation}
which doesn't match with the experimental data.

\end{description}

\subsection{\label{subsec:vacuum_alignment}Vacuum alignment}

%%%%%%%%%%%%%%%%%%%%%%
\begin{table} [hptb!]
\begin{center}
\begin{tabular}{|c||c|c|c|c|c|c|c|c|c|c|c|c|c|}
\hline\hline

{\tt Field}    & $D^{\prime 0}_{l}$ &  $D^{\prime\prime0}_{l}$ & $\varphi^{0}_l$  &  $\phi^{0}_l$  &   $\varphi^{0}$  &  $\phi^{0}$   &  $\chi^{0}$   &  $\Delta^{0}$  &  $\xi^{0}$  &  $\sigma^{0}\left(\sigma^{\prime0},\sigma^{\prime\prime0}\right)$
\\ \hline

&   &   &   &   &    &    &    &   &    &          \\ [-0.18in]

$\Delta(48)$  & $\mathbf{1}^{\prime}$  & $\mathbf{1}^{\prime\prime}$  &   $\mathbf{3{}^{\prime}}$   &   $\mathbf{3}$ & $\mathbf{\overline{3}}$  &  $\mathbf{\widetilde{3}}$ &  $\mathbf{3}^{\prime}$  & $\mathbf{\widetilde{3}}$  &   $\mathbf{1}$   &   $\mathbf{1}$  \\ 
\hline

$Z_2$  & 1  & 1   & 1  &   1   &  $1$   &  1  &   $1$   &   1   &   1   &   1  \\\hline

$Z_5$  & $\omega^2_5$ &  $\omega^2_5$  &  $\omega^3_5$   &  $\omega^2_5$   &  1   &   1   &    1  &    1    &  1   &  1  \\\hline

$Z_6$  & $\omega^3_6$ &  $\omega^3_6$   &  $\omega^4_6$ &  $\omega^3_6$   &   $\omega^2_6$  &   $\omega^2_6$    &   $\omega^4_6$   &    $\omega^4_6$    &  $\omega^2_6$    &  $\omega^4_6$   \\\hline

$U(1)_R$   &   2   &   2    &   2    &   2    &   2    &   2  &   2    &   2  &   2    &   2 \\\hline\hline

\end{tabular}
\caption{\label{tab:driving2_new_model}Driving fields and their
transformation properties under the family symmetry $\Delta(48) \times Z_2\times Z_5\times Z_6$ and $U(1)_R$.}
\end{center}
\end{table}
%%%%%%%%%%%%%%%%%%

We exploit the supersymmetric driving
field method to solve the vacuum alignment problem~\cite{Altarelli:2005yx}. This approach generally introduces a continuous $U(1)_R$ symmetry under which matter superfields carry charge $+1$, the Higgs and flavon fields are uncharged and the so-called driving fields indicated with the superscript ``0'' carry charges $+2$. In the limit of unbroken supersymmetry, the $F-$terms of the driving fields should vanish such that vacuums of flavons get aligned. The driving field content and transformation properties of these fields are shown in Table~\ref{tab:driving2_new_model}. The most general driving superpotential invariant under the family symmetry $\Delta(48) \times Z_2\times Z_5\times Z_6$ is given by
\begin{equation}
w_d=w^{l}_d+w^{\nu}_d\,,
\end{equation}
with
\begin{eqnarray}
\nonumber w^{l}_d&=&g_1D^{\prime 0}_{l}\left(\rho_{l}\varphi_l\right)_{\mathbf{1}^{\prime\prime}}+g_2D^{\prime\prime 0}_{l}\left(\rho_{l}\varphi_l\right)_{\mathbf{1}^{\prime}}+M_{\varphi_{l}}\left(\varphi^{0}_l\varphi_l\right)_{\mathbf{1}}
+g_3\left(\varphi^{0}_l\left(\rho_l\rho_l\right)_{\overline{\mathbf{3}}{}^{\prime}_S}\right)_{\mathbf{1}}\\
&& +M_{\phi_{l}}\left(\phi^{0}_{l}\phi_{l}\right)_{\mathbf{1}}+g_4\left(\phi^{0}_{l}\left(\rho_{l}\varphi_{l}\right)_{\mathbf{\overline{3}}}\right)_{\mathbf{1}}\,,\\
\nonumber  w^{\nu}_d &=&f_1 \big( \varphi^0 (\varphi\varphi)_{\mathbf{3}_S} \big)_\mathbf{1}+M_{\phi}\left(\phi^{0}\phi\right)_{\mathbf{1}}+f_2\left(\phi^{0}\left(\varphi\varphi\right)_{\mathbf{\widetilde{3}}}\right)_{\mathbf{1}}
+f_3\left(\chi^{0}\left(\chi\chi\right)_{\overline{\mathbf{3}}{}^{\prime}_S}\right)_\mathbf{1}\\
\nonumber&&+h_1\left(\Delta^{0}\left(\phi\phi\right)_{\mathbf{\widetilde{3}}_S}\right)_{\mathbf{1}}+h_2\left(\Delta^{0}\phi\right)_\mathbf{1}\xi
+h_3\left(\Delta^{0}\left(\chi\chi\right)_{\mathbf{\widetilde{3}}}\right)_{\mathbf{1}}
+M_{\xi}\xi^0\xi+k\xi^0\sigma^2 \\
\nonumber&& +M_{\sigma}\sigma^0\sigma+k_1\sigma^0\xi^2+k_{2}\sigma^0\rho^{2}+k_3\sigma^{0}(\phi\phi)_{\mathbf{1}}
+M^{\prime}_{\sigma}\sigma^{\prime0}\sigma+k^{\prime}_1\sigma^{\prime0}\xi^2+k^{\prime}_{2}\sigma^{\prime0}\rho^{2}\\
\label{eq:driving}&&+k^{\prime}_3\sigma^{\prime0}(\phi\phi)_{\mathbf{1}}+M^{\prime\prime}_{\sigma}\sigma^{\prime\prime0}\sigma+k^{\prime\prime}_1\sigma^{\prime\prime0}\xi^2
+k^{\prime\prime}_{2}\sigma^{\prime\prime0}\rho^{2}+k^{\prime\prime}_3\sigma^{\prime\prime0}(\phi\phi)_{\mathbf{1}}\,.
\end{eqnarray}
Here all the coupling constants and mass parameters are real due to the imposed generalised CP symmetry. The vacuum alignment associated to the charged lepton sector is determined by the $F-$term conditions of the driving fields $D^{\prime 0}_{l}$, $D^{\prime\prime 0}_{l}$, $\varphi^0_l$ and $\phi^0_l$, i.e.,
\begin{eqnarray}
&&\frac{\partial w^{l}_{d}}{\partial D^{\prime0}_{l}}=g_1\left(\rho_{l,1}\varphi_{l,2}+\rho_{l,2}\varphi_{l,3}+\rho_{l,3}\varphi_{l,1}\right)=0\,,\nonumber\\
&&\frac{\partial w^{l}_{d}}{\partial D^{\prime\prime0}_{l}}=g_2\left(\rho_{l,1}\varphi_{l,3}+\rho_{l,2}\varphi_{l,1}+\rho_{l,3}\varphi_{l,2}\right)=0\,,\nonumber\\
&&\frac{\partial w^{l}_{d}}{\partial\varphi^{0}_{l,1}}=M_{\varphi_{l}}\varphi_{l,1}+2g_3\left(\rho^2_{l,1}-\rho_{l,2}\rho_{l,3}\right)=0\,,\nonumber\\
&&\frac{\partial w^{l}_{d}}{\partial\varphi^{0}_{l,2}}=M_{\varphi_{l}}\varphi_{l,2}+2g_3\left(\rho^2_{l,2}-\rho_{l,1}\rho_{l,3}\right)=0\,,\nonumber\\
&&\frac{\partial w^{l}_{d}}{\partial\varphi^{0}_{l,3}}=M_{\varphi_{l}}\varphi_{l,3}+2g_3\left(\rho^2_{l,3}-\rho_{l,1}\rho_{l,2}\right)=0\,,\nonumber\\
&&\frac{\partial w^{l}_{d}}{\partial\phi^{0}_{l,1}}=M_{\phi_{l}}\phi_{l,1}+g_4\left(\rho_{l,1}\varphi_{l,1}+\omega\rho_{l,3}\varphi_{l,3}+\omega^2\rho_{l,2}\varphi_{l,2}\right)=0\,,\nonumber\\
&&\frac{\partial w^{l}_{d}}{\partial\phi^{0}_{l,2}}=M_{\phi_{l}}\phi_{l,2}+g_4\left(\rho_{l,2}\varphi_{l,3}+\omega\rho_{l,1}\varphi_{l,2}+\omega^2\rho_{l,3}\varphi_{l,1}\right)=0\,,\nonumber\\
&&\frac{\partial w^{l}_{d}}{\partial\phi^{0}_{l,3}}=M_{\phi_{l}}\phi_{l,3}+g_4\left(\rho_{l,3}\varphi_{l,2}+\omega\rho_{l,2}\varphi_{l,1}+\omega^2\rho_{l,1}\varphi_{l,3}\right)=0\,.
\end{eqnarray}
The solution to the above equation is given by
\begin{equation}
\label{eq:vev_lepton}\langle\phi_l\rangle=\left(\begin{array}{c}
1\\
0\\
0
\end{array}\right)v_{\phi_l},\quad
\langle\varphi_l\rangle=\left(\begin{array}{c}
0\\
1\\
0
\end{array}\right)v_{\varphi_l},\quad
\langle\rho_l\rangle=\left(\begin{array}{c}
0\\
1\\
0
\end{array}\right)v_{\rho_l}\,.
\end{equation}
The VEVs $v_{\phi_l}$, $v_{\varphi_l}$ and $v_{\rho_l}$ are related by
\begin{equation}
v_{\varphi_l}=-2g_3\frac{v^2_{\rho_l}}{M_{\varphi_{l}}},\qquad v_{\phi_l}=2\omega^2g_3g_4\frac{v^3_{\rho_l}}{M_{\varphi_{l}}M_{\phi_{l}}}\,.
\end{equation}
In the neutrino sector, the vacuum alignment problem is more complicated, since we need to realize not only the desired $Z_2$ residual family symmetry but also the residual CP symmetry after symmetry breaking. In other words, we have to handle the phases of the VEVs very carefully. In the following, we will solve this problem step by step from the driving superpotential $w^{\nu}_d$. The $F-$term conditions of the driving fields $\varphi^{0}$ and $\phi^{0}$ take the following form
\begin{eqnarray}
\nonumber&&\frac{\partial w^{\nu}_{d}}{\partial\varphi^{0}_1}=2f_1\left(\varphi^2_{1}-\varphi_2\varphi_3\right)=0\,,\\
\nonumber&&\frac{\partial w^{\nu}_{d}}{\partial\varphi^{0}_2}=2f_1\left(\varphi^2_{2}-\varphi_3\varphi_1\right)=0\,,\\
\nonumber&&\frac{\partial w^{\nu}_{d}}{\partial\varphi^{0}_3}=2f_1\left(\varphi^2_{3}-\varphi_1\varphi_2\right)=0\,,\\
\nonumber&&\frac{\partial w^{\nu}_d}{\partial\phi^{0}_1}=M_{\phi}\phi_1+f_2\left(\varphi^2_1+2\varphi_2\varphi_3\right)=0,\\
\nonumber&&\frac{\partial w^{\nu}_d}{\partial\phi^{0}_2}=M_{\phi}\phi_3+f_2\left(\varphi^2_3+2\varphi_1\varphi_2\right)=0,\\
&&\frac{\partial w^{\nu}_d}{\partial\phi^{0}_3}=M_{\phi}\phi_2+f_2\left(\varphi^2_2+2\varphi_1\varphi_3\right)=0\,.
\end{eqnarray}
we can straightforwardly get vacuums of $\varphi$ and $\phi$ as
\begin{equation}
\label{eq:vev_varphi_phi}\langle\varphi\rangle=\left(\begin{array}{c}
1\\
1\\
1
\end{array}\right)v_{\varphi},\qquad \langle\phi\rangle=\left(\begin{array}{c}
1\\
1\\
1
\end{array}\right)v_{\phi} \,,
\end{equation}
where
\begin{equation}
\label{eq:VEV_relation_clepton}v^2_{\varphi}=-\frac{M_{\phi}v_{\phi}}{3f_2}\,.
\end{equation}
Furthermore, the vacuum of $\chi$ is determined by
\begin{eqnarray}
\nonumber&&\frac{\partial w^{\nu}_d}{\partial\chi^{0}_1}=2f_3\left(\chi^2_1-\chi_2\chi_3\right)=0,\\
\nonumber&&\frac{\partial w^{\nu}_d}{\partial\chi^{0}_2}=2f_3\left(\chi^2_2-\chi_1\chi_3\right)=0,\\
\nonumber&&\frac{\partial w^{\nu}_d}{\partial\chi^{0}_3}=2f_3\left(\chi^2_3-\chi_1\chi_2\right)=0,\\
\nonumber&&\frac{\partial w^{\nu}_d}{\partial\Delta^{0}_1}=2h_1\left(\phi^2_1-\phi_2\phi_3\right)+h_2\phi_1\xi+h_3\left(\chi^2_1+2\chi_2\chi_3\right)=0,\\
\nonumber&&\frac{\partial w^{\nu}_d}{\partial\Delta^{0}_2}=2h_1\left(\phi^2_2-\phi_1\phi_3\right)+h_2\phi_3\xi+h_3\left(\chi^2_2+2\chi_1\chi_3\right)=0,\\
&&\frac{\partial w^{\nu}_d}{\partial\Delta^{0}_3}=2h_1\left(\phi^2_3-\phi_1\phi_2\right)+h_2\phi_2\xi+h_3\left(\chi^2_3+2\chi_1\chi_2\right)=0\,.
\end{eqnarray}
Given the alignment of $\phi$ in Eq.~\eqref{eq:vev_varphi_phi}, the vacuum of the flavon $\chi$ is derived as
\begin{equation}
\label{eq:vev_chi}\langle\chi\rangle=\left(
\begin{array}{c}
1\\
1\\
1
\end{array}
\right)v_{\chi}\,,
\end{equation}
with
\begin{equation}
\label{eq:VEV_relation_neutrino}v^2_{\chi}=-\frac{h_2}{3h_3}v_{\phi}v_{\xi}\,.
\end{equation}
Finally, the $F-$term conditions obtained from the singlet driving fields $\xi^{0}$, $\sigma^{0}$, $\sigma^{\prime0}$ and $\sigma^{\prime\prime0}$ are of the form
\begin{eqnarray}
\nonumber&&\frac{\partial w^{\nu}_d}{\partial\xi^{0}}=M_{\xi}\xi+k\sigma^{2}=0\,,\\
\nonumber&&\frac{\partial w^{\nu}_d}{\partial\sigma^{0}}=M_{\sigma}+k_1\xi^2+k_2\rho^2+k_3\left(\phi^2_1+2\phi_2\phi_3\right)=0\,,\\
\nonumber&&\frac{\partial w^{\nu}_d}{\partial\sigma^{\prime0}}=M^{\prime}_{\sigma}+k^{\prime}_1\xi^2+k^{\prime}_2\rho^2+k^{\prime}_3\left(\phi^2_1+2\phi_2\phi_3\right)=0\,,\\
&&\frac{\partial w^{\nu}_d}{\partial\sigma^{\prime\prime0}}=M^{\prime\prime}_{\sigma}+k^{\prime\prime}_1\xi^2+k^{\prime\prime}_2\rho^2+k^{\prime\prime}_3\left(\phi^2_1+2\phi_2\phi_3\right)=0\,.
\end{eqnarray}
The solution to this equation is
\begin{equation}
v_{\sigma}=\left(\frac{M^2_{\xi}M_1}{k^2}\right)^{\frac{1}{3}},\quad v_{\xi}=\pm\left(\frac{M^2_{\xi}M^4_1}{k^2}\right)^{\frac{1}{6}},\quad v^2_{\rho}=M_2v_{\sigma},\quad  3v^2_{\phi}=M_{3}v_{\sigma}\,,
\end{equation}
where $v_{\xi}=\langle\xi\rangle$, $v_{\sigma}=\langle\sigma\rangle$, $v_{\rho}=\langle\rho\rangle$, and the mass parameters $M_{1,2,3}$ are defined as
\begin{eqnarray}
\nonumber&&M_1\equiv\frac{\left(k^{\prime}_3k^{\prime\prime}_2-k^{\prime}_2k^{\prime\prime}_3\right)M_{\sigma}+\left(k_2k^{\prime\prime}_3-k_3k^{\prime\prime}_2\right)M^{\prime}_{\sigma}
+\left(k_3k^{\prime}_2-k_2k^{\prime}_3\right)M^{\prime\prime}_{\sigma}}{k_1\left(k^{\prime}_2k^{\prime\prime}_3-k^{\prime}_3k^{\prime\prime}_2\right)+k_2\left(k^{\prime}_3k^{\prime\prime}_1-k^{\prime}_1k^{\prime\prime}_3\right)+
k_3\left(k^{\prime}_1k^{\prime\prime}_2-k^{\prime}_2k^{\prime\prime}_1\right)}\,,\\
\nonumber&&M_2\equiv\frac{\left(k^{\prime}_1k^{\prime\prime}_3-k^{\prime}_3k^{\prime\prime}_1\right)M_{\sigma}+\left(k_3k^{\prime\prime}_1-k_1k^{\prime\prime}_3\right)M^{\prime}_{\sigma}
+\left(k_1k^{\prime}_3-k_3k^{\prime}_1\right)M^{\prime\prime}_{\sigma}}{k_1\left(k^{\prime}_2k^{\prime\prime}_3-k^{\prime}_3k^{\prime\prime}_2\right)+k_2\left(k^{\prime}_3k^{\prime\prime}_1-k^{\prime}_1k^{\prime\prime}_3\right)+
k_3\left(k^{\prime}_1k^{\prime\prime}_2-k^{\prime}_2k^{\prime\prime}_1\right)}\,,\\
&&M_3\equiv\frac{\left(k^{\prime}_2k^{\prime\prime}_1-k^{\prime}_1k^{\prime\prime}_2\right)M_{\sigma}+\left(k_1k^{\prime\prime}_2-k_2k^{\prime\prime}_1\right)M^{\prime}_{\sigma}
+\left(k_2k^{\prime}_1-k_1k^{\prime}_2\right)M^{\prime\prime}_{\sigma}}{k_1\left(k^{\prime}_2k^{\prime\prime}_3-k^{\prime}_3k^{\prime\prime}_2\right)+k_2\left(k^{\prime}_3k^{\prime\prime}_1-k^{\prime}_1k^{\prime\prime}_3\right)+
k_3\left(k^{\prime}_1k^{\prime\prime}_2-k^{\prime}_2k^{\prime\prime}_1\right)}\,.
\end{eqnarray}
For the case $M_1M_2>0$ and $M_1M_3<0$, we have
\begin{equation}
v_{\rho}=\pm\left(\frac{M^2_{\xi}M_1M^3_2}{k^2}\right)^{\frac{1}{6}},\qquad v_{\phi}=\pm i\left(-\frac{M^2_{\xi}M_1M^3_3}{k^2}\right)^{\frac{1}{6}}\,.
\end{equation}
For the opposite case $M_1M_2<0$ and $M_1M_3>0$, the VEVs $v_{\rho}$ and $v_{\phi}$ would become purely imaginary and real, respectively. However, we do not consider this option in this paper. Taking into account the relations among the different VEVs shown in Eq.~\eqref{eq:VEV_relation_clepton} and Eq.~\eqref{eq:VEV_relation_neutrino}, we have
\begin{eqnarray}
v_{\varphi}=\pm\left(\frac{M^{6}_{\phi}M^2_{\xi}M_1M^3_3}{3^9f^6_2k^2}\right)^{\frac{1}{12}}e^{\pm i\frac{\pi}{4}},\qquad v_{\chi}=\pm\left(\frac{h^6_2M^4_{\xi}M^5_1M^3_3}{3^9h^6_3k^4}\right)^{\frac{1}{12}}e^{\pm i\frac{\pi}{4}}\,.
\end{eqnarray}
Thus we have elaborated that the desired vacuums in Eqs.~\eqref{eq:vev} and \eqref{eq:VEV_phases} can be achieved.

\subsection{Ultraviolet completion}

\begin{table} [hptb!]
\begin{center}
\begin{tabular}{|c||c|c|c|c|c|c|c|c|c|c|}
\hline\hline
{\tt Field}    &  $\Theta_1$   &   $\Theta^{c}_1$  &  $\Theta_2$   &   $\Theta^{c}_2$    &  $\Theta_3$   &   $\Theta^{c}_3$    & $\Sigma_1$   &   $\Sigma^{c}_1$  &  $\Sigma_2$   &   $\Sigma^{c}_2$  \\ \hline

  &   &   &   &   &   &   &  &  &  &  \\ [-0.18in]

$\Delta(48)$  &  $\mathbf{\overline{3}}$   &  $\mathbf{3}$   &  $\mathbf{\overline{3}}$   &  $\mathbf{3}$   & $\mathbf{\overline{3}}$   &  $\mathbf{3}$  &  $\mathbf{\overline{3}}$   &  $\mathbf{3}$  &  $\mathbf{\overline{3}}$   &  $\mathbf{3}$  \\ \hline

$Z_2$  &   1   &  1   &   1   &  1    &   1   &  1   &   $-1$   &  $-1$  &  1    &   1  \\  \hline

$Z_5$  &   1   &  1  &  $\omega^3_5$  &    $\omega^2_5$  &  $\omega_5$  &  $\omega^4_5$   &   1   &  1  &  1    &   1  \\  \hline

$Z_6$  &   1  &  1    &  $\omega^4_6$   & $\omega^2_6$    & $\omega^2_6$   & $\omega^4_6$  &  $\omega^4_6$   & $\omega^2_6$ & $\omega^2_6$   & $\omega^4_6$   \\  \hline

$U(1)_{R}$  &  1  &  1   &  1  &  1 &  1  &  1 &  1  &  1  &  1  &  1 \\  \hline\hline

\end{tabular}
\caption{\label{tab:messenger_model}Messenger fields and their
transformation rules under the family symmetry $\Delta(48) \times Z_2\times Z_5\times Z_6$ and $U(1)_R$.}
\end{center}
\end{table}

Having completed the model construction and its phenomenological implication analysis at LO, the question of the higher order corrections arises. At the purely effective level, all the higher dimensional (non-renormalisable) operators compatible with the imposed symmetries should be taken into account. As a consequence, the successful LO results generally tend to be erased partly by the large numbers of higher order contributions. Thus a purely effective formulation would leave room for different physical predictions. In order to remove any such ambiguity within our model, we would like to formulate an ultraviolet (UV) completion of the above effective model. In such UV-completed model, the non-renormalisable terms of the effective theory arise by integrating out the heavy messenger fields.

\begin{figure}[t!]
\begin{center}
\includegraphics[width=0.8\textwidth]{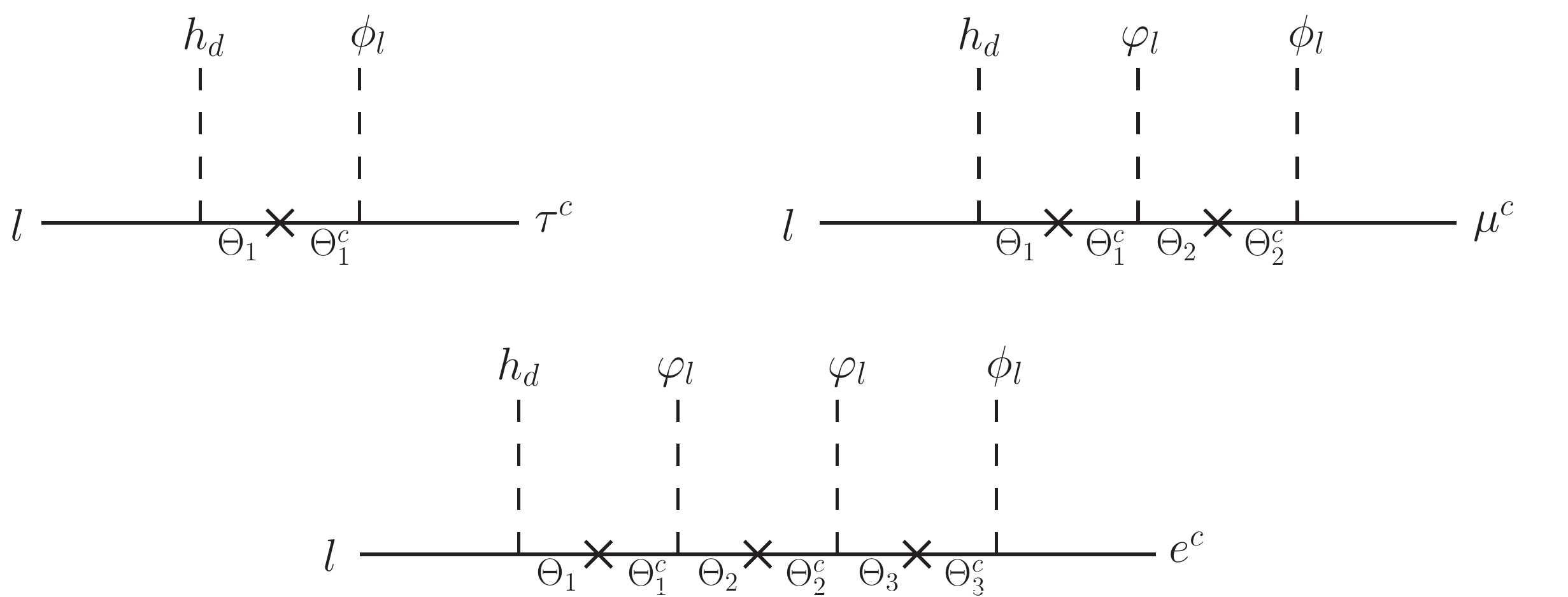}
\caption{\label{fig:charged_gene} Diagrams which generate the effective operators for the charged lepton masses, where a cross denotes the mass insertion of a fermion.}
\end{center}
\end{figure}

\begin{figure}[t!]
\begin{center}
\includegraphics[width=0.8\textwidth]{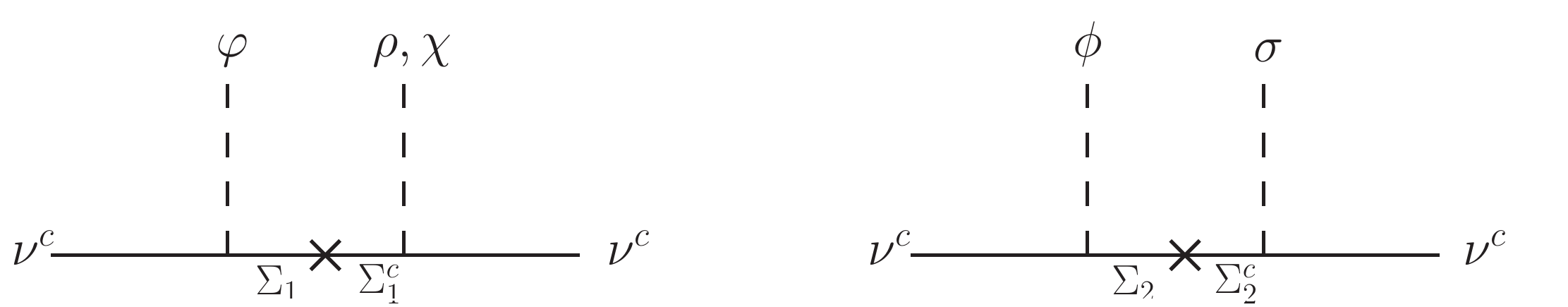}
\caption{\label{fig:neutrino_gene} Diagrams for generating the effective operators of the right-handed neutrino masses, where a cross indicates a fermionic mass insertion.}
\end{center}
\end{figure}

The driving superpotential $w_d$ in Eq.~\eqref{eq:driving} which produces the required vacuum alignment is already fully renormalisable, and its existence is not subject to the presence of the messenger fields. As a result, the vacuum alignments of flavon fields shown in Eqs.~\eqref{eq:vev_lepton}, \eqref{eq:vev_varphi_phi}, and \eqref{eq:vev_chi} remain intact. We come to the non-renormalisable
superpotential $w^{\text{eff}}_{l}$ in Eq.~\eqref{eq:w_charged_lepton} which is responsible for the charged lepton masses. To generate these terms, we add three pairs of heavy fields $\Theta_i$ and $\Theta^{c}_i$ with $i=1, 2, 3$. Notice that these messenger fields are chiral superfields with non-vanishing hypercharge: 2 (+2) for $\Theta_i$ ($\Theta^{c}_i$). Similar to the lepton fields, they carry a charge $+1$ under the $U(1)_{R}$ symmetry. Their transformation properties under the family symmetry $\Delta(48)\times Z_2\times Z_5\times Z_6$ can be found from Table~\ref{tab:messenger_model}. With the particle content and charge assignments collected in Table~\ref{tab:matter_flavon_new_model} and Table~\ref{tab:messenger_model}, the renormalisable superpotential for the charged lepton sector reads
\begin{eqnarray}
\nonumber \hskip-0.2in w_{l}&=&z_1\left(l\Theta_{1}\right)_{\mathbf{1}}h_d+z_2\left(\Theta^{c}_1\phi_{l}\right)_{\mathbf{1}}\tau^{c}+z_3\left(\left(\Theta^{c}_1\Theta_2\right)_{\mathbf{3^{\prime}}}\varphi_{l}\right)_{\mathbf{1}}
+z_4\left(\Theta^{c}_2\phi_{l}\right)_{\mathbf{1}}\mu^{c}+z_5\left(\left(\Theta^{c}_2\Theta_3\right)_{\mathbf{3^{\prime}}}\varphi_{l}\right)_{\mathbf{1}}\\
\hskip-0.2in&&+z_6\left(\Theta^{c}_3\phi_{l}\right)_{\mathbf{1}}e^{c}+M_{\Theta_1}\left(\Theta_1\Theta^{c}_1\right)_{\mathbf{1}}
+M_{\Theta_2}\left(\Theta_2\Theta^{c}_2\right)_{\mathbf{1}}+M_{\Theta_3}\left(\Theta_3\Theta^{c}_3\right)_{\mathbf{1}}\,,
\end{eqnarray}
where the generalised CP invariance again implies that all the coupling constants $z_i$ (i=1\ldots6) and the messenger masses $M_{\Theta_i}$ ($i=1, 2, 3$) are real. This superpotential gives rise to the Feynman diagrams shown in Fig.~\ref{fig:charged_gene}. After integrating out the messenger pairs $\Theta_i$ and $\Theta^{c}_i$, we obtain the following effective operators:
\begin{eqnarray}
\nonumber w^{\text{eff}}_{l}&=&-\frac{z_1z_2}{M_{\Theta_1}}\left(l\phi_{l}\right)_{\mathbf{1}}\tau^{c}h_d+\frac{z_1z_3z_4}{M_{\Theta_1}M_{\Theta_2}}\left(l\left(\phi_{l}\varphi_{l}\right)_{\mathbf{\overline{3}}}\right)_{\mathbf{1}}\mu^{c}h_d\\
&&-\frac{z_1z_3z_5z_6}{2M_{\Theta_1}M_{\Theta_2}M_{\Theta_3}}\left(l\left(\phi_{l}\left(\varphi_{l}\varphi_{l}\right)_{\mathbf{3}{}^{\prime}_S}\right)_{\mathbf{\overline{3}}}\right)_{\mathbf{1}}e^{c}h_d\,.
\end{eqnarray}
Note that the term $\left(l\left(\rho_l\left(\phi_l\phi_l\right)_{\mathbf{\widetilde{3}}}\right)_{\mathbf{\overline{3}}}\right)_{\mathbf{1}}e^{c}h_d$ in Eq.~\eqref{eq:w_charged_lepton} is not generated. The reason is that here we consider the minimal UV completion which has the least number of extra messenger fields and the fewest number of associated (renormalisable) couplings, and there is no messenger field to mediate this non-renormalisable term. However, this doesn't affect the low-energy observables, as all the entries are produced at the same order as in the original effective theory. Inserting the flavon VEVs $\langle\phi_{l}\rangle=\left(v_{\phi{_l}}, 0, 0\right)$ and $\langle\varphi_{l}\rangle=\left(0, v_{\varphi_{l}}, 0\right)$ in Eq.~\eqref{eq:vev_lepton}, a diagonal charged lepton mass matrix is obtained with
\begin{equation}
m_{\tau}=-z_{1}z_{2}\frac{v_{\phi_{l}}}{M_{\Theta_1}}v_d,~~m_{\mu}=\omega^2z_1z_3z_4\frac{v_{\phi_{l}}v_{\varphi_{l}}}{M_{\Theta_1}M_{\Theta_2}}v_d,~~
m_{e}=-\omega^2z_1z_3z_5z_6\frac{v_{\phi_{l}}v^{2}_{\varphi_{l}}}{M_{\Theta_1}M_{\Theta_2}M_{\Theta_3}}v_d\,.
\end{equation}
For the effective neutrino superpotential $w^{\text{eff}}_{\nu}$ in Eq.~\eqref{eq:w_neutrino}, the neutrino Dirac coupling term $\left(l\nu^{c}\right)_{\mathbf{1}}h_{u}$ is already renormalisable. The Majorana mass terms for right-handed neutrinos couple with two flavons, and they can be generated by introducing two new pairs of messengers: $\Sigma_i$ and $\Sigma^{c}_i$ ($i=1, 2$) which are chiral superfields with vanishing hypercharges. The renormalisable neutrino superpotential of the minimal completion giving rise to the effective potential $w^{\text{eff}}_{\nu}$ is
\begin{eqnarray}
\nonumber w_{\nu}&=&y\left(l\nu^{c}\right)_{\mathbf{1}}h_u+q_1\left(\left(\varphi\nu^{c}\right)_{\mathbf{3}_S}\Sigma_1\right)_{\mathbf{1}}+q_2\left(\left(\varphi\nu^{c}\right)_{\mathbf{3}_A}\Sigma_1\right)_{\mathbf{1}}
+q_3\left(\nu^{c}\Sigma^{c}_1\right)_{\mathbf{1}}\rho+q_4\left(\left(\nu^{c}\chi\right)_{\mathbf{\overline{3}}}\Sigma^{c}_1\right)_{\mathbf{1}}\\
&& +q_5\left(\left(\nu^{c}\phi\right)_{\mathbf{3}}\Sigma_2\right)_{\mathbf{1}}+q_{6}\left(\nu^{c}\Sigma^{c}_2\right)_{\mathbf{1}}\sigma
+M_{\Sigma_{1}}\left(\Sigma_1\Sigma^{c}_1\right)_{\mathbf{1}}+M_{\Sigma_{2}}\left(\Sigma_2\Sigma^{c}_2\right)_{\mathbf{1}}\,.
\end{eqnarray}
With this superpotential, the Feynman diagrams shown in Fig.~\ref{fig:neutrino_gene}  can be constructed. Integrating out the messengers, we obtain
\begin{eqnarray}\label{eq:UV_completion1}
\nonumber w^{\text{eff}}_{\nu}&=&-\frac{q_1q_3}{M_{\Sigma_1}}\left(\left(\nu^{c}\nu^{c}\right)_{\mathbf{3}_S}\varphi\right)_{\mathbf{1}}\rho
-\frac{\left(\sqrt{3}\;q_1-iq_2\right)q_4}{2\sqrt{3}\;M_{\Sigma_1}}\left(\left(\nu^{c}\nu^{c}\right)_{\mathbf{3}_S}\left(\varphi\chi\right)_{\mathbf{\overline{3}}}\right)_{\mathbf{1}}\\
&&-\frac{\left(\sqrt{3}\;q_1+iq_2\right)q_4}{\sqrt{3}\;M_{\Sigma_1}}\left(\left(\nu^c\nu^c\right)_{\mathbf{\widetilde{3}}}\left(\varphi\chi\right)_{\mathbf{\widetilde{3}}}\right)_{\mathbf{1}}
-\frac{q_5q_6}{M_{\Sigma_2}}\left(\left(\nu^{c}\nu^{c}\right)_{\mathbf{\widetilde{3}}}\phi\right)_{\mathbf{1}}\sigma\,.
\end{eqnarray}
We see that all the non-renormalisable terms in Eq.~\eqref{eq:w_neutrino} are reproduced here. We note that there are additional terms which also appear at the renormalisable level involving the messengers
\begin{equation}
p_1\left(\left(\Sigma_1\Sigma_1\right)_{\mathbf{\widetilde{3}}}\phi\right)_{\mathbf{1}}+p_2\left(\Sigma_1\Sigma^{c}_2\right)_{\mathbf{1}}\rho
+p_3\left(\left(\Sigma_1\Sigma^{c}_2\right)_{\mathbf{\overline{3}^{\prime}}}\chi\right)_{\mathbf{1}}
+p_4\left(\left(\Sigma^c_2\Sigma^{c}_2\right)_{\mathbf{\widetilde{3}}}\phi\right)_{\mathbf{1}}\,.
\end{equation}
These terms can also be taken into account when the heavy fields are integrated out.  They give rise to subleading contributions of the form $\left(\nu^c\nu^c\Phi^3_{\nu}\right)_{\mathbf{1}}/M^2_{\Sigma}$, where $\Phi_{\nu}=\left\{\varphi, \phi, \chi, \rho, \sigma\right\}$ and $M_{\Sigma}$ denote the generic messenger mass $M_{\Sigma_1}$ or $M_{\Sigma_2}$. Since these subleading operators are not contaminated by the charged lepton flavons $\phi_{l}$, $\varphi_{l}$ and $\rho_{l}$, the results for the mixing parameters remain unchanged, whereas the light neutrino masses acquire corrections which are suppressed by $\langle\Phi_{\nu}\rangle/M_{\Sigma}$.

\section{\label{sec:conclusions}Conclusions}

In this work, we perform a comprehensive analysis of the $\Delta(48)$ family symmetry combined with the generalized CP symmetry $H_{\rm{CP}}$. The generalized CP transformation is a necessary extension of the canonical CP transformation $\phi\to\phi^*$ in the presence of a non-abelian discrete family symmetry. It has been established that each generalised CP transformation corresponds to an automorphism of the family group. The automorphism group of $\Delta(48)$ is somewhat complex: $\mathrm{Aut}(\Delta(48))\cong\left(\left(\left(\left(Z_4\times Z_4\right)\rtimes Z_3\right)\rtimes Z_4\right)\rtimes Z_2\right)$, and its order is 384. Hence $\Delta(48)$ family symmetry provides much more choices for admissible generalised CP transformations than some popular family groups such as $A_4$, $S_4$ etc. As a consequence, different results for mixing angles and CP phases arise.

We have performed a systematic and model-independent analysis of lepton mixing within $\Delta(48)\rtimes H_{\rm{CP}}$, where neutrinos are taken to be Majorana particles, and the generalized CP and family symmetries are assumed to be broken to different subgroups in the charged lepton and neutrino sectors, respectively. Totally, we have found 10 cases, as shown in Table~\ref{tab:caseI} and Table~\ref{tab:caseII_VI_and_IV_VIII}. The predictions for neutrino masses and lepton mixing parameters are presented in detail for each case. All mixing angles and CP phases are determined in terms of a single real parameter $\theta$. In order to assess to what extent the experimental data can be explained, a $\chi^2$ analysis is performed. The measured lepton mixing angles can be accommodated rather well for certain values of the parameter $\theta$ except cases III and V  which predict a slightly large $\theta_{13}$. In particular, we find a new mixing pattern in which all CP phases are nontrivial functions of the parameter $\theta$, as shown in Table~\ref{tab:caseII_VI_and_IV_VIII}. The excellent agreement with the experimental data can be achieved by a proper choice of $\theta$, and the Dirac-type CP violation is neither maximal nor vanishing. This mixing pattern can be tested in future long-baseline neutrino oscillation experiments.

Motivated by the general analysis, we construct an effective supersymmetric model based on $\Delta(48)$ family symmetry and generalised CP symmetry, and the auxiliary symmetry $Z_2\times Z_5\times Z_6$ is introduced to eliminated dangerous operators. In our model, the observed charged lepton mass hierarchy is reproduced, since the tau, muon and electron mass terms involve one flavon, two flavons and three flavons, respectively. The symmetry $\Delta(48)\rtimes H_{\rm{CP}}$ is spontaneously broken down to $Z^{c^2}_2\times CP$ in the neutrino sector, and the new interesting mixing pattern in Table~\ref{tab:caseII_VI_and_IV_VIII} is naturally realized. At leading order, the light neutrino mass matrix depends on three parameters which can be fixed by the measured mass-squared differences $\Delta m^2_{12}$, $\Delta m^2_{31}$ for NO (or $\Delta m^2_{32}$ for IO ), and the mixing angle $\theta_{13}$. Therefore our model have definite predictions for mixing angles, CP phases and the absolute neutrino mass scale.  In addition, we have shown that the desired vacuum alignment can be achieved in the driving field approach. Furthermore, the UV completion of the model is formulated in order to remove the ambiguity caused by higher dimensional operators allowed by the symmetry, in which non-renormalisable operators arise from integrating out the heavy messenger fields.

One of the major physical goals of the future experimental neutrino physics is to measure the CP violation. Combining family symmetry with generalized CP symmetry may shed new light on the origin of lepton mixing and CP violation in the lepton sector. Most models based on this idea usually predict that the Dirac CP phase $\delta_{\rm{CP}}$ takes some specific values $\delta_{\rm{CP}}=0,~\pi$ or $\pm\pi/2$. However, as what happened in the quark sector, the Dirac CP-violating phase in the lepton sector might not be such regular values. If that is the case, $\Delta(48)$ family symmetry and the associated generalised CP could be a useful alternative.

\section*{Acknowledgements}

This work was supported in part by the National Natural Science Foundation of China under Grant Nos. 11275188, 11179007, and 11135009.

\newpage

\section*{Appendices}
\cleqn

\begin{appendix}

\section{\label{sec:appendix_A_group_theory}Group theory of $\Delta(48)$}
\cleqn

$\Delta(48)$ is a non-Abelian finite subgroup of $SU(3)$ of order $48$, and it belongs to the well-known $\Delta(3n^2)$ series with $n=4$. The group theory of $\Delta(3n^2)$ has been extensively discussed in Ref.~\cite{Luhn:2007uq}. $\Delta(48)$ is isomorphic to $(Z_4\times Z_4)\rtimes Z_3$, and it can be generated by three generators $a$, $c$ and $d$ satisfying the following multiplication rules:
\begin{eqnarray}
&\nonumber a^3=c^4=d^4=1,\quad cd=dc,\\
\label{eq:group_rules}&aca^{-1}=c^{-1}d^{-1},\quad ada^{-1}=c
\end{eqnarray}
where $a$ generates $Z_3$ and $c$ and $d$ are generators of $Z_4\times Z_4$. Since $d$ can be expressed as $d=a^{-1}ca$, only two generators $a$ and $c$ are independent. Any group element $g$ of $\Delta(48)$ can be written as a product of powers of $a$, $c$ and $d$
\begin{eqnarray}
g=a^k c^m d^n
\end{eqnarray}
where $k=0,~1,~2$ and $m,~n=0,~1,~2,~3$. The structure of the $\Delta(48)$ group is somewhat complex. It has three $Z_2$ subgroups, sixteen $Z_3$ subgroups, six $Z_4$ subgroups and one $K_4\cong Z_2\times Z_2$ subgroup, which can be expressed in terms of the generators $a$, $c$ and $d$ as follows:
\begin{itemize}
  \item {$Z_2$ subgroups}
  \begin{equation}
  Z^{c^2}_2=\left\{1, c^2\right\},\quad Z^{d^2}_2=\left\{1, d^2\right\},\quad Z^{c^2d^2}_2=\left\{1, c^2d^2\right\}\,.
  \end{equation}
All the three $Z_3$ subgroups are conjugate to each other as follows
\begin{equation}
\label{eq:z2_conjugation}aZ^{c^2}_2a^{-1}=Z^{c^2d^2}_2,\qquad a^2Z^{c^2}_2a^{-2}=Z^{d^2}_2\,.
\end{equation}

\item{$Z_3$ subgroups}
  \begin{equation}
  Z^{(x,y)}_3=\left\{1, ac^{x}d^{y}, a^2c^{x-y}d^{x}\right\},\qquad x,y=0, 1, 2, 3\,.
  \end{equation}
  All these $Z_3$ subgroups are related to each other by group conjugation:
  \begin{eqnarray}
  \nonumber&&\left[c^{2x-y}d^{x+y}\right]Z^{(0,0)}_3\left[c^{2x-y}d^{x+y}\right]^{-1}=Z^{(x,y)}_3,\\
  \nonumber&&\left[ac^{-x-y}d^{x-2y}\right]Z^{(0,0)}_3\left[ac^{-x-y}d^{x-2y}\right]^{-1}=Z^{(x,y)}_3,\\
  \label{eq:z3_conjugation}&&\left[a^2c^{-x+2y}d^{-2x+y}\right]Z^{(0,0)}_3\left[a^2c^{-x+2y}d^{-2x+y}\right]^{-1}=Z^{(x,y)}_3\,.
  \end{eqnarray}

  \item{$Z_4$ subgroups}
  \begin{eqnarray}
  \nonumber&&\hskip-0.3in Z^{c}_4=\left\{1, c, c^2, c^3\right\},\quad Z^{d}_4=\left\{1, d, d^2, d^3\right\},\quad Z^{cd}_4=\left\{1, cd, c^2d^2, c^3d^3\right\}\,\\
  &&\hskip-0.3in Z^{cd^2}_4=\left\{1, cd^2, c^2, c^3d^2\right\},\quad Z^{c^2d}_4=\left\{1, c^2d, d^2, c^2d^3\right\},\quad Z^{cd^3}_4=\left\{1, cd^3, c^2d^2, c^3d\right\}\,.
  \end{eqnarray}
  The first three $Z_4$ subgroups are conjugate to each other via
  \begin{equation}
  aZ^{c}_4a^{-1}=Z^{cd}_4,\qquad a^2Z^{c}_4a^{-2}=Z^{d}_4\,.
  \end{equation}
The last three $Z_4$ subgroups are conjugate to each other as well
  \begin{equation}
  aZ^{cd^2}_4a^{-1}=Z^{cd^3}_4,\qquad a^2Z^{cd^2}_4a^{-2}=Z^{c^2d}_4\,.
  \end{equation}
  \item{$K_4$ subgroup}
  \begin{equation}
  K_4=\left\{1, c^2, d^2, c^2d^2\right\}\,.
  \end{equation}

\end{itemize}

$\Delta(48)$ has 8 conjugacy classes as follows:
\begin{eqnarray}
&&1C_1=\{1\},\nonumber\\
&&3C_2=\{c^2, d^2, c^2d^2\},\nonumber\\
&&3C_4=\{c, d, c^3d^3\},\nonumber\\
&&3C^{\prime}_4=\{c^3, d^3, cd\},\nonumber\\
&&3C^{\prime\prime}_4=\{cd^2, cd^3, c^2d^3 \},\nonumber\\
&&3C^{\prime\prime\prime}_4=\{c^2d, c^3d, c^3d^2\},\nonumber\\
&&16C_3=\{ac^xd^y|x,y=0,1,2,3 \},\nonumber\\
&&16C^{\prime}_3=\{a^2c^xd^y|x,y=0,1,2,3 \},
\end{eqnarray}
The number of irreducible representations of a group is equal to the number of its conjugacy class, therefore $\Delta(48)$ has eight irreducible representations: three one-dimensional representations $\mathbf{1}$, $\mathbf{1^{\prime}}$ and $\mathbf{1^{\prime\prime}}$, five three-dimensional representation $\mathbf{3}$, $\mathbf{\overline{3}}$, $\mathbf{3^{\prime}}$, $\mathbf{\overline{3}^{\prime}}$ and $\mathbf{\widetilde{3}}$. The representation matrices for the generators $a$ and $c$ in different irreducible representations are listed in Table~\ref{tab:rep_matrix}.
The character table of $\Delta(48)$ follows immediately as shown in Table~\ref{tab:character}. From this character table, the Kronecker products between different irreducible representations can be easily obtained:
\begin{eqnarray}
&&\hspace{-0.5cm}\mathbf{1^{\prime}}\otimes\mathbf{1^{\prime}}=\mathbf{1^{\prime\prime}}, ~~\mathbf{1^{\prime\prime}}\otimes\mathbf{1^{\prime\prime}}=\mathbf{1^{\prime}}, ~~\mathbf{1^{\prime}}\otimes\mathbf{1^{\prime\prime}}=\mathbf{1}, \nonumber\\
&&\hspace{-0.5cm}\mathbf{1^{\prime}} \otimes \mathbf{3}^{m}= \mathbf{3}^{m}, ~~\mathbf{1^{\prime}} \otimes \mathbf{\overline{3}}{}^{m}= \mathbf{\overline{3}}{}^{m},~~
\mathbf{1^{\prime}} \otimes \mathbf{\widetilde{3}}= \mathbf{\widetilde{3}}, ~~\mathbf{1^{\prime\prime}} \otimes \mathbf{3}^{m}= \mathbf{3}^{m}, ~~\mathbf{1^{\prime\prime}} \otimes \mathbf{\overline{3}}{}^{m}= \mathbf{\overline{3}}{}^{m}, ~~\mathbf{1^{\prime\prime}} \otimes \mathbf{\widetilde{3}}= \mathbf{\widetilde{3}}, \nonumber\\
&&\hspace{-0.5cm}\mathbf{3}^m\otimes\mathbf{3}^m=\mathbf{\overline{3}}{}^m_S\oplus\mathbf{\overline{3}}{}^m_A\oplus\mathbf{\widetilde{3}}, ~~\mathbf{\overline{3}}{}^m\otimes\mathbf{\overline{3}}{}^m=\mathbf{3}{}^m_S\oplus\mathbf{3}{}^m_A\oplus\mathbf{\widetilde{3}}, ~~\mathbf{3}{}^m\otimes\mathbf{\overline{3}}{}^m=\mathbf{1}\oplus\mathbf{1^{\prime}}\oplus\mathbf{1^{\prime\prime}}\oplus\mathbf{3}{}^n\oplus\mathbf{\overline{3}}{}^n, \nonumber\\
&&\hspace{-0.5cm}\mathbf{3}^m\otimes\mathbf{3}^n=\mathbf{3}{}^m\oplus\mathbf{3}{}^n\oplus\mathbf{\widetilde{3}},
~~\mathbf{\overline{3}}{}^m\otimes\mathbf{\overline{3}}{}^n=\mathbf{\overline{3}}{}^m\oplus\mathbf{\overline{3}}{}^n\oplus\mathbf{\widetilde{3}}, ~~\mathbf{3}^m\otimes\mathbf{\overline{3}}{}^n=\mathbf{3}{}^m\oplus\mathbf{\overline{3}}{}^n\oplus\mathbf{\widetilde{3}}, \nonumber\\
&&\hspace{-0.5cm}\mathbf{3}{}^m\otimes\mathbf{\widetilde{3}}=\mathbf{\overline{3}}{}^m\oplus\mathbf{3}^n\oplus\mathbf{\overline{3}}{}^n, ~~\mathbf{\overline{3}}{}^m\otimes\mathbf{\widetilde{3}}=\mathbf{3}{}^m\oplus\mathbf{3}^n\oplus\mathbf{\overline{3}}{}^n, ~~\mathbf{\widetilde{3}}\otimes\mathbf{\widetilde{3}}=\mathbf{1}\oplus\mathbf{1^{\prime}}\oplus\mathbf{1^{\prime\prime}}\oplus\mathbf{\widetilde{3}}_S\oplus\mathbf{\widetilde{3}}_A
\end{eqnarray}
where the superscript $m, n=0, 1$ ($m\neq n$) count the number of primes on their corresponding representation, and the subscript $S(A)$ denotes symmetric (antisymmetric) combinations.

%%%%%%%%%%%%%%%
\begin{table}[t!]
\begin{center}
\begin{tabular}{|c|c|c|}
\hline\hline
   & $a$ & $c$  \\\hline
$\mathbf{1}$ & 1 & 1 \\
$\mathbf{1^{\prime}}$ & $\omega$ & 1 \\
$\mathbf{1^{\prime\prime}}$ & $\omega^2$ & 1  \\

$\mathbf{3}$ &  $\left(\begin{array}{ccc}
 1 & 0 & 0 \\
 0 & \omega  & 0 \\
 0 & 0 & \omega ^2
\end{array}\right)$ &  $\displaystyle\frac{1}{3}
\left(\begin{array}{ccc}
 1 & 1-\sqrt{3} & 1+\sqrt{3} \\
 1+\sqrt{3} & 1 & 1-\sqrt{3} \\
 1-\sqrt{3} & 1+\sqrt{3} & 1
\end{array}\right)$  \\

  &       &       \\ [-0.15in]

$\mathbf{\overline{3}}$ &  $
\left(\begin{array}{ccc}
 1 & 0 & 0 \\
 0 & \omega ^2 & 0 \\
 0 & 0 & \omega
\end{array}\right)$ & $\displaystyle\frac{1}{3}
\left(\begin{array}{ccc}
 1 & 1-\sqrt{3} & 1+\sqrt{3} \\
 1+\sqrt{3} & 1 & 1-\sqrt{3} \\
 1-\sqrt{3} & 1+\sqrt{3} & 1
\end{array}\right)$ \\

  &       &          \\ [-0.15in]

$\mathbf{3^{\prime}}$ &  $\left(\begin{array}{ccc}
 1 & 0 & 0 \\
 0 & \omega  & 0 \\
 0 & 0 & \omega^2
\end{array}\right)$ & $\displaystyle\frac{1}{3}
\left(\begin{array}{ccc}
 -1+2 i & -1-i & -1-i \\
 -1-i & -1+2 i & -1-i \\
 -1-i & -1-i & -1+2 i
\end{array}\right)$  \\

  &       &        \\ [-0.15in]

$\mathbf{\overline{3}^{\prime}}$ &  $\left(\begin{array}{ccc}
 1 & 0 & 0 \\
 0 & \omega^2 & 0 \\
 0 & 0 & \omega
\end{array}\right)$ & $\displaystyle\frac{1}{3}\left(\begin{array}{ccc}
 -1-2 i & -1+i & -1+i \\
 -1+i & -1-2 i & -1+i \\
 -1+i & -1+i & -1-2 i
\end{array}\right)$ \\

  &       &         \\ [-0.15in]

$\mathbf{\widetilde{3}}$ &  $\left(\begin{array}{ccc}
 1 & 0 & 0 \\
 0 & \omega  & 0 \\
 0 & 0 & \omega^2
\end{array}\right)$ &  $\displaystyle\frac{1}{3}\left(\begin{array}{ccc}
 -1 & 2 & 2 \\
 2 & -1 & 2 \\
 2 & 2 & -1
\end{array}\right)$  \\ [0.25in]  \hline\hline

\end{tabular}
\caption{\label{tab:rep_matrix} The representation matrices for the $\Delta(48)$ generators $a$ and $c$ in our chosen basis, where $\omega$ is the cube root of unit $\omega=e^{2\pi i/3}$, and the representation matrix of $d$ is given by $d=a^{-1}ca$.}
\end{center}
\end{table}
%%%%%%%%%%%%%%%%%%%%

\begin{table}[h!]
\begin{center}
\begin{tabular}{|c| c| c| c| c| c| c| c| c| c|}
\hline\hline
& $1C_1$ & $3C_2$ & $3C_4$ & $3C^{\prime}_4$
& $3C^{\prime\prime}_4$ & $3C^{\prime\prime\prime}_4$ & $16C_3$
& $16C^{\prime}_3$  \\ \hline

$G$  &  1  &  $c^2$ &  $c$   &   $c^3$  &   $cd^2$  &  $c^2d$  &  $a$  &  $a^2$   \\ \hline

$\mathbf{1}$ & 1 & 1 & 1 & 1 & 1 & 1 & 1 & 1 \\

$\mathbf{1^{\prime}}$ & $1$ & $1$ & $1$ & $1$ & $1$ & $1$ & $\omega$ & $\omega^2$ \\

$\mathbf{1^{\prime\prime}}$ & $1$ & $1$ & $1$ & $1$ & $1$ & $1$ & $\omega^2$ & $\omega$ \\

$\mathbf{3}$ & 3 & $-1$ & 1 &  1 & $-1+2i$ &  $-1-2i$ &  0 & 0  \\

$\mathbf{\overline{3}}$ & 3 &   $-1$  & 1  &  1 &  $-1-2i$ &  $-1+2i$  &  0 & 0  \\

$\mathbf{3^{\prime}}$  &  3 &  $-1$  & $-1+2i$  & $-1-2i$ & 1 & 1 & 0 & 0  \\

$\mathbf{\overline{3}^{\prime}}$  & 3  & $-1$ &   $-1-2i$  &  $-1+2i$  & 1  & 1  & 0 & 0  \\

$\mathbf{\widetilde{3}}$  & 3  & 3  & $-1$  & $-1$ &   $-1$ &  $-1$  & 0  & 0
\\
\hline
\hline
\end{tabular}
\caption{\label{tab:character}Character table of $\Delta(48)$, where $c_1C_{c_2}$ denotes a conjugacy class with $c_1$ elements which have order $c_2$, and $G$ is a representative element of the $c_1C_{c_2}$ class.}
\end{center}
\end{table}

%%%%%%%%%%%%%%%%%%%%%%%%%%%%%%
%%%%%%%%%%%%%%%%%%%%%%%%%%%%%%

Starting from the representation matrix shown in Table~\ref{tab:rep_matrix}, we can straightforwardly calculate the Clebsch-Gordan (CG) coefficients of $\Delta(48)$ even though it is somewhat lengthy. In Tables~\ref{tab:CGcoefficients1} and \ref{tab:CGcoefficients2}, we present the complete set of CG coefficients of the $\Delta(48)$ group, and we use $\alpha_i$ to denote the elements of the first representation and $\beta_j$ to indicate those of the second representation of the product.

%%%%%%%%%%%%%%%%%%%%%%%%%%%%%%
%%%%%%%%%%%%%%%%%%%%%%%%%%%%%%
\begin{table}[]
\begin{center}
\begin{tabular}{| lL{7cm} | lL{7cm} |}
%%%%%%
\hline\hline
&\multicolumn{3}{c|}{}\\[-0.15in]
$\bullet$ &
\multicolumn{3}{l|}{
$\mathbf{1^{\prime}}\otimes\mathbf{1^{\prime}}=\mathbf{1^{\prime\prime}}$, $\mathbf{1^{\prime\prime}}\otimes\mathbf{1^{\prime\prime}}=\mathbf{1^{\prime}}$, $\mathbf{1^{\prime}}\otimes\mathbf{1^{\prime\prime}}=\mathbf{1}$ }
%%%%%%%%%%%
\\
%%%%%%%%%%%
&
\multicolumn{3}{l|}{\hspace{0.25cm}
$\mathbf{1^{\prime\prime}}, ~\mathbf{1^{\prime}}, ~\mathbf{1} \sim \alpha_1\beta_1$}%1x1

%%%%%%%%%%%%%%%%%%%%%%
\\[-0.15in]&\multicolumn{3}{c|}{}\\\hline&&&\\[-0.15in]
%%%%%%%%%%%%%%%%%%%%%%

%%%%%%%%%%%%
$\bullet$&
$\mathbf{1^{\prime}} \otimes \mathbf{3}^{m}= \mathbf{3}^{m}$, $\mathbf{1^{\prime\prime}} \otimes \mathbf{\overline{3}}{}^{m}= \mathbf{\overline{3}}{}^{m}$, $\mathbf{1^{\prime}} \otimes \mathbf{\widetilde{3}}= \mathbf{\widetilde{3}}$
&
$\bullet$&
$\mathbf{1^{\prime\prime}} \otimes \mathbf{3}^{m}= \mathbf{3}^{m}$, $\mathbf{1^{\prime}} \otimes \mathbf{\overline{3}}{}^{m}= \mathbf{\overline{3}}{}^{m}$, $\mathbf{1^{\prime\prime}} \otimes \mathbf{\widetilde{3}}= \mathbf{\widetilde{3}}$
%%%%%%%%%%%
\\
%%%%%%%%%%%
&
\hspace{0.25cm}
$\mathbf{3}^{m}, ~\mathbf{\overline{3}}{}^{m}, ~\mathbf{\widetilde{3}} \sim
\left(\begin{array}{ccc}
 \alpha_1 \beta_3 \\
 \alpha_1 \beta_1 \\
 \alpha_1 \beta_2
\end{array}\right)$
&
&
\hspace{0.25cm}
$\mathbf{3}^{m}, ~\mathbf{\overline{3}}{}^{m}, ~\mathbf{\widetilde{3}} \sim
\left(\begin{array}{ccc}
 \alpha_1 \beta_2 \\
 \alpha_1 \beta_3 \\
 \alpha_1 \beta_1
\end{array}\right)$
%%%%%%%%%%%%%%%%%%%%%%
\\[-0.15in]&&&\\\hline&&&\\[-0.15in]
%%%%%%%%%%%%%%%%%%%%%%
$\bullet$&
$\mathbf{3}\otimes\mathbf{3}=\mathbf{\overline{3}}{}_S\oplus\mathbf{\overline{3}}{}_A\oplus\mathbf{\widetilde{3}}$
&
$\bullet$&
$\mathbf{3}{}^{\prime}\otimes\mathbf{3}{}^{\prime}=\mathbf{\overline{3}}{}^{\prime}_S\oplus\mathbf{\overline{3}}{}^{\prime}_A\oplus\mathbf{\widetilde{3}}$
%%%%%%%%%%%
\\
%%%%%%%%%%%
&
\hspace{0.25cm}
$\begin{array}{l}
\mathbf{\overline{3}}{}_S\sim
\left(\begin{array}{c}
 2 \alpha_1 \beta_1-\alpha_2 \beta_3-\alpha_3 \beta_2 \\
 2 \alpha_2 \beta_2-\alpha_1 \beta_3-\alpha_3 \beta_1 \\
 2 \alpha_3 \beta_3-\alpha_1 \beta_2-\alpha_2 \beta_1
\end{array}\right) \\[-0.15in]\\
\mathbf{\overline{3}}{}_A\sim
\left(\begin{array}{c}
 \alpha_2 \beta_3-\alpha_3 \beta_2 \\
 \alpha_3 \beta_1-\alpha_1 \beta_3 \\
 \alpha_1 \beta_2-\alpha_2 \beta_1
\end{array}\right) \\[-0.15in]\\
\mathbf{\widetilde{3}}~\,\sim
\left(\begin{array}{c}
 \alpha_1 \beta_1+\alpha_2 \beta_3+\alpha_3 \beta_2 \\
 \alpha_3 \beta_3+\alpha_1 \beta_2+\alpha_2 \beta_1 \\
 \alpha_2 \beta_2+\alpha_1 \beta_3+\alpha_3 \beta_1
\end{array}\right)
\end{array}$%3x3
&
&
\hspace{0.25cm}
$\begin{array}{l}
\mathbf{\overline{3}}{}^{\prime}_S\sim
\left(\begin{array}{c}
 2 \alpha_1 \beta_1-\alpha_2 \beta_3-\alpha_3 \beta_2 \\
 2 \alpha_2 \beta_2-\alpha_1 \beta_3-\alpha_3 \beta_1 \\
 2 \alpha_3 \beta_3-\alpha_1 \beta_2-\alpha_2 \beta_1
\end{array}\right) \\[-0.15in]\\
\mathbf{\overline{3}}{}^{\prime}_A\sim
\left(\begin{array}{c}
 \alpha_2 \beta_3-\alpha_3 \beta_2 \\
 \alpha_3 \beta_1-\alpha_1 \beta_3 \\
 \alpha_1 \beta_2-\alpha_2 \beta_1
\end{array}\right)\\[-0.15in]\\
\mathbf{\widetilde{3}}~\,\sim
\left(\begin{array}{c}
 \alpha_1 \beta_1+\alpha_2 \beta_3+\alpha_3 \beta_2 \\
 \alpha_3 \beta_3+\alpha_1 \beta_2+\alpha_2 \beta_1 \\
 \alpha_2 \beta_2+\alpha_1 \beta_3+\alpha_3 \beta_1
\end{array}\right)
\end{array}$%3'x3'
%%%%%%%%%%%%%%%%%%%%%%
\\[-0.15in]&&&\\\hline&&&\\[-0.15in]
%%%%%%%%%%%%%%%%%%%%%%
$\bullet$&
$\mathbf{\overline{3}}\otimes\mathbf{\overline{3}}=\mathbf{3}{}_S\oplus\mathbf{3}{}_A\oplus\mathbf{\widetilde{3}}$
&
$\bullet$&
$\mathbf{\overline{3}}{}^{\prime}\otimes\mathbf{\overline{3}}{}^{\prime}=\mathbf{3}{}^{\prime}_S\oplus\mathbf{3}{}^{\prime}_A\oplus\mathbf{\widetilde{3}}$
%%%%%%%%%%%
\\
%%%%%%%%%%%
&
\hspace{0.25cm}
$\begin{array}{l}
\mathbf{3}{}_S\sim
\left(\begin{array}{c}
 2 \alpha_1 \beta_1-\alpha_2 \beta_3-\alpha_3 \beta_2 \\
 2 \alpha_2 \beta_2-\alpha_1 \beta_3-\alpha_3 \beta_1 \\
 2 \alpha_3 \beta_3-\alpha_1 \beta_2-\alpha_2 \beta_1
\end{array}\right)\\[-0.15in]\\
\mathbf{3}{}_A\sim
\left(\begin{array}{c}
 \alpha_2 \beta_3-\alpha_3 \beta_2 \\
 \alpha_3 \beta_1-\alpha_1 \beta_3 \\
 \alpha_1 \beta_2-\alpha_2 \beta_1
\end{array}\right)
\\[-0.15in]\\
\mathbf{\widetilde{3}}~\,\sim
\left(\begin{array}{c}
 \alpha_1 \beta_1+\alpha_2 \beta_3+\alpha_3 \beta_2 \\
 \alpha_2 \beta_2+\alpha_1 \beta_3+\alpha_3 \beta_1 \\
 \alpha_3 \beta_3+\alpha_1 \beta_2+\alpha_2 \beta_1
\end{array}\right)
\end{array}$%3bx3b
&
&
\hspace{0.25cm}
$\begin{array}{l}
\mathbf{3}{}^{\prime}_S\sim
\left(\begin{array}{c}
 2 \alpha_1 \beta_1-\alpha_2 \beta_3-\alpha_3 \beta_2 \\
 2 \alpha_2 \beta_2-\alpha_1 \beta_3-\alpha_3 \beta_1\\
 2 \alpha_3 \beta_3-\alpha_1 \beta_2-\alpha_2 \beta_1
\end{array}\right) \\[-0.15in]\\
\mathbf{3}{}^{\prime}_A\sim
\left(\begin{array}{c}
 \alpha_2 \beta_3-\alpha_3 \beta_2 \\
 \alpha_3 \beta_1-\alpha_1 \beta_3 \\
 \alpha_1 \beta_2-\alpha_2 \beta_1
\end{array}\right) \\[-0.15in]\\
\mathbf{\widetilde{3}}~\,\sim
\left(\begin{array}{c}
 \alpha_1 \beta_1+\alpha_2 \beta_3+\alpha_3 \beta_2 \\
 \alpha_2 \beta_2+\alpha_1 \beta_3+\alpha_3 \beta_1 \\
 \alpha_3 \beta_3+\alpha_1 \beta_2+\alpha_2 \beta_1
\end{array}\right)
\end{array}$%3'bx3'b
%%%%%%%%%%%
\\[-0.15in]&&&\\\hline&&&\\[-0.15in]
%%%%%%%%%%%
$\bullet$&
$\mathbf{3}\otimes\mathbf{\overline{3}}=\mathbf{1}\oplus\mathbf{1^{\prime}}\oplus\mathbf{1^{\prime\prime}}\oplus\mathbf{3}{}^{\prime}\oplus\mathbf{\overline{3}}{}^{\prime}$,%3x3b

$\mathbf{3}{}^{\prime}\otimes\mathbf{\overline{3}}{}^{\prime}=\mathbf{1}\oplus\mathbf{1^{\prime}}\oplus\mathbf{1^{\prime\prime}}\oplus\mathbf{3}\oplus\mathbf{\overline{3}}$%3'x3'b
&
$\bullet$&
$\mathbf{\widetilde{3}}\otimes\mathbf{\widetilde{3}}=\mathbf{1}\oplus\mathbf{1^{\prime}}\oplus\mathbf{1^{\prime\prime}}\oplus\mathbf{\widetilde{3}}_S\oplus\mathbf{\widetilde{3}}_A$%3tx3t
%%%%%%%%%%%
\\
%%%%%%%%%%%
&
\hspace{0.25cm}
$\begin{array}{l}
\mathbf{1}~~\sim ~
\alpha_1 \beta_1+\alpha_2 \beta_2+\alpha_3 \beta_3 \\[-0.15in]\\
\mathbf{1^{\prime}}\,~\sim ~
\alpha_1 \beta_3+\alpha_2 \beta_1+\alpha_3 \beta_2 \\[-0.15in]\\
\mathbf{1^{\prime\prime}}~\sim ~
\alpha_1 \beta_2+\alpha_2 \beta_3+\alpha_3 \beta_1 \\[-0.15in]\\
\mathbf{3}{}^{\prime},\mathbf{3}\sim \hspace{-0.1cm}
\left(\begin{array}{c}
 \alpha_1 \beta_1+\omega  \alpha_2 \beta_2+\omega ^2 \alpha_3 \beta_3 \\
 \alpha_3 \beta_2+\omega  \alpha_1 \beta_3+\omega ^2 \alpha_2 \beta_1 \\
 \alpha_2 \beta_3+\omega  \alpha_3 \beta_1+\omega ^2 \alpha_1 \beta_2
\end{array}\right)\hspace{-0.1cm}
\\[-0.15in]\\
\mathbf{\overline{3}}{}^{\prime},\mathbf{\overline{3}}\sim \hspace{-0.1cm}
\left(\begin{array}{c}
 \alpha_1 \beta_1+\omega  \alpha_3 \beta_3+\omega ^2 \alpha_2 \beta_2 \\
 \alpha_2 \beta_3+\omega  \alpha_1 \beta_2+\omega ^2 \alpha_3 \beta_1 \\
 \alpha_3 \beta_2+\omega  \alpha_2 \beta_1+\omega ^2 \alpha_1 \beta_3
\end{array}\right)\hspace{-0.1cm}
\end{array}$%3x3b&3'x3'b
&
&
\hspace{0.25cm}
$\begin{array}{l}
\mathbf{1}~~\sim ~
\alpha_1 \beta_1+\alpha_2 \beta_3+\alpha_3 \beta_2 \\[-0.15in]\\
\mathbf{1^{\prime}}\,~\sim ~
\alpha_3 \beta_3+\alpha_1 \beta_2+\alpha_2 \beta_1 \\[-0.15in]\\
\mathbf{1^{\prime\prime}}~\sim ~
\alpha_2 \beta_2+\alpha_1 \beta_3+\alpha_3 \beta_1 \\[-0.15in]\\
\mathbf{\widetilde{3}}_S\sim
\left(\begin{array}{c}
 2 \alpha_1 \beta_1-\alpha_2 \beta_3-\alpha_3 \beta_2 \\
 2 \alpha_3 \beta_3-\alpha_1 \beta_2-\alpha_2 \beta_1 \\
 2 \alpha_2 \beta_2-\alpha_1 \beta_3-\alpha_3 \beta_1
\end{array}\right)
\\[-0.15in]\\
\mathbf{\widetilde{3}}_A\sim
\left(\begin{array}{c}
 \alpha_2 \beta_3-\alpha_3 \beta_2 \\
 \alpha_1 \beta_2-\alpha_2 \beta_1 \\
 \alpha_3 \beta_1-\alpha_1 \beta_3
\end{array}\right)
\end{array}$%3tx3t
%%%%%%%%%%%
%%%%%%%%%%%
\\[-0.15in]&&&\\\hline\hline
%%%%%%%%%%%%%%%%%%%%%%
\end{tabular}

%\end{tabular}
%\end{supertabular}
\caption{\label{tab:CGcoefficients1} Clebsch-Gordan coefficients of $\Delta(48)$ in our basis: Part I.}
\end{center}
\end{table}

%%%%%%%%%%%%%%%%%%%%%%%%%%%%
%%%%%%%%%%%%%%%%%%%%%%%%%%%%
%%%%%%%%%%%%%%%%%%%%%%%%%%%%
\begin{table}[]
\begin{center}
\begin{tabular}{| lL{7cm} | lL{7cm} |}

\hline\hline&&&\\[-0.15in]

$\bullet$&
$\mathbf{3}\otimes\mathbf{3}{}^{\prime}=\mathbf{3}\oplus\mathbf{3}{}^{\prime}\oplus\mathbf{\widetilde{3}}$
&
$\bullet$&
$\mathbf{3}\otimes\mathbf{\overline{3}}{}^{\prime}=\mathbf{3}\oplus\mathbf{\overline{3}}{}^{\prime}\oplus\mathbf{\widetilde{3}}$
%%%%%%%%%%%
\\
%%%%%%%%%%%
&
\hspace{0.25cm}
$\begin{array}{l}
\mathbf{3}~\sim
\left(\begin{array}{c}
 \alpha_1 \beta_1+\omega  \alpha_2 \beta_3+\omega ^2 \alpha_3 \beta_2 \\
 \alpha_3 \beta_3+\omega  \alpha_1 \beta_2+\omega ^2 \alpha_2 \beta_1 \\
 \alpha_2 \beta_2+\omega  \alpha_3 \beta_1+\omega ^2 \alpha_1 \beta_3
\end{array}\right) \\[-0.15in]\\
\mathbf{3}{}^{\prime}\sim
\left(\begin{array}{c}
 \alpha_1 \beta_1+\omega  \alpha_3 \beta_2+\omega ^2 \alpha_2 \beta_3 \\
 \alpha_3 \beta_3+\omega  \alpha_2 \beta_1+\omega ^2 \alpha_1 \beta_2 \\
 \alpha_2 \beta_2+\omega  \alpha_1 \beta_3+\omega ^2\alpha_3 \beta_1
\end{array}\right)\\[-0.15in]\\
\mathbf{\widetilde{3}}~\sim
\left(\begin{array}{c}
 \alpha_1 \beta_1+\alpha_2 \beta_3+\alpha_3 \beta_2 \\
 \omega ^2(\alpha_1\beta_2+\alpha_2\beta_1+\alpha_3 \beta_3) \\
 \omega (\alpha_1\beta_3+\alpha_2 \beta_2+\alpha_3 \beta_1)
\end{array}\right)
\end{array}$%3x3'
&
&
\hspace{0.25cm}
$\begin{array}{l}
\mathbf{3}\,\sim
\left(\begin{array}{c}
 \alpha_1 \beta_1+\omega  \alpha_3 \beta_3+\omega ^2 \alpha_2 \beta_2 \\
 \alpha_3 \beta_2+\omega  \alpha_2 \beta_1+\omega ^2 \alpha_1 \beta_3 \\
 \alpha_2 \beta_3+\omega  \alpha_1 \beta_2+\omega ^2 \alpha_3 \beta_1
\end{array}\right)  \\[-0.15in]\\
\mathbf{\overline{3}}{}^{\prime}\sim
\left(\begin{array}{c}
 \alpha_1 \beta_1+\omega  \alpha_2 \beta_2+\omega ^2 \alpha_3 \beta_3 \\
 \alpha_2 \beta_3+\omega  \alpha_3 \beta_1+\omega ^2 \alpha_1 \beta_2 \\
 \alpha_3 \beta_2+\omega  \alpha_1 \beta_3+\omega ^2 \alpha_2 \beta_1
\end{array}\right) \\[-0.15in]\\
\mathbf{\widetilde{3}}\,\sim
\left(\begin{array}{c}
 \alpha_1 \beta_1+\alpha_2 \beta_2+\alpha_3 \beta_3 \\
 \omega (\alpha_1 \beta_3+\alpha_2 \beta_1+\alpha_3 \beta_2) \\
 \omega ^2(\alpha_1 \beta_2+\alpha_2 \beta_3+\alpha_3 \beta_1)
\end{array}\right)
\end{array}$%3x3'b
%%%%%%%%%%%%%%%%%%%%%%
\\[-0.15in]&&&\\\hline&&&\\[-0.15in]
%%%%%%%%%%%%%%%%%%%%%%
$\bullet$&
$\mathbf{\overline{3}}\otimes\mathbf{\overline{3}}{}^{\prime}=\mathbf{\overline{3}}\oplus\mathbf{\overline{3}}{}^{\prime}\oplus\mathbf{\widetilde{3}}$
&
$\bullet$&
$\mathbf{\overline{3}}\otimes\mathbf{3}^{\prime}=\mathbf{\overline{3}}\oplus\mathbf{3}{}^{\prime}\oplus\mathbf{\widetilde{3}}$
%%%%%%%%%%%
\\
%%%%%%%%%%%
&
\hspace{0.25cm}
$\begin{array}{l}
\mathbf{\overline{3}}\,\sim
\left(\begin{array}{c}
 \alpha_1 \beta_1+\omega  \alpha_3 \beta_2+\omega ^2 \alpha_2 \beta_3 \\
 \alpha_3 \beta_3+\omega  \alpha_2 \beta_1+\omega ^2 \alpha_1 \beta_2 \\
 \alpha_2 \beta_2+\omega  \alpha_1 \beta_3+\omega ^2\alpha_3 \beta_1
\end{array}\right)  \\[-0.15in]\\
\mathbf{\overline{3}}{}^{\prime}\sim
\left(\begin{array}{c}
 \alpha_1 \beta_1+\omega  \alpha_2 \beta_3+\omega ^2 \alpha_3 \beta_2 \\
 \alpha_3 \beta_3+\omega  \alpha_1 \beta_2+\omega ^2 \alpha_2 \beta_1 \\
 \alpha_2 \beta_2+\omega  \alpha_3 \beta_1+\omega ^2 \alpha_1 \beta_3
\end{array}\right)  \\[-0.15in]\\
\mathbf{\widetilde{3}}\,\sim
\left(\begin{array}{c}
 \alpha_1 \beta_1+\alpha_2 \beta_3+\alpha_3 \beta_2 \\
 \omega ^2(\alpha_2 \beta_2+\alpha_1 \beta_3+\alpha_3 \beta_1) \\
 \omega (\alpha_3 \beta_3+\alpha_1 \beta_2+\alpha_2 \beta_1)
\end{array}\right)
\end{array}$% 3bx3'b
&
&
\hspace{0.25cm}
$\begin{array}{l}
\mathbf{\overline{3}}\sim
\left(\begin{array}{c}
 \alpha_1 \beta_1+\omega  \alpha_2 \beta_2+\omega ^2 \alpha_3 \beta_3 \\
 \alpha_3 \beta_2+\omega  \alpha_1 \beta_3+\omega ^2 \alpha_2 \beta_1 \\
 \alpha_2 \beta_3+\omega  \alpha_3 \beta_1+\omega ^2 \alpha_1 \beta_2
\end{array}\right)  \\[-0.15in]\\
\mathbf{3}^{\prime}\sim
\left(\begin{array}{c}
 \alpha_1 \beta_1+\omega  \alpha_3 \beta_3+\omega ^2 \alpha_2 \beta_2 \\
 \alpha_2 \beta_3+\omega  \alpha_1 \beta_2+\omega ^2 \alpha_3 \beta_1 \\
 \alpha_3 \beta_2+\omega  \alpha_2 \beta_1+\omega ^2 \alpha_1 \beta_3
\end{array}\right)  \\[-0.15in]\\
\mathbf{\widetilde{3}}\sim
\left(\begin{array}{c}
 \alpha_1 \beta_1+\alpha_2 \beta_2+\alpha_3 \beta_3 \\
 \omega (\alpha_1 \beta_2+\alpha_2 \beta_3+\alpha_3 \beta_1) \\
 \omega ^2(\alpha_1 \beta_3+\alpha_2 \beta_1+\alpha_3 \beta_2)
\end{array}\right)
\end{array}$%3'x3t
%%%%%%%%%%%%%%%%%%%%%%
\\[-0.15in]&&&\\\hline&&&\\[-0.15in]
%%%%%%%%%%%%%%%%%%%%%%
$\bullet$&
$\mathbf{3}\otimes\mathbf{\widetilde{3}}=\mathbf{\overline{3}}\oplus\mathbf{3}{}^{\prime}\oplus\mathbf{\overline{3}}{}^{\prime}$
&
$\bullet$&
$\mathbf{3}{}^{\prime}\otimes\mathbf{\widetilde{3}}=\mathbf{3}\oplus\mathbf{\overline{3}}\oplus\mathbf{\overline{3}}{}^{\prime}$
%%%%%%%%%%%
\\
%%%%%%%%%%%
&
\hspace{0.25cm}
$\begin{array}{l}
\mathbf{\overline{3}}~\sim
\left(\begin{array}{c}
 \alpha_1 \beta_1+\alpha_2 \beta_3+\alpha_3\beta_2 \\
 \alpha_2 \beta_2+\alpha_1 \beta_3+\alpha_3\beta_1 \\
 \alpha_3 \beta_3+\alpha_1 \beta_2+\alpha_2\beta_1
\end{array}\right)  \\[-0.15in]\\
\mathbf{3}{}^{\prime}\sim
\left(\begin{array}{c}
 \alpha_1 \beta_1+\omega  \alpha_2 \beta_3+\omega ^2 \alpha_3 \beta_2 \\
 \alpha_2 \beta_1+\omega  \alpha_3 \beta_3+\omega ^2 \alpha_1 \beta_2 \\
 \alpha_3 \beta_1+\omega  \alpha_1 \beta_3+\omega ^2 \alpha_2 \beta_2
\end{array}\right)  \\[-0.15in]\\
\mathbf{\overline{3}}{}^{\prime}\sim
\left(\begin{array}{c}
 \alpha_1 \beta_1+\omega  \alpha_3 \beta_2+\omega ^2 \alpha_2 \beta_3 \\
 \alpha_3 \beta_1+\omega  \alpha_2 \beta_2+\omega ^2 \alpha_1 \beta_3 \\
 \alpha_2 \beta_1+\omega  \alpha_1 \beta_2+\omega ^2 \alpha_3 \beta_3
\end{array}\right)
\end{array}$%3x3t
&
&
\hspace{0.25cm}
$\begin{array}{l}
\mathbf{3}\,\sim
\left(\begin{array}{c}
 \alpha_1 \beta_1+\omega  \alpha_2 \beta_3+\omega ^2 \alpha_3 \beta_2 \\
 \alpha_2 \beta_1+\omega  \alpha_3 \beta_3+\omega ^2 \alpha_1 \beta_2 \\
 \alpha_3 \beta_1+\omega  \alpha_1 \beta_3+\omega ^2 \alpha_2 \beta_2
\end{array}\right)  \\[-0.15in]\\
\mathbf{\overline{3}}\,\sim
\left(\begin{array}{c}
 \alpha_1 \beta_1+\omega  \alpha_3 \beta_2+\omega ^2 \alpha_2 \beta_3 \\
 \alpha_3 \beta_1+\omega  \alpha_2 \beta_2+\omega ^2 \alpha_1 \beta_3 \\
 \alpha_2 \beta_1+\omega  \alpha_1 \beta_2+\omega ^2 \alpha_3 \beta_3
\end{array}\right)  \\[-0.15in]\\
\mathbf{\overline{3}}{}^{\prime}\sim
\left(\begin{array}{c}
 \alpha_1 \beta_1+\alpha_2 \beta_3+\alpha_3 \beta_2 \\
 \alpha_2 \beta_2+\alpha_1 \beta_3+\alpha_3 \beta_1 \\
 \alpha_3 \beta_3+\alpha_1 \beta_2+\alpha_2 \beta_1
\end{array}\right)
\end{array}$%3'x3t
%%%%%%%%%%%%%%%%%%%%%%
\\[-0.15in]&&&\\\hline&&&\\[-0.15in]
%%%%%%%%%%%%%%%%%%%%%%
$\bullet$&
$\mathbf{\overline{3}}\otimes\mathbf{\widetilde{3}}=\mathbf{3}\oplus\mathbf{3}{}^{\prime}\oplus\mathbf{\overline{3}}{}^{\prime}$
&
$\bullet$&
$\mathbf{\overline{3}}{}^{\prime}\otimes\mathbf{\widetilde{3}}=\mathbf{3}\oplus\mathbf{\overline{3}}\oplus\mathbf{3}{}^{\prime}$%3'px3t
%%%%%%%%%%%
\\
%%%%%%%%%%%
&
\hspace{0.25cm}
$\begin{array}{l}
\mathbf{3}~\sim
\left(\begin{array}{c}
 \alpha_1 \beta_1+\alpha_2 \beta_2+\alpha_3 \beta_3 \\
 \alpha_1 \beta_2+\alpha_2 \beta_3+\alpha_3 \beta_1 \\
 \alpha_1 \beta_3+\alpha_2 \beta_1+\alpha_3 \beta_2
\end{array}\right)  \\[-0.15in]\\
\mathbf{3}{}^{\prime}\sim
\left(\begin{array}{c}
 \alpha_1 \beta_1+\omega  \alpha_2 \beta_2+\omega ^2 \alpha_3 \beta_3 \\
 \alpha_3 \beta_1+\omega  \alpha_1 \beta_2+\omega ^2\alpha_2 \beta_3 \\
 \alpha_2 \beta_1+\omega  \alpha_3 \beta_2+\omega ^2\alpha_1 \beta_3
\end{array}\right)  \\[-0.15in]\\
\mathbf{\overline{3}}{}^{\prime}\sim
\left(\begin{array}{c}
 \alpha_1 \beta_1+\omega  \alpha_3 \beta_3+\omega ^2 \alpha_2 \beta_2 \\
 \alpha_2 \beta_1+\omega  \alpha_1 \beta_3+\omega ^2 \alpha_3 \beta_2 \\
 \alpha_3 \beta_1+\omega  \alpha_2 \beta_3+\omega ^2 \alpha_1 \beta_2
\end{array}\right)
\end{array}$%3bx3t
&
&
\hspace{0.25cm}
$\begin{array}{l}
\mathbf{3}\,\sim
\left(\begin{array}{c}
 \alpha_1 \beta_1+\omega  \alpha_2 \beta_2+\omega ^2 \alpha_3 \beta_3 \\
 \alpha_3 \beta_1+\omega  \alpha_1 \beta_2+\omega ^2\alpha_2 \beta_3 \\
 \alpha_2 \beta_1+\omega  \alpha_3 \beta_2+\omega ^2\alpha_1 \beta_3
\end{array}\right)
\\[-0.15in]\\
\mathbf{\overline{3}}\,\sim
\left(\begin{array}{c}
 \alpha_1 \beta_1+\omega  \alpha_3 \beta_3+\omega ^2 \alpha_2 \beta_2 \\
 \alpha_2 \beta_1+\omega  \alpha_1 \beta_3+\omega ^2 \alpha_3 \beta_2 \\
 \alpha_3 \beta_1+\omega  \alpha_2 \beta_3+\omega ^2 \alpha_1 \beta_2
\end{array}\right)  \\[-0.15in]\\
\mathbf{3}{}^{\prime}\sim
\left(\begin{array}{c}
 \alpha_1 \beta_1+\alpha_2 \beta_2+\alpha_3 \beta_3 \\
 \alpha_1 \beta_2+\alpha_2 \beta_3+\alpha_3 \beta_1 \\
 \alpha_1 \beta_3+\alpha_2 \beta_1+\alpha_3 \beta_2
\end{array}\right)
\end{array}$%3'px3t
%%%%%%%%%%%
\\[-0.15in]&&&\\\hline\hline
%%%%%%%%%%%

\end{tabular}
%\end{supertabular}
\caption{\label{tab:CGcoefficients2} Clebsch-Gordan coefficients of $\Delta(48)$ in our basis: Part II.}
\end{center}
\end{table}

\section{\label{sec:appendix_B}Vacuum invariant under the remnant symmetries }

The mixing pattern predicted in cases II, IV, VI, and VIII is a very interesting new mixing texture. Its predictions for the lepton mixing angles and CP violation phases are displayed in Table~\ref{tab:caseII_VI_and_IV_VIII}. Excellent agreement with the present experimental data can be achieved, and in particular the lepton CP phases do not take regular values such as $0$, $\pm\pi/2$ or $\pi$ anymore. The best-fit value of the Dirac CP phase $\delta_{\rm{CP}}$ is given by $\left|\sin\delta_{\rm{CP}}(\theta_{\text{bf}})\right|=0.725$ which is compatible with the present $1\sigma$ preferred range $0.9\pi\leq\delta_{\rm{CP}}\leq2.0\pi$~\cite{GonzalezGarcia:2012sz}. In the charged lepton diagonal basis, this mixing pattern is dictated by the remnant symmetry $G_{\nu}\cong Z^{c^2}_2\times H^{\nu}_{\rm{CP}}$ in the neutrino sector. We would like to spontaneously break the full symmetry group $\Delta(48)\rtimes H_{\rm{CP}}$ down to $G_{\nu}$ in the neutrino sector in order to derive this mixing pattern. Hence it is convenient to list the most general form of the VEVs that flavon fields in different representations of $\Delta(48)$ can take and which leave $G_{\nu}$ invariant. As shown in Table~\ref{tab:constraints_by_GCP}, the concrete value of the remnant CP symmetry $H^{\nu}_{\rm{CP}}$ depends on which $\Delta(48)$ triplets ($\mathbf{3}$, $\mathbf{3^{\prime}}$, $\mathbf{\overline{3}}$ or $\mathbf{\overline{3}^{\prime}}$) the three generations of the left-handed lepton doublets $l$ are embedded into. Here we assign $l$ to be a triplet $\mathbf{3}$ without loss of generality. The vacuum invariant under the residual symmetry in the relevant four cases II, IV, VI and VIII are of the following forms.
\begin{itemize}
  \item {Case II:~ $H^{\nu}_{\rm{CP}}=\left\{\rho(d)X(\mathfrak{h_1}), \rho(c^2d)X(\mathfrak{h_1})\right\}$}

      \begin{equation}
      \label{eq:vacuum_caseII}\begin{array}{lc}

        \varphi_{\nu}\sim\mathbf{3}~~ \text{or}~~ \varphi_{\nu}\sim\mathbf{\overline{3}^{\prime}}:  &  \langle\varphi_{\nu}\rangle=e^{-i\frac{\pi}{4}}\left(\begin{array}{c}
         1\\
         1\\
         1
        \end{array}\right) v,\quad v\in\mathbb{R},  \\ [0.30in]

        \varphi_{\nu}\sim\mathbf{3^{\prime}}~~ \text{or}~~ \varphi_{\nu}\sim\mathbf{\overline{3}}:  &  \langle\varphi_{\nu}\rangle=e^{i\frac{\pi}{4}}\left(\begin{array}{c}
         1\\
         1\\
         1
        \end{array}\right) v^{\prime},\quad v^{\prime}\in\mathbb{R},  \\[0.30in]

         \varphi_{\nu}\sim\mathbf{\widetilde{3}}:  &  \langle\varphi_{\nu}\rangle=\left(\begin{array}{c}
        u+i\widetilde{v}\\
       w+i\left(\sqrt{3}\;w+\widetilde{v}\right)\\
       -u-w+i\left(\sqrt{3}\;u+\sqrt{3}\;w+\widetilde{v}\right)
        \end{array}\right) ,\\ [0.25in]

         & \hskip-1.1in u\in\mathbb{R},\widetilde{v}\in\mathbb{R}, w\in\mathbb{R}\,.

      \end{array}
      \end{equation}

     Note that here $v$, $v^{\prime}$, $\widetilde{v}$, $u$ and $w$ are arbitrary real parameters. For the special values $u=w=0$, the VEV of $\varphi_{\nu}\sim\mathbf{\widetilde{3}}$ reduces to
     \begin{equation}
      \langle\varphi_{\nu}\rangle=\left(\begin{array}{c}
        i\\
       i\\
       i
       \end{array}\right) \widetilde{v},\qquad \widetilde{v}\in\mathbb{R}\,.
     \end{equation}

  \item {Case IV:~$H^{\nu}_{\rm{CP}}=\left\{\rho(d^3)X(\mathfrak{h_1}), \rho(c^2d^3)X(\mathfrak{h_1})\right\}$}

   \begin{equation}
      \label{eq:vacuum_caseIV}\begin{array}{lc}

       \varphi_{\nu}\sim\mathbf{3}~~ \text{or}~~ \varphi_{\nu}\sim\mathbf{\overline{3}^{\prime}}:  &  \langle\varphi_{\nu}\rangle=e^{i\frac{\pi}{4}}\left(\begin{array}{c}
         1\\
         1\\
         1
        \end{array}\right) v,\quad v\in\mathbb{R},  \\ [0.30in]

        \varphi_{\nu}\sim\mathbf{3^{\prime}}~~ \text{or}~~ \varphi_{\nu}\sim\mathbf{\overline{3}}:  &  \langle\varphi_{\nu}\rangle=e^{-i\frac{\pi}{4}}\left(\begin{array}{c}
         1\\
         1\\
         1
        \end{array}\right) v^{\prime},\quad v^{\prime}\in\mathbb{R},  \\[0.30in]

         \varphi_{\nu}\sim\mathbf{\widetilde{3}}:  &  \langle\varphi_{\nu}\rangle=\left(\begin{array}{c}
        u+i\widetilde{v}\\
       w+i\left(\sqrt{3}\;w+\widetilde{v}\right)\\
       -u-w+i\left(\sqrt{3}\;u+\sqrt{3}\;w+\widetilde{v}\right)
        \end{array}\right) ,\\ [0.25in]

         & \hskip-1.1in u\in\mathbb{R},\widetilde{v}\in\mathbb{R}, w\in\mathbb{R}\,.

      \end{array}
   \end{equation}

  Notice that the VEV of $\varphi_{\nu}\sim\mathbf{\widetilde{3}}$ takes the same form as the one of case II.

  \item {Case VI:~$H^{\nu}_{\rm{CP}}=\left\{\rho(cd)X(\mathfrak{h_1}), \rho(c^3d)X(\mathfrak{h_1})\right\}$}

      \begin{equation}
      \label{eq:vacuum_caseVI}\begin{array}{lc}

       \varphi_{\nu}\sim\mathbf{3}~~ \text{or}~~ \varphi_{\nu}\sim\mathbf{3^{\prime}}:  &  \langle\varphi_{\nu}\rangle=e^{-i\frac{\pi}{4}}\left(\begin{array}{c}
         1\\
         1\\
         1
        \end{array}\right) v,\quad v\in\mathbb{R},  \\ [0.30in]

        \varphi_{\nu}\sim\mathbf{\overline{3}}~~ \text{or}~~ \varphi_{\nu}\sim\mathbf{\overline{3}^{\prime}}:  &  \langle\varphi_{\nu}\rangle=e^{i\frac{\pi}{4}}\left(\begin{array}{c}
         1\\
         1\\
         1
        \end{array}\right) \overline{v},\quad \overline{v}\in\mathbb{R},  \\[0.30in]

         \varphi_{\nu}\sim\mathbf{\widetilde{3}}:  &  \langle\varphi_{\nu}\rangle=\left(\begin{array}{c}
           u+i\widetilde{v}\\
           w+i\left(-\sqrt{3}\;w+\widetilde{v}\right)\\
           -u-w+i\left(-\sqrt{3}\;u-\sqrt{3}\;w+\widetilde{v}\right)
           \end{array}\right) ,\\ [0.25in]

         & \hskip-1.1in u\in\mathbb{R},\widetilde{v}\in\mathbb{R}, w\in\mathbb{R}\,.

      \end{array}
   \end{equation}

For the case of $u=w=0$, the VEV of $\varphi_{\nu}\sim\mathbf{\widetilde{3}}$ becomes
\begin{equation}
\langle\varphi_{\nu}\rangle=\left(\begin{array}{c}
           i\\
           i\\
           i
           \end{array}\right) \widetilde{v},\qquad \widetilde{v}\in\mathbb{R}\,.
\end{equation}

  \item {Case VIII:~$H^{\nu}_{\rm{CP}}=\left\{\rho(cd^3)X(\mathfrak{h_1}), \rho(c^3d^3)X(\mathfrak{h_1})\right\}$}

       \begin{equation}
      \label{eq:vacuum_caseVIII}\begin{array}{lc}

       \varphi_{\nu}\sim\mathbf{3}~~ \text{or}~~ \varphi_{\nu}\sim\mathbf{3^{\prime}}:  &  \langle\varphi_{\nu}\rangle=e^{i\frac{\pi}{4}}\left(\begin{array}{c}
         1\\
         1\\
         1
        \end{array}\right) v,\quad v\in\mathbb{R},  \\ [0.30in]

        \varphi_{\nu}\sim\mathbf{\overline{3}}~~ \text{or}~~ \varphi_{\nu}\sim\mathbf{\overline{3}^{\prime}}:  &  \langle\varphi_{\nu}\rangle=e^{-i\frac{\pi}{4}}\left(\begin{array}{c}
         1\\
         1\\
         1
        \end{array}\right) \overline{v},\quad \overline{v}\in\mathbb{R},  \\[0.30in]

         \varphi_{\nu}\sim\mathbf{\widetilde{3}}:  &  \langle\varphi_{\nu}\rangle=\left(\begin{array}{c}
           u+i\widetilde{v}\\
           w+i\left(-\sqrt{3}\;w+\widetilde{v}\right)\\
           -u-w+i\left(-\sqrt{3}\;u-\sqrt{3}\;w+\widetilde{v}\right)
           \end{array}\right) ,\\ [0.25in]

         & \hskip-1.1in u\in\mathbb{R},\widetilde{v}\in\mathbb{R}, w\in\mathbb{R}\,,

      \end{array}
   \end{equation}

where the VEV of $\varphi_{\nu}\sim\mathbf{\widetilde{3}}$ is of the same form as that of case VI.

\end{itemize}

\end{appendix}

\end{document}